\documentclass[aps,twocolumn,floats,prx,nofootinbib,superscriptaddress,10pt]{revtex4-2}

\usepackage{hyperref}

\usepackage[dvips]{graphicx} %
\usepackage{graphicx,amsmath,amsfonts,amssymb,slashed}
\usepackage{wasysym}
\usepackage{graphicx}
\usepackage{natbib}
\usepackage[usenames,dvipsnames]{xcolor}

\usepackage{soul}

\definecolor{RedWine}{rgb}{0.743,0,0}
\definecolor{RoyalBlue}{rgb}{0.25,.41,.88}

\setstcolor{Blue}

\newcommand{\jcap}{JCAP}
\newcommand{\mnras}{MNRAS}
\newcommand{\aj}{AJ}
\newcommand{\pasp}{PASP}
\newcommand{\apjl}{ApJL}

\newcommand{\VEV}[1]{\left<{1}\right>}

\newcommand{\bea}{\begin{eqnarray}}
\newcommand{\eea}{\end{eqnarray}}
\newcommand{\be}{\begin{equation}}
\newcommand{\ee}{\end{equation}}

\begin{document}

\title{Impact of a Midband Gravitational Wave Experiment On Detectability of Cosmological Stochastic Gravitational Wave Backgrounds}

\author{Barry C. Barish}
\email{E-mail: barry.barish@ucr.edu}
\affiliation{LIGO Laboratory, California Institute of Technology, Pasadena, California 91125, USA}
\affiliation{Department of Physics and Astronomy, University of California Riverside, Riverside, CA 90217, USA}
\author{Simeon Bird}
\email{E-mail: sbird@ucr.edu}
\affiliation{Department of Physics and Astronomy, University of California Riverside, Riverside, CA 90217, USA}
\author{Yanou Cui}
\email{E-mail: yanou.cui@ucr.edu}
\affiliation{Department of Physics and Astronomy, University of California Riverside, Riverside, CA 90217, USA}

\begin{abstract}
We make forecasts for the impact a future ``midband'' space-based gravitational wave experiment, most sensitive to $10^{-2}- 10 $ Hz, could have on potential detections of cosmological stochastic gravitational wave backgrounds (SGWBs). Specific proposed midband experiments considered are TianGo, B-DECIGO and AEDGE. We propose a combined power-law integrated sensitivity (CPLS) curve combining GW experiments over different frequency bands, which shows the midband improves sensitivity to SGWBs by up to two orders of magnitude at $10^{-2} - 10$ Hz. We consider GW emission from cosmic strings and phase transitions as benchmark examples of cosmological SGWBs. We explicitly model various astrophysical SGWB sources, most importantly from unresolved black hole mergers. Using Markov Chain Monte Carlo, we demonstrated that midband experiments can, when combined with LIGO A+ and LISA, significantly improve sensitivities to cosmological SGWBs and better separate them from astrophysical SGWBs.
In particular, we forecast that a midband experiment improves sensitivity to cosmic string tension $G\mu$ by up to a factor  of $10$, driven by improved component separation from astrophysical sources. For phase transitions, a midband experiment can detect signals peaking at $0.1 - 1$ Hz, which for our fiducial model corresponds to early Universe temperatures of $T_*\sim 10^4 - 10^6$ GeV,  generally beyond the reach of LIGO and LISA. The midband closes an energy gap and better captures characteristic spectral shape information. It thus substantially improves measurement of the properties of phase transitions at lower energies of $T_* \sim O(10^3)$ GeV, potentially relevant to new physics at the electroweak scale, whereas in this energy range LISA alone will detect an excess but not effectively measure the phase transition parameters. Our modeling code and chains are publicly available.\footnote{\url{https://github.com/sbird/grav_midband}}
\end{abstract}
%\date{\today}

\maketitle

\section{Introduction}
{\small LIGO} recently ushered in the era of gravitational wave (GW) physics by detecting a binary black hole merger \cite{Abbott:2016}. Around 2034, ground-based detectors are expected to be supplemented by the space-based {\small LISA}  satellite constellation.
{\small LISA}, with an interferometer arm length of $2.5 \times 10^9$ m, is most sensitive to GWs in the frequency range $10^{-5}$ to $10^{-2}$ Hz, with some sensitivity from $10^{-7}$ to $10$ Hz \cite{LISASRD}. The ground based LIGO, limited by low-frequency oscillations of the Earth, is sensitive to signals in the $10-5000$ Hz range \cite{Abbott:2009ws}. There is thus a frequency gap between the two detectors, from $10^{-2} - 10$ Hz, known as the \textit{midband}. Several GW experiments have recently been proposed to close this gap, based on laser- or atomic-interferometer techniques, including B-DECIGO, TianGo, TianQin, MAGIS and AEDGE \cite{DECIGO2011,TianQin, TianGo, AEDGE, Graham:2017pmn, Graham:2016plp}.

GW detectors are sensitive not just to resolved sources, but also to unresolved coherent stochastic gravitational wave backgrounds (SGWB).
An important source of SGWB are cosmological signals. Among the many well-motivated cosmogenic SGWB sources (for a review see e.g. \cite{Caprini:2018}), we will focus on two well-motivated examples: GW emission from cosmic strings and phase transitions. Discovering such cosmogenic SGWBs would elucidate the dynamics of the very early Universe and reveal new particle physics beyond the Standard Model (SM).

Cosmic strings \cite{Vilenkin:1981bx, Turok:1984cn,Vachaspati:1984gt, Burden:1985md,Olum:1999sg,Moore:2001px}, are one-dimensional topological defects which can arise from e.g. superstring theory or a $U(1)$ symmetry breaking in the early Universe \cite{Nielsen:1973cs, Kibble:1976sj, Jackson:2004zg,Tye:2005fn, Dubath:2007mf, Figueroa:2020}. Phase transitions arise from first-order electroweak symmetry breaking or a dark sector \cite{Caprini:2009fx, Schwaller:2015tja, Helmboldt:2019pan,Hall:2019ank}. In both scenarios, the observation of GWs serves as a probe of other potential new physics, such as those related to  dark matter, mechanisms addressing the long-standing matter-antimatter puzzle, unification of forces and the Universe's dynamics prior to big bang nucleosynthesis \cite{Cohen:1990it, Anderson:1991zb, Cui:2017ufi, Caldwell:2018giq, Cui:2018, Cui:2019kkd, Chang:2019mza, Dror:2019syi, Buchmuller:2019gfy, Gouttenoire:2019kij, Gouttenoire:2019rtn, Hall:2019ank, Dev:2019njv}.

Both sources are speculative at present, yet are well-motivated and represent fairly minimal extensions to the SM of particle physics. They can also produce strong signals that are within the reach of current/near future GW detectors and are amongst the primary targets of SGWB searches by the {\small LIGO} and {\small LISA} collaborations \cite{Abbott:2009ws, Abbott:2017mem, Caprini:2019egz, Auclair:2019wcv}.
Intriguingly, the NANOGrav pulsar timing experiment recently detected an excess signal \cite{NANOGRAV}. This signal could be explained by a SGWB originating from cosmic strings or a dark phase transition \cite{Ellis:2020ena, Addazi:2020zcj,Ratzinger:2020koh,Blasi:2020mfx, Buchmuller:2020lbh, Samanta:2020cdk, Nakai:2020oit, Neronov:2020qrl}, although the lack of a quadrupole correlation prevents a claim of GW detection with current data.

The typical broadband nature of SGWB signatures makes it feasible to boost sensitivity by simultaneously utilizing data from multiple experiments. Here we investigate the potential of a future midband experiment, taking TianGo and B-DECIGO as examples, to improve sensitivities to cosmological SGWB signals from cosmic strings and phase transitions. We pay particular attention to potential astrophysical sources of a SGWB, as one of the possible benefits of a midband experiment is breaking degeneracies between astrophysical and cosmological signals. Our analysis is at the power spectrum level, but a full analysis of the astrophysical sources would make use of the information available in higher order statistics. (e.g.~\cite{Smith:2017vfk,Bartolo:2018qqn,Ginat:2019aed}) We create simulated signals with astrophysical SGWB sources and both with and without a cosmological source component. Using Markov Chain Monte Carlo (MCMC), we forecast satellite mission sensitivities to cosmogenic SGWBs.

Different SGWB sources produce signals with different power law indices, allowing component separation (e.g.~\cite{Cui:2018rwi}). Bayesian stochastic background detection techniques have been considered by Refs.~\cite{Romano:2017, Romano:2019}. Various separation techniques have also been considered for {\small LISA} \citep{Cutler:2006, Pan:2019, Pieroni:2020, Boileau:2020rpg}. Ref.~\cite{Sedda:2019} mentioned that a midband experiment could improve detectability of a SGWB from a phase transition near the electroweak symmetry breaking scale of $\sim 100$ GeV, assuming that the SGWB from lower redshift black hole mergers could be completely subtracted. Here we improve these estimates by explicitly modeling relevant astrophysical and cosmological backgrounds and using Bayesian techniques to marginalise the amplitude of each one. This allows us to compute the extent to which a midband experiment improves cosmological detectability.

We first propose a generalization of power-law integrated sensitivity curves \cite{Thrane:2013oya}, commonly derived for individual experiments, to combinations of multiple experiments covering different frequency bands. We then present our likelihood analysis and results with benchmark cosmological and astrophysical source models, demonstrating ways that a midband GW experiment can boost the discovery prospect for a cosmological SGWB. Finally we summarize and conclude.

\section{Combined Sensitivity Curve Incorporating Midband Data}

Below, we demonstrate how midband data would enhance sensitivity to cosmological SGWBs when marginalising over astrophysical sources. Here we present an analytical approach to illustrate this improvement, the combined power-law sensitivity curve. The discussion here focuses on distinguishing an SGWB from experimental noise, and does not yet address issues of separability into astrophysical and cosmological sources.

\subsection{Combined Power-Law Sensitivity to SGWB}

An individual GW experiment has an effective characteristic strain noise amplitude $h_n(f)$ and an effective strain noise spectral density $S_n(f)=h_n^2(f)/f$\footnote{See Ref.~\cite{Moore:2015} for a discussion of the different GW sensitivity conventions in use.}. For SGWB searches the energy density sensitivity,
\begin{equation}
 \Omega_s(f)\equiv\frac{4\pi^2}{3H_0^2}f^3 S_n(f)\,,
\end{equation}
is usually introduced to characterize noise level. $H_0$ is the current-day Hubble expansion rate (we assume $H_0 = 70$ km/s/Mpc).
The corresponding GW energy density for signals is defined as \cite{Caprini:2019}
\be
\Omega_{\rm GW}(f)\equiv\frac{1}{\rho_c}\frac{d\rho_{\rm GW}}{d\ln f}=\frac{1}{3H_0^2M_p^2}\frac{d\rho_{\rm GW}}{d\ln f},
\ee
where $M_p$ is the reduced Planck mass. $\Omega_{\rm GW}(f)$ can be detected with signal to noise ratio (SNR) $\rm SNR > 1$ if $\Omega_s(f) < \Omega_{\rm GW}(f)$. Thus, $\Omega_s(f)$ is an estimate of the sensitivity to a SGWB signal in a single narrow frequency bin. However, in practice the sensitivity to a SGWB will be much better: the signal is generally expected to be spread over a wide frequency range and static throughout the observational time window. A more realistic estimate of SNR integrates over all observations and scales as $\sqrt{T\Delta f}$ \cite{Thrane:2013oya} for observation time $T$ and frequency $f$.
For a frequency-dependent signal, SNR is defined as
\be
{\rm SNR}(f, B)=\sqrt{T\int^{f_{\rm max}}_{f_{\rm min}}df\left(\frac{\Omega_{\rm GW}(f, B)}{\Omega_s(f)} \right)^2}. \label{eq: SNR0}
\ee

Ref.~\cite{Thrane:2013oya} introduced a modification, the integrated power-law sensitivity (PLS) curve, which describes the sensitivity to a general signal with a piece-wise power-law dependence on $f$.
For a given power law signal $\Omega_{\rm GW}(f, B)= (f/f_\mathrm{ref})^B$, with index $B$ and reference frequency $f_\mathrm{ref}$, the sensitivity $\Omega_s(f)$ is defined so that SNR$(f, B)$ from Eq.~\ref{eq: SNR0} is equal to the target threshold $\rm SNR^{\rm thr}$. The PLS for $\Omega(f)$ is then defined by maximising over $B$
\begin{equation}
    \Omega_{PLS}(f) = \mathrm{max}_B\left[ \left(\frac{f}{f_\mathrm{ref}}\right)^B \frac{\mathrm{SNR}^{\rm thr}}{\mathrm{SNR}(f, B)}\right]\,,
    \label{eq:omegapls}
\end{equation}
where we take the maximum over all integer $B$ from $-8$ to $8$. Note that $\Omega_{PLS}(f)$ is independent of $f_\mathrm{ref}$.

In Ref.~\cite{Thrane:2013oya}, PLS curves are drawn for individual experiments. Here we propose that they can be further generalized to combine data from GW experiments designed for different frequency ranges, such as {\small LISA} and a midband experiment. \textit{We can consider the combination of these different GW experiments as one big experiment for GW measurements}, even if their running times do not overlap: the SGWB is expected to be static over the relevant $5-10$ year observational time window. Labeling different experiments with $i$, we can define the combined SNR for a given SGWB as:
\be
{\rm SNR^{comb}}(f, B)=\sqrt{\sum_i T_i \int^{f^i_{ \rm max}}_{f^i_{\rm min}}df\left(\frac{\Omega_{\rm GW}(f, B)}{\Omega^i_{s}(f)} \right)^2}.
\label{eq:plsens}
\ee
and then substitute SNR$^\mathrm{comb}(f, B)$ into Eq.~\ref{eq:omegapls} to define the combined experimental PLS, $\Omega_{GW}^\mathrm{comb}(f)$. We can use this CPLS to give an estimate for the improvement in SGWB measurements expected from a midband experiment, in advance of our likelihood results later in the paper.

\begin{figure}
  \includegraphics[width=0.48\textwidth]{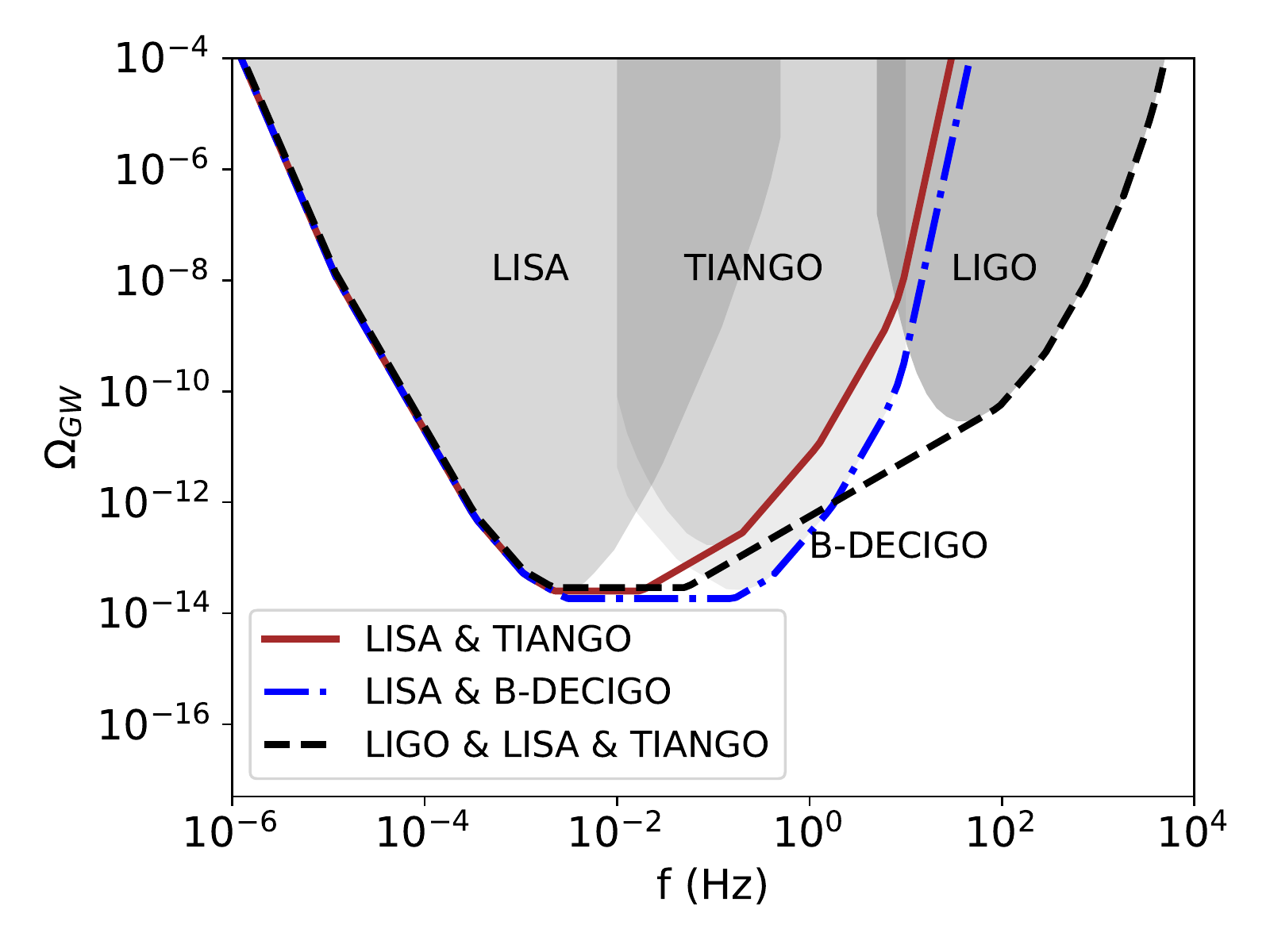}
\caption{The power-law sensitivity curves for SGWB from LIGO (the A+ detector), {\small LISA}, TianGo and {\small B-DECIGO}.  We assume a nominal $4$ year observation time for {\small LISA}, TianGo and {\small B-DECIGO}, $3$ years for LIGO in A+ mode and SNR$^{\rm thr} = 1$. {\small B-DECIGO} overlaps TianGo strongly and is shown as the lighter grey band extending to lower $\Omega_{\rm GW}$. We also show several \textit{combined} PLS (CPLS) curves, computed as explained in the text, which demonstrate the notable improvement in the transitional frequency bands compared to the PLS curves based on individual experiments.}
  \label{fig:combinedSNR}
\end{figure}

\subsection{Expected Strain Sensitivities}

In this section we detail our assumed models for the gravitational wave detector landscape around $2035$, the timescale of the full {\small LISA} mission. We model {\small LISA} and the funded {\small LIGO} A+ detector becoming operational in the $2020$s. We also discuss the impact of the proposed third generation ground-based detector network, Cosmic Explorer and Einstein Telescope \cite{CosmicExplorer, EinsteinTelescope}. The first phase of this network, improving by a factor of $\sim 5$ in sensitivity to strain and $25$ in sensitivity to $\Omega_\mathrm{GW}$ over {\small LIGO} A+, could begin operations by $2035$, a similar timeframe to {\small LISA} \cite{CosmicExplorer}.

The midband landscape is substantially more uncertain, including several space-based designs and atomic interferometers. We choose to focus on two space missions, TianGo and B-DECIGO, where B-DECIGO is more ambitious. Our results for B-DECIGO are also relevant for the atomic interferometer AEDGE \cite{AEDGE}, which has similar sensitivity. We do not consider the Taiji mission \cite{Taiji} as its constraining power is similar to {\small LISA} and it is thus of limited interest. We also neglect the earlier TianQin mission concept \cite{TianQin}, which reaches into the midband, but to a lesser extent than TianGo. Other missions in a similar frequency range are possible, but should give similar results for realistic error budgets.

Figure~\ref{fig:combinedSNR} shows the power law sensitivity curves for our three main experiments, given our assumed $S_n(f)$ models, as well as the power law sensitivity for the combination. In the transitional region between {\small LIGO} and {\small LISA}, the midband experiment TianGo improves sensitivity by several orders of magnitude.

\textit{LISA:}
We use the noise model from \cite{Caprini:2019}, which is based on the {\small LISA} science requirements document \cite{LISASRD}. This assumes a single detector channel\footnote{See \cite{Flauger:2020qyi} for a model with a complete set of detector channels.}. We are moderately more ambitious by assuming $4$ years of observational data in a $5$ year mission (thus setting $T$ in Eq.~\ref{eq:plsens}). The noise budget at high frequencies is dominated by the ``optical metrology system'' noise $P_\mathrm{oms}(f,P)$ and at low frequencies by the ``mass acceleration'' noise $P_\mathrm{acc}(f,A)$, where $P$ and $A$ are dimensionless accuracy constants (see also \cite{Larson:2000, Cornish:2003}). For arm length, $L_0$, the shape of the noise curve is
\begin{align}
 P_\mathrm{oms}(f, P) &= P^2 \left[ 1 + \left(\frac{2\mathrm{mHz}}{f}\right)^4\right] \left( \frac{2 \pi f}{c}\right)^2\,, \label{eq:omsnoise}\\
 P_\mathrm{acc}(f, A) &= \frac{A^2}{\left( 2 \pi f c\right)^2} \left[ 1 + \left(\frac{0.4 \mathrm{mHz}}{f}\right)^2\right] P_\mathrm{disp}\,, \\
 P_\mathrm{disp} &= \left[1+\left(\frac{f}{f_\mathrm{disp}} \frac{2.5\times 10^9 m}{L_0}\right)^4\right]\,.
 \label{eq:noises}
\end{align}
$P_\mathrm{disp}$, which relaxes the sensitivity at high frequencies, comes from white noise displacement of the test masses converted into acceleration. The constant $f_\mathrm{disp}$ is $8 \mathrm{mHz}$ for {\small LISA}. For other missions we have assumed it scales linearly with arm length, as it becomes important when frequency is comparable to round-trip laser time.

We combine Eqs.~\ref{eq:omsnoise}-\ref{eq:noises} with the gravitational wave transfer function $R(w)$ to give
\begin{equation}
 S_n = \frac{1}{R(w)} \left[ P_\mathrm{oms}(f, P) + \left(3 + \cos\left(w\right) \right)P_\mathrm{acc}(f, A)\right]\,.
  \label{eq:noisepower}
\end{equation}
$w = 2 \pi f L_0 /c$ and the transfer function $R(w)$ is
\begin{equation}
 R(w) = \frac{3}{10}(w)^2\left[1 + 0.6 \left(w\right)^2\right]^{-1}\,.
\end{equation}
Here $L_0$ is the length of the satellite arms, $f$ is the frequency in Hz, $c$ is the speed of light, $A$ is the residual acceleration noise and $P$ is the position noise.
For {\small LISA}, we set $L_0 = 2.5\times 10^6$ m, acceleration noise of $A = 3.0\times 10^{-15}$ m s$^{-2}$ Hz$^{-1/2}$ and position noise of $P = 1.5 \times 10^{-11}$ m Hz$^{-1/2}$, sensitive to frequencies between $3 \times 10^{-5}$ Hz and $0.5$ Hz, following Ref.~\cite{Caprini:2019}.

\textit{TianGo}: We use the sensitivity curve from Ref.~\cite{TianGo}. This can be derived from Eq.~\ref{eq:noisepower} by assuming three satellites sensitive to a frequency range of $10^{-2} - 10 $ Hz with an arm length of $L_0 = 10^5$ m, acceleration noise of $A = 1.4 \times 10^{-17}$ m s$^{-2}$ Hz$^{-1/2}$ and position noise $P = 2\times 10^{-22}$ m Hz$^{-1/2}$. The template includes extra noise at $f < 0.03$ Hz due to gravity gradient.

\textit{DECIGO}: The DECIGO experiment has two components: an initial mission, B-DECIGO, which comprises three drag-free satellites in a geocentric orbit with an arm length of $L_0 = 10^5$ m, and the full DECIGO mission, which is a constellation of four sets of three drag-free satellites at three different points in a heliocentric orbit. The science target of DECIGO is the detection of the stochastic background from inflation \cite{DECIGO2011, Sato_2017, PreDECIGO}. Here we consider B-DECIGO, as the next generation satellite mission expected to launch in the 2030s.
B-DECIGO is expected to be sensitive to frequencies from $10^{-2} - 100$ Hz. The satellites of B-DECIGO have $30$ kg test masses with a force noise of $10^{-16} N /Hz^{1/2}$ and thus acceleration noise of $3 \times 10^{-18}$ m s$^{-2}$ Hz$^{-1/2}$. We assume position noise of $P = 2 \times 10^{-23} L_0 = 2\times 10^{-18}$ m Hz$^{-1/2}$\cite{PreDECIGO}.

\textit{{\small AEDGE}}: {\small AEDGE} is an alternative design for a satellite experiment using a detector based on cold atom interferometry, also capable of probing the midband. The sensitivity curves for {\small AEDGE} are similar to those for {\small B-DECIGO}, although achieved with only two satellites \cite{AEDGE}. Our conclusions for {\small B-DECIGO} are thus also applicable to {\small AEDGE}.

\textit{{\small LIGO/VIRGO} ground-based detectors:} The operational ground-based detector network (including LIGO, VIRGO, KAGRA and LIGO India) in $2035$ is expected to be well developed. We have conservatively used the presently funded A+ detector \cite{Adetector}, although there are proposals \cite{CosmicExplorer,EinsteinTelescope} for detectors with an order of magnitude better sensitivity.
We assume the $A+$ experiment will obtain $3$ years of data and use the public forecast sensitivity curve obtained from the LIGO website\footnote{\url{https://dcc.ligo.org/LIGO-T1800042/public}}. %We use the Cosmic Explorer first generation sensitivity obtained from \footnote{\url{https://dcc.cosmicexplorer.org/cgi-bin/DocDB/ShowDocument?docid=T2000017}}.

\section{Analysis for Benchmark Cosmological Sources}
\label{sec:mcmc}

\subsection{Cosmological Stochastic Gravitational Wave Backgrounds}
We consider two representative cosmological sources of SGWB: cosmic strings and phase transitions. These two new physics scenarios are also being probed by other experimental means. For example, the cosmic microwave background constrains cosmic strings. The Large Hadron Collider and or future collider experiments could probe a Higgs sector capable of producing a strong electroweak (EW) phase transition through precision measurements of Higgs couplings. However, {\small LISA} is several orders of magnitude more sensitive to a cosmic string network than current or future microwave background experiments, and can complement related collider searches for an extended Higgs sector \cite{Huang:2016cjm, Gould:2019qek}.

\subsubsection{Cosmic Strings}
Cosmic strings are one-dimensional topological defects, generically predicted by particle physics theories beyond the standard model. Examples include fundamental strings in superstring theory and vortex-like solutions in field theories with a spontaneously broken $U(1)$ symmetry. At macroscopic scales the string properties are characterized by energy per unit length (tension), $\mu$. The string network forms in the early Universe, composed of a few long strings per horizon volume and copious, unstable string loops (formed upon long string intersections), tracing the background energy by a fraction $\sim G\mu$. For many cosmic string models GW production is usually considered the dominant radiation mode for the oscillating string loops \footnote{Although Ref.~\cite{Vincent:1997cx, Bevis:2006mj} argue that particle emission dominates for gauge strings.}, and yields a SGWB from the accumulation of these decaying string loops. In this work we calculate the SGWB from strings following Ref.~\cite{Cui:2018rwi, Auclair:2019wcv}, which incorporates the simulation results for loop distribution from Ref.~\cite{Blanco-Pillado:2013qja} and an analytical derivation based on a velocity-dependent one scale (VOS) model \footnote{This loop distribution is widely accepted, but other possibilities are discussed in Refs.~\cite{Ringeval:2005kr, Auclair:2019wcv}}.

The shape of the SGWB spectrum from strings is sensitive to the cosmic expansion history, and a number of recent papers have explored how an early matter domination or kination period may imprint such a spectrum \cite{Cui:2017ufi, Cui:2018rwi, DEramo:2019tit, Chang:2019mza, Cui:2017ufi,Cui:2019kkd, Dror:2019syi, Buchmuller:2019gfy,Blasi:2020wpy}.
For the purpose of this work we consider only the case with a standard cosmology: the post-inflationary Universe is radiation dominated until $z \sim 3500$, when it transitions to matter domination. More complex cosmologies we defer to future work.

The cosmic string SGWB spans a wide range of frequency with a nearly flat plateau towards high $f$. As we specify the cosmic history and the loop distribution, the SGWB signal is parametrized by one parameter, the cosmic string tension $G\mu$. We sample string tensions up to the upper limit from EPTA~\cite{EPTA}, $G\mu = 2\times 10^{-11}$, which is several orders of magnitude larger than {\small LISA}'s detection limit. The excess noise in NANOGrav, if interpreted as a detection of cosmic strings, would imply $G\mu = 4\times 10^{-11} - 10^{-10}$ \cite{Ellis:2020ena}. The exact upper limit we assume has no effect on our results, as LISA alone is able to detect a cosmic string tension many orders of magnitude lower.

\subsubsection{Phase Transitions}

A strong first order phase transition (PT) may occur in the early Universe, associated with, for example, electroweak symmetry breaking, generation of a matter-antimatter asymmetry or the formation of dark matter \cite{Caprini:2016}. Notably, with simple extensions to the Higgs sector, in the SM the electroweak symmetry breaking phase transition may be first order, and so trigger electroweak baryogenesis. Such a phase transition can generate a SGWB with a peaky structure \cite{Caprini:2009fx, Alanne:2020, Schmitz:2020}.

The gravitational wave signal from phase transitions arises from three major effects: collisions between bubbles, long lasting sound waves, and possibly turbulence \cite{Caprini:2009fx, Caprini:2018, Caprini:2020}. Each of these three effects produce a component of gravitational wave spectrum which follow a broken power law, peaking around a frequency which roughly scales as the average bubble size (e.g. \cite{Caprini:2020}). The specific amplitude, power laws and peak location depend on the underlying phase transition model. Recent studies show that the GW component from bubble collisions is generally sub-dominant in many particle physics models, such as the $H^6$ extension of the SM for the electroweak phase transition. It can however be important in special cases such as a classically scale-invariant $U(1)_{B-L}$ extension of the SM \cite{Ellis:2019oqb, Hindmarsh:2020}. The signal from turbulence is currently uncertain, as it may only be derived from numerical simulations, which are challenging in the strongly turbulent regime \cite{Cutting:2020}. We will therefore consider the sound wave component only, neglecting other sources. As described below, we focus on parameter regions where this is likely to be a good approximation (e.g. away from extreme supercooling \cite{Ellis:2019oqb, Caprini:2019egz}).

The SGWB spectra from a first order PT is determined by four independent parameters: the bubble wall velocity $v_w$, the temperature $T_*$ at which the transition occurs, the strength of the transition $\alpha$, and the duration of the transition $\beta/H_*$ (which we refer to as $\beta$ hereafter). For any given particle physics model $T_*$, $\beta$ and $\alpha$ can be computed from the field Lagrangian, although $v_w$ requires detailed simulation. As our focus is on detectability using a midband experiment we do not choose a specific particle physics model and instead marginalise over these phenomenological parameters.

The emitted gravitational wave spectrum may be computed from these parameters using the formulae derived in \cite{Hindmarsh:2017, Caprini:2020, Guo:2020, Hindmarsh:2020}. For a phase transition at temperature $T_*$, with Hubble expansion rate $H_* = H(T_*)$ and bubble size $R_*$ at the percolation time, we have\footnote{Ref.~\cite{Caprini:2020} uses max$(v_w, c_s)$, where $c_s$ is the sound speed instead of $v_w$, but see \cite{Hindmarsh:2017, Guo:2020, PTPlot}.}
\begin{equation}
 R_* = \frac{(8 \pi)^{1/3}}{H_* \beta v_w}\,.
 \label{eq:bubble}
\end{equation}
The gravitational wave spectrum peaks at a frequency proportional to $R_*$, which today becomes
\begin{equation}
 f_{p,0} = \frac{2.6 \times 10^{-5}}{H_* R_*} \left(\frac{T_*}{100\;\mathrm{GeV}}\right)\left(\frac{g_*}{100}\right)^{1/6} \; \mathrm{Hz}\,.
\end{equation}
$g_*$ is the number of degrees of freedom at the phase transition which for $T_* \gtrsim 200$ GeV is $106.75$, assuming particle content as in the SM.

The gravitational wave spectrum today is \cite{Hindmarsh:2017, Caprini:2019, Hindmarsh:2020}:
\begin{equation}
 \frac{d \Omega_\mathrm{GW,0}}{d \ln f} = 2.061 F_\mathrm{GW,0} K^2 H_* R_* \tilde{\Omega}_\mathrm{GW} C\left(\frac{f}{f_{p,0}}\right)\,.
 \label{eq:omegagw}
\end{equation}
The normalisation $\tilde{\Omega}_\mathrm{GW}$ comes from fitting to the numerical simulations of Ref.~\cite{Hindmarsh:2017}.\footnote{An erratum was issued for their eq. 39. We use the corrected equation. However, at the time of writing the correction has not propagated to the equivalent equation (eq. 29) of Ref.~\cite{Caprini:2020}, with which we disagree by a factor of $\sqrt{3}$.}
Here subscript $0$ denotes the present day and subscript $*$ denotes the time of the phase transition. $F_\mathrm{GW,0}$ evolves $\Omega_\mathrm{GW,*}$ into $\Omega_\mathrm{GW,0}$ and is given by
\begin{equation}
 F_\mathrm{GW,0} = 1.65 \times 10^{-5} h^{-2} \left(\frac{100}{g_*}\right)^{1/3}\,.
\end{equation}
$h$ is the reduced Hubble parameter, which we assume to be $0.679$ in agreement with Planck \cite{Planck2018}. We neglect for simplicity the possibility of an early matter dominated phase induced by a very strong phase transition \cite{Ellis:2020}. The shape function $C(s)$ is chosen to fit numerical simulations \cite{Hindmarsh:2015,Caprini:2016}:
\begin{equation}
 C(s) = s^3 \left(\frac{7}{4 + 3 s^2}\right)^{7/2}\,.
\end{equation}
The numerical factor Eq.~\ref{eq:omegagw} comes from $3 \int^\infty_0  C(s) d \ln s$. This shape function overestimates power at small $s$ and underestimates it at large $s$. Its domain of validity is $\alpha < 0.1$, $0.4 < v_w < 0.5$ \cite{Hindmarsh:2020}. We are particularly interested in this regime, as it includes the upper limit on $\alpha$ for  well-constrained phase transition energies.  The factor $\tilde{\Omega}_\mathrm{GW} = 0.012$ is numerically determined \cite{Hindmarsh:2017}. $K$ is the kinetic energy fraction in the fluid, given by
\begin{align}
 K &= \kappa \frac{\alpha}{1 + \alpha}\,,\\
 \kappa &= \frac{\alpha}{0.73 + 0.083 \sqrt{\alpha} + \alpha} \,.
\end{align}
As shown by \cite{Ellis:2019, Guo:2020} when the phase transition is slow the gravitational wave amplitude decays by a factor proportional to the optical depth, due to shock formation \cite{Ellis:2020a}
\begin{equation}
 H_* \tau_{sh} \sim \frac{H_* R_*}{\sqrt{4/3 K}}\,,
 \label{eq:taush}
\end{equation}
so that Eq.~\ref{eq:omegagw} is multiplied by $H_* \tau_{sh}$ when $H_* \tau_{sh} < 1$, which is the generic case as noted by Ref.~\cite{Ellis:2019}.\footnote{We define $\tau_{sh}$ following Ref.~\cite{Ellis:2020}, but older models, omit the factor of $\sqrt{4/3}$ \cite{Caprini:2020}.} Thus when $H_* \tau_{sh} < 1$, the final equation is
\begin{equation}
 \frac{d \Omega_\mathrm{GW,0}}{d \ln f} = 1.785 F_\mathrm{GW,0} K^{3/2} (H_* R_*)^2 \tilde{\Omega}_\mathrm{GW} C\left(\frac{f}{f_{p,0}}\right)\,.
 \label{eq:omegagw2}
\end{equation}
Eq.~\ref{eq:omegagw2} applies in practice to all our phase transition predictions.

To summarize, this model includes four free parameters. First, the strength of the phase transition, $\alpha$, which controls the amplitude of the gravitational wave signal. Second, $T_*$, the energy density of the phase transition which controls the frequency of the emitted gravitational waves. Third, the speed of the phase transition, $\beta / H_*$. Finally, the speed of the bubbles, $v_w$. As $v_w$ occurs only in Eq.~\ref{eq:bubble}, it is observationally degenerate with $\beta$. We therefore fix $v_w = 0.5$, a regime where the equations above are accurate. For any given particle physics model for the phase transition, $\beta / H_*$ correlates with $\alpha$ (e.g.~\cite{Ellis:2019}), and is observationally degenerate with a combination of $\alpha$ and $T_*$. For the purposes of our parameter constraints we fix $\beta / H_* = 40$ as a fiducial value for which the above equations are valid. We have confirmed explicitly by running dedicated chains that varying $\beta / H_*$ produces a three-way parameter degeneracy. We will therefore vary only $T_*$ and $\alpha$ in our analysis.

We scan $T_*$ over the range of $100~ {\rm GeV} < T_* < 10^7$ GeV, the region most relevant for observation with a midband experiment. This includes the $\sim 100$ GeV energy range generally expected for the electroweak phase transition as well as possible more energetic phase transitions associated with, for example, EW PT in Randall-Sundrum models \cite{Agashe:2020lfz}, supersymmetry breaking \cite{Craig:2009zx} or a dark sector \cite{Schwaller:2015tja,Hall:2019ank}. We choose to limit $\alpha < 0.8$ in our chains, which generally ensures that the PT can be completed \cite{Ellis:2019}.

\subsection{Astrophysical Stochastic Gravitational Wave Backgrounds}

Gravitational waves have been detected from mergers of compact objects: black holes and neutron stars. These objects also contribute to the SGWB. The unresolved signals that make it up are merger events which are too far away to be detectable, and the early inspiral phase of ultimately observable mergers. The latter emit weakly at low frequencies and thus may last much longer at low frequencies than the mission time of LISA. Coalescing compact objects emit GWs with a spectral energy density $dE/df_\mathrm{s}$, where $f_\mathrm{s}$ is the frequency in the source frame. The background energy density is then
\begin{align}
 \Omega_\mathrm{GW} (f_\mathrm{obs}) &= \frac{f_\mathrm{obs}}{\rho_c}\frac{d\rho_\mathrm{GW}}{d f_\mathrm{obs}} \nonumber \\
 &= \frac{f_\mathrm{obs}}{c^2 \rho_c} \int_0^{10} dz \, \frac{R_m(z)}{(1+z) H(z)} \frac{dE}{df_\mathrm{s}}\,.
 \label{eq:sgwb}
\end{align}
Here $\rho_c = 8.5\times 10^{-27}$ kg m$^{-3}$ is the critical density, $f_\mathrm{obs} = f_\mathrm{s} / (1+z)$ is the frequency in the observed frame, $H(z)$ is the Hubble expansion rate and $R_m(z)$ is the merger rate in Gpc$^{-3}$ yr$^{-1}$. For all astrophysical backgrounds we integrate redshift from $z=0$ to $z=10$, approximately the time of formation of the earliest black hole binaries. We have checked that our results are insensitive to the upper redshift limit.

In the below Section, we discuss a variety of astrophysical SGWB sources. The most important are: the unresolved inspiral phases of the already detected {\small LIGO} mergers, which we call Stellar Mass Binary Black Holes (StMBBH), mergers from putative intermediate mass ratio inspirals (IMRIs), and, in the {\small LISA} band, extreme mass ratio inspirals (EMRIs). We discuss, and conclude to be subdominant, SGWB signals from supermassive black holes, white dwarf mergers and type 1a supernovae. The SGWB from StMBBH and IMRI can be approximated as a power law with index $2/3$. The shape of the EMRI SGWB is more complex, but can be approximated by a power law with index $-1/3$ for $3\times 10^{-3} - 3\times 10^{-2}$ Hz. These astrophysical sources are summarized in Figure~\ref{fig:sgwb}.

\subsubsection{Stellar Mass Black Hole Binary Mergers}
\label{sec:smbbh}

Mergers detected in the LIGO band emit GWs at lower frequencies during their inspiral phase \cite{Finn:2000}. We model the signal from these stellar mass binary black hole (StMBBH) mergers following~\cite{Cholis:2016, LIGOSGWB}. We neglect neutron star mergers as they are subdominant and degenerate with the overall merger rate, which we marginalize over. By allowing the merger rate to vary we include possible signals from as-yet undetected sources such as primordial black holes \cite{Mandic:2016lcn}. We note that there is still considerable uncertainty in even the shape of the mass function of binary black hole mergers, and that future LIGO merger data may still shift the preferred power law indices \cite{LIGOO2pop}. However, the power law index of the SGWB at lower frequencies is dominated by the emission in the inspiral phase and is actually relatively well-characterised, at least compared to other potential SGWB sources.

We compute $dE/df_\mathrm{s}$ using separate templates for the merger and inspiral phases from \cite{Ajith:2008}. For the inspiral phase
\begin{align}
 \frac{d E_\mathrm{insp}}{df_\mathrm{s}} = \frac{1}{3}\left(\frac{\pi^2 G^2}{ f_\mathrm{s}}\right)^{1/3}
 \frac{m_1 m_2 }{\left(m_1 + m_2\right)^{1/3}}\,.
\end{align}
$m_1$ and $m_2$ are the masses of the two merging objects and $G$ is the gravitational constant. During the inspiral phase the emission varies over a wide frequency range.
For the merger phase
\begin{align}
 \frac{d E_\mathrm{merg}}{df_\mathrm{s}} = \frac{1}{3}\left(\pi^2 G^2\right)^{1/3} \frac{f_\mathrm{s}^{2/3}}{f^\mathrm{StBBH}_\mathrm{merg}}\frac{m_1 m_2 }{\left(m_1 + m_2\right)^{1/3}}\,.
\end{align}
$f^\mathrm{StBBH}_\mathrm{merg}$ is the GW frequency at merger in the source frame:
\begin{align}
f^\mathrm{StBBH}_\mathrm{merg} = 0.04 \frac{c^3}{G(m_1 + m_2)}\,.
\end{align}
We neglect the subdominant signal from ringdown, and so set $dE/df = 0 $ for $f > f_\mathrm{ring}$, the source frame ringdown frequency:
\begin{align}
 f_\mathrm{ring} = \frac{0.915 (1-0.63)(1-0.67)^{0.3} c^3 }{2 \pi G (m_1 + m_2)}\,.
\end{align}
Thus $dE/df_\mathrm{s}$, the total energy emitted as a function of frequency, is the sum of the signals from merger and inspiral, integrated over the mass distributions, $m_1$ and $m_2$.

In the {\small LISA} band the stochastic signal is dominated by the low-frequency inspiral phases, while the merger phase is important only in the LIGO band. We assume mergers occur for masses $5 < m_1, m_2 < 50 M_\odot$. $m_1$ has a power law mass distribution $m_1^{-2.3}$ and $m_2$ is uniformly distributed. We note that the best fit to the latest LIGO data is a slightly steeper power law with an index of $-2.6$ and a separate Gaussian peak at $33 M_\odot$ \cite{LIGOO2pop}, which differs moderately from our model. However, our assumed model is only moderately disfavoured at present.

We assume that the merger rate evolves with redshift following an empirical fit to the star formation rate:
\begin{align}
 R_m(z) \sim \frac{a \exp\left[b (z-z_m)\right] }{a + b \left(\exp\left[a (z- z_m)\right]- 1\right)}\,.
\end{align}
We take $a = 1.92$, $b = 1.5$, $z_m = 2.6$ and we define a normalizing constant $R_0$ to specify the rate at $z=0$, which we leave as a free parameter in our Markov chains. The shape of $R_m(z)$ and the values of $a$ and $b$ are currently uncertain. However, in the midband region the signal is dominated by the early inspiral phase of relatively low redshift binaries, so we found that for reasonable values of these parameters they were degenerate with the total merger rate. For similar reasons we have not attempted to remove the contribution for merger events resolved by LIGO, which is also degenerate with the overall merger rate.

\subsubsection{Extreme Mass Ratio Inspirals}
\label{sec:emri}

{\small LISA} will be sensitive to extreme mass ratio inspirals (EMRIs), mergers between stellar mass and supermassive black holes (SuMBH) \cite{Amaro:2007, Babak:2017, Amaro:2018}. The merger frequency of these objects is approximately
\begin{equation}
 f^{\rm EMRI}_\mathrm{merg} = 0.01 \left(\frac{M_\mathrm{SuMBH}}{10^6 M_\odot}\right)^{-1} \;\mathrm{Hz}\,.
\end{equation}
The non-detection of a black hole in M33 \cite{Gebhardt:2001} suggests that a reasonable guess for a lower limit on the SuMBH mass is $2\times 10^6 M_\odot$, while cosmological simulations use a seed mass around $5.6\times 10^5 M_\odot$.
The EMRI signal thus lies within the {\small LISA} band, and would not be detected by a midband experiment.
A typical EMRI signal lasts $\sim 1$ year and includes up to $10^5$ orbits \cite{Babak:2017}. A fiducial merger rate is $\sim 1$ Gpc$^{-3}$ year$^{-1}$, or $300$ {\small LISA} detections year$^{-1}$~\cite{Gair:2004}. Although these signals are faint, the mock {\small LISA} data challenge \cite{Babak:2010} demonstrated that they are detectable in the datastream due to the high number of orbits.

Modeling the overall signal from EMRIs is complex, as they have a large range of possible parameters, including both black hole masses, eccentricity and black hole spin. We use the EMRI population model from Ref.~\cite{Bonetti:2020}, based on the fiducial population model (M1) of Ref.~\cite{Babak:2017}, with detected sources removed.
We calculate $\Omega_{\rm GW}$ using
\begin{equation}
\Omega_{\rm GW}(f) = \frac{4 \pi^2 f^2}{3 H_0^2} h_c(f)^2
\end{equation}
where $h_c^2(f)$ is the EMRI SGWB characteristic strain. When making forecasts, we leave the overall rate of EMRI mergers as a free parameter to model uncertainty in the EMRI population \cite{Amaro:2011, Babak:2017}.

\subsubsection{Supermassive Black Holes}
\label{sec:subbh}

{\small LISA} will also be sensitive to mergers between two SuMBH of masses $ 10^4 M_\odot - 10^7 M_\odot$.
We do not consider the stochastic background from these objects as {\small LISA} is sensitive enough to detect essentially all such mergers for $z<8$.
At higher redshifts the expected number of supermassive black hole mergers is reduced exponentially, following the number density of halos and the expected timescale for SuMBH formation. SuMBH with $M> 10^7 M_\odot$, when they occur, would merge in a timescale too short to be resolved from LISA's data stream \cite{Salcido:2016}. As these objects are rare, brief, transients, they are better treated as glitches rather than a SGWB and so we do not include them.

\subsubsection{Intermediate Mass Ratio Inspirals}
\label{sec:imbh}

Between stellar mass and supermassive black hole populations lies a hypothetical population of intermediate mass black holes (IMBH) with $10^2 - 10^4 M_\odot$ \cite[e.g~]{Amaro:2007}. The best candidate for their production is dense star clusters which may produce a runaway merger \cite{Ebisuzaki:2001,Miller:2005}.
Only one IMBH has yet been observed, indirectly as the outcome of GW190521 \cite{GW190521}, although some may be accessible with {\small LIGO} \cite{Ezquiaga:2020tns}.

We can postulate Intermediate Mass Ratio Inspirals (IMRIs) with a mass ratio of $10^2 - 10^4 M_\odot$, resulting from the merger of stellar mass black holes and IMBHs. Such a merger would be observable by a midband experiment at $\sim 1$ Hz \cite{AmaroSeoaneIMRI}. Like EMRIs, the merger rate would depend on a variety of uncertain parameters, including the dynamics inside star clusters and the spin distribution of the IMBH.

These mergers would produce a corresponding SGWB. However, the shape of the merger has not yet been computed in the literature. We therefore model the IMRI signal using the same model as we used for stellar mass binary black holes, modifying only the mass distribution of the IMBH and the fiducial merger rate. We assume for the IMBH a uniform mass distribution with a range $10^3 - 10^4 M_\odot$. Ref.~\cite{Amaro:2007} predicted the inspiral phase of $1-10$ IMRIs could be observed by {\small LISA}, implying a merger rate of $10^{-3} - 10^{-2}$ Gpc$^{-3}$ year$^{-1}$. We thus choose a fiducial merger rate for our IMRI SGWB model of $5\times 10^{-3}$ Gpc$^{-3}$ year$^{-1}$.

At this rate the SGWB from IMRIs in the {\small LISA} band is similar, but subdominant to, the SGWB from stellar mass binaries merging in the LIGO band. At low frequencies the shape of the signal is completely degenerate with the lower mass objects, with the degeneracy being broken only by the signal from the merger phase in the midband.

Our modeling of the IMRI SGWB is simplistic and likely to be incorrect in detail. However, we suspect that the broad picture of a SGWB component, moderately subdominant to stellar mass binary black holes, degenerate during inspiral and distinguishable during mergers, is likely to be upheld by more detailed future modeling.

\subsubsection{White Dwarf Mergers}
\label{sec:whitedwarf}

{\small LISA} is sensitive to gravitational wave emission from white dwarf mergers, weak unresolved instances of which would also produce a stochastic gravitational wave background \cite{Bender:1997}. However, as the emission from these objects is weak, {\small LISA}'s sensitivity is limited to mergers in the Milky Way.
The stochastic signal from these objects would thus be highly anisotropic, both in space and in time (due to the earth's rotation around the Sun). We assume that the stochastic signal can be successfully decomposed using angular harmonics, and all but the isotropic component discarded, effectively allowing the white dwarf background to be neglected \cite{Thrane:2009, Adams:2014, Pieroni:2020}.

\subsubsection{Slowly Rotating Neutron Stars}
\label{sec:neutronstars}

Non-axisymmetric neutron stars are expected to produce gravitational waves \cite{Press:1972, Riles:2012yw}. These gravitational waves arise from the rotation of a small deviation from spherical symmetry and have a frequency twice the rotational frequency of the neutron star. The stochastic background from this source thus peaks at high frequencies, reaching perhaps $\Omega_{GW} \sim 10^{-8}$ at $f = 1000$ Hz and dropping to $\Omega_{GW} \sim 10^{-16}$ by $f = 10$ Hz \cite{Marassi:2010wj, Rosado:2012bk}, although these amplitude estimates are uncertain. These gravitational waves may thus be marginally detectable by LIGO detectors, but sensitivity is likely to be limited to the Milky Way \cite{Christensen:2018iqi} and can thus be separated from other SGWB sources through their angular harmonics, as with white dwarf mergers.

\subsubsection{Type 1a Supernovae}
\label{sec:sn1a}

A source of gravitational waves unique to a midband experiment is type 1a supernovae, whose GW signal peaks in the $1$ Hz range~\cite{Seitenzahl:2015}. There is no inspiral phase to this event, as the supernovae are assumed to originate from white dwarfs reaching the Chandrasekhar mass by accretion. The events are also faint, with a peak energy of $dE/df_\mathrm{s} = 10^{39}$ erg/Hz in a frequency range of $0.5 - 1.5$ Hz. If we approximate $R_m(z)$ in Eq.~\ref{eq:sgwb} as $R_{SN} \delta(z=0)$, then we have
\begin{equation}
 \Omega_{GW} = \frac{f_\mathrm{obs}}{c^2 \rho_c H_0} \frac{dE}{df_\mathrm{s}} R_{SN}\,.
\end{equation}
For a cosmological type 1a rate of $R_{SN} = 10^5$ yr$^{-1}$ Gpc$^{-3}$ \cite{Bonaparte:2013}, and $f_\mathrm{obs} \sim 1$ Hz, this evaluates to $\Omega_{GW} = 5\times 10^{-21}$, small enough that we can safely neglect it.

Note that this result is physically due to the lack of an inspiral phase. The full GW energy is released in $1-2$ s and thus produces detectable events without contributing significantly to a SGWB.

\subsection{Forecast Generation}

For our analysis, we first consider signals with fiducial astrophysical SGWB models only. We generate forecasts to show how a midband experiment can improve constraints on cosmogenic SGWB signals. The upper $2-\sigma$ confidence limits on these parameters provides an estimate of the level at which we could rule out the cosmological signal with the provided set of detectors.
We then sample a likelihood function which allows for a non-zero cosmic string or phase transition GW signal. To investigate discovery potential, we separately estimate our ability to extract parameters from models containing cosmological SGWB sources, both a phase transition and a cosmic string background.

Our likelihood function is derived from the overall sensitivity curves of each experiment, and is defined similarly to the squared power-law sensitivity of Eq.~\ref{eq:plsens} as
\begin{equation}
 \log \mathcal{L}(p) = - \Sigma_i T \int df \left(\frac{M_i(f,p) - D_i(f)}{S^i_n(f)}\right)^2\,.
\end{equation}
Here $T$ is the length of each experiment and $M(f,p)$ is the model prediction for a SGWB signal with frequency $f$ and parameters $p$. $S_n(f)$ is the noise spectral density for each experiment, computed using Eq.~\ref{eq:noisepower}. $D(f)$ is the mock data, generated without detector noise\footnote{Detector noise is not necessary to forecast the experimental covariances.} using the default parameters of our astrophysical model. This was a stellar mass BH merger rate of $56$ Gpc$^{-3}$ yr$^{-1}$, an IMBH merger rate of $5\times 10^{-3}$ Gpc$^{-3}$ yr$^{-1}$ and an EMRI merger rate matching the fiducial choices of Ref.~\cite{Bonetti:2020}. We perform separate chains where $D(f)$ includes a cosmological signal. For cosmic strings we include a SGWB with $G\mu = 10^{-16}$, and for a phase transition we use $T_* = 5\times 10^3$ GeV and $\alpha = 0.2$, which peaks at $f\sim 1$ Hz, in the midband region.

The summation $\Sigma_i$ denotes a summation over $i$ experiments. Since we are interested in the extra constraining power of a midband experiment we compare $i = ($ {\small LISA}, LIGO $)$ to constraints from chains which also include a midband experiment, either {\small B-DECIGO} or TianGo. We thus generated multiple chains using different experiments.

Markov chains were sampled using {\small EMCEE} \cite{ForemanMackey:2013}, a widely used affine-invariant sampler. We ran the sampler using $100$ walkers for $6\times 10^4$ samples each.
The walkers were initialized at randomly chosen positions in a ball in the middle of parameter space and moved for $600$ samples each. These samples were then discarded and the position of the walkers used as the initial positions for the main sampling run. Acceptance fractions after burn-in were $\sim 0.3$.

To summarize our parameters, they were: 1) The overall merger rate of stellar mass black holes. 2) The overall merger rate of intermediate mass ratio black holes. 3) The overall rate of EMRI mergers. Depending on the cosmological model we then had: 4) The cosmic string tension $G\mu$, or 4) the phase transition temperature scale $T_*$ and 5) the phase transition strength $\alpha$.

\section{Results}
\subsection{Astrophysical SGWB sources}
\begin{figure}
  \includegraphics[width=0.5\textwidth]{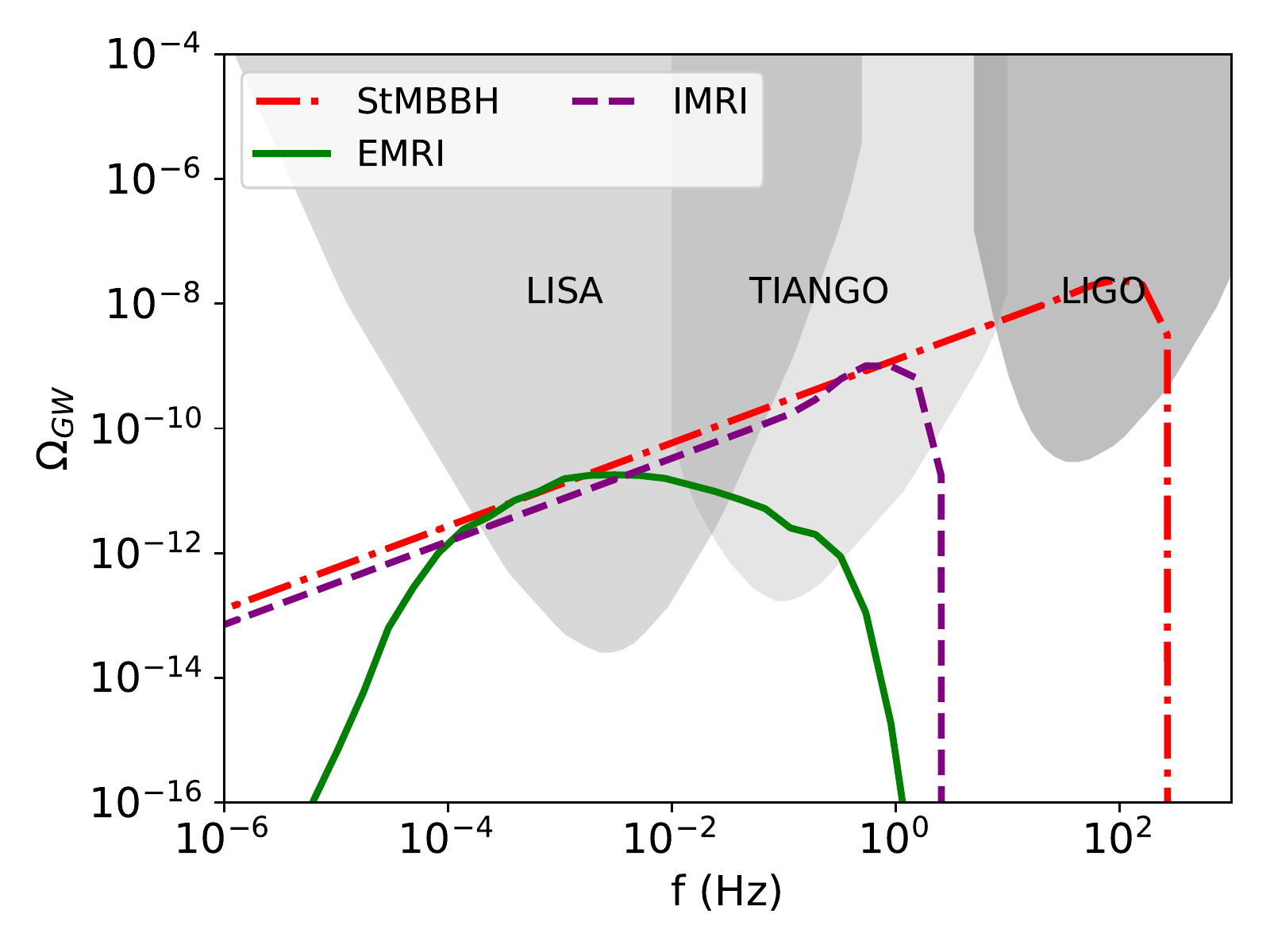}
\caption{Example stochastic gravitational wave backgrounds as a function of frequency. Shown are the astrophysical backgrounds from stellar mass binary black hole mergers with our fiducial merger rate of $56$ yr$^{-1}$ Gpc$^{-3}$ (StMBBH, dot-dashed), EMRI mergers with the fiducial merger rate of \protect\cite{Bonetti:2020} (EMRI, solid) and IMBHs with a merger rate of $4 \times 10^{-3}$ yr$^{-1}$ Gpc$^{-3}$ (IMRI, dashed). We have offset this curve from the fiducial merger rate of $5\times 10^{-3}$ yr$^{-1}$ Gpc$^{-3}$ for clarity: the similar amplitudes for our fiducial model assumptions are largely coincidental. Grey shaded regions show experimental power law sensitivity curves with SNR$=1$.}
  \label{fig:sgwb}
\end{figure}

Figure~\ref{fig:sgwb} shows example signals from the astrophysical SGWB signals. We show for comparison the PLS for {\small LIGO}, {\small LISA} and TianGo. Midband experiments improve sensitivity in the region between $0.01$ Hz and $10$ Hz. In addition to TianGo, we have run chains with {\small B-DECIGO}, which has roughly a factor of two higher sensitivity.

The astrophysical signal from StMBBH and IMRIs is dominated by the inspiral phase until near the peak amplitude. These two astrophysical signals have similar shapes and we have chosen the (uncertain) fiducial merger rate of the IMBH SGWB so that the amplitude of the GW signal is similar to the fiducial StMBBH signal. They are thus extremely degenerate in the {\small LISA} and midband frequency channels, although this degeneracy is broken by the high frequency measurements of {\small LIGO} and (somewhat) by the signal from the merger phase at $f \sim 1$ Hz.
The shape of the EMRI signal differs substantially, as explained in \cite{Bonetti:2020}. That the overall amplitude is similar in the {\small LISA} band to the fiducial StMBBH merger rate is largely a coincidence and sensitive to our assumptions about how many EMRI mergers are resolvable.

\subsection{Cosmic Strings}
\subsubsection{Constraints}
\begin{figure*}
 \includegraphics[width=0.90\textwidth]{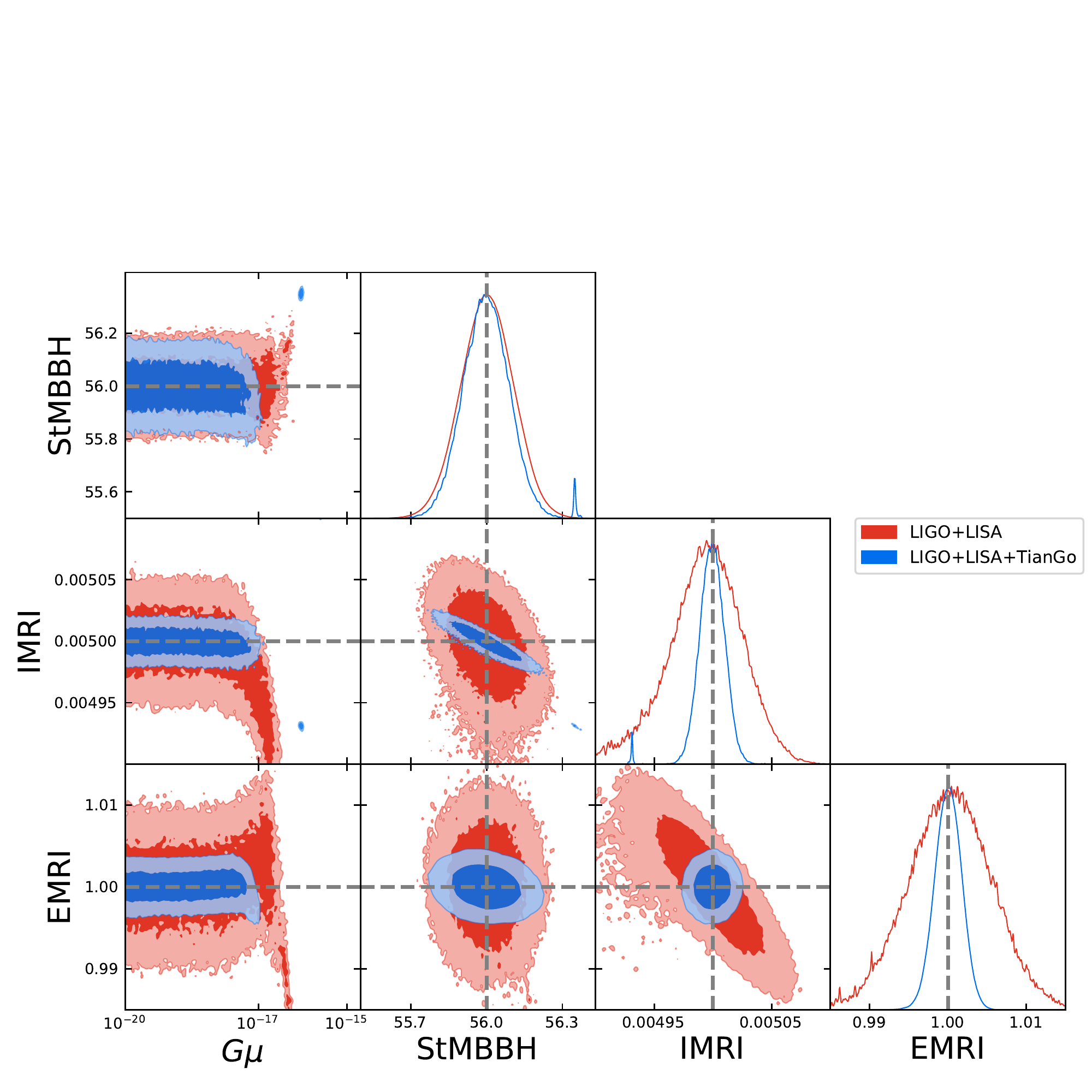}
 \caption{Posterior likelihood contours for signal input with astrophysical SGWB sources only, with which we attempt to constrain the cosmic string tension. Red:  Including {\small LISA} and LIGO but no midband. Blue: Including {\small LISA}, LIGO and the TianGo midband experiment. IMRI and StMBBH merger rates are shown in units of yr$^{-1}$ Gpc$^{-3}$. The EMRI SGWB amplitude parameter is given as a fraction of the fiducial model. $G\mu$ is dimensionless. Dashed lines show the true parameters of the mock astrophysical model. Line plots show marginalised one-dimensional likelihoods, while the 2D shaded regions show $1-\sigma$ and $2-\sigma$ marginalised confidence interval contours for each two-parameter combination.}
 \label{fig:stringsample}
\end{figure*}

\begin{figure}
  \includegraphics[width=0.5\textwidth]{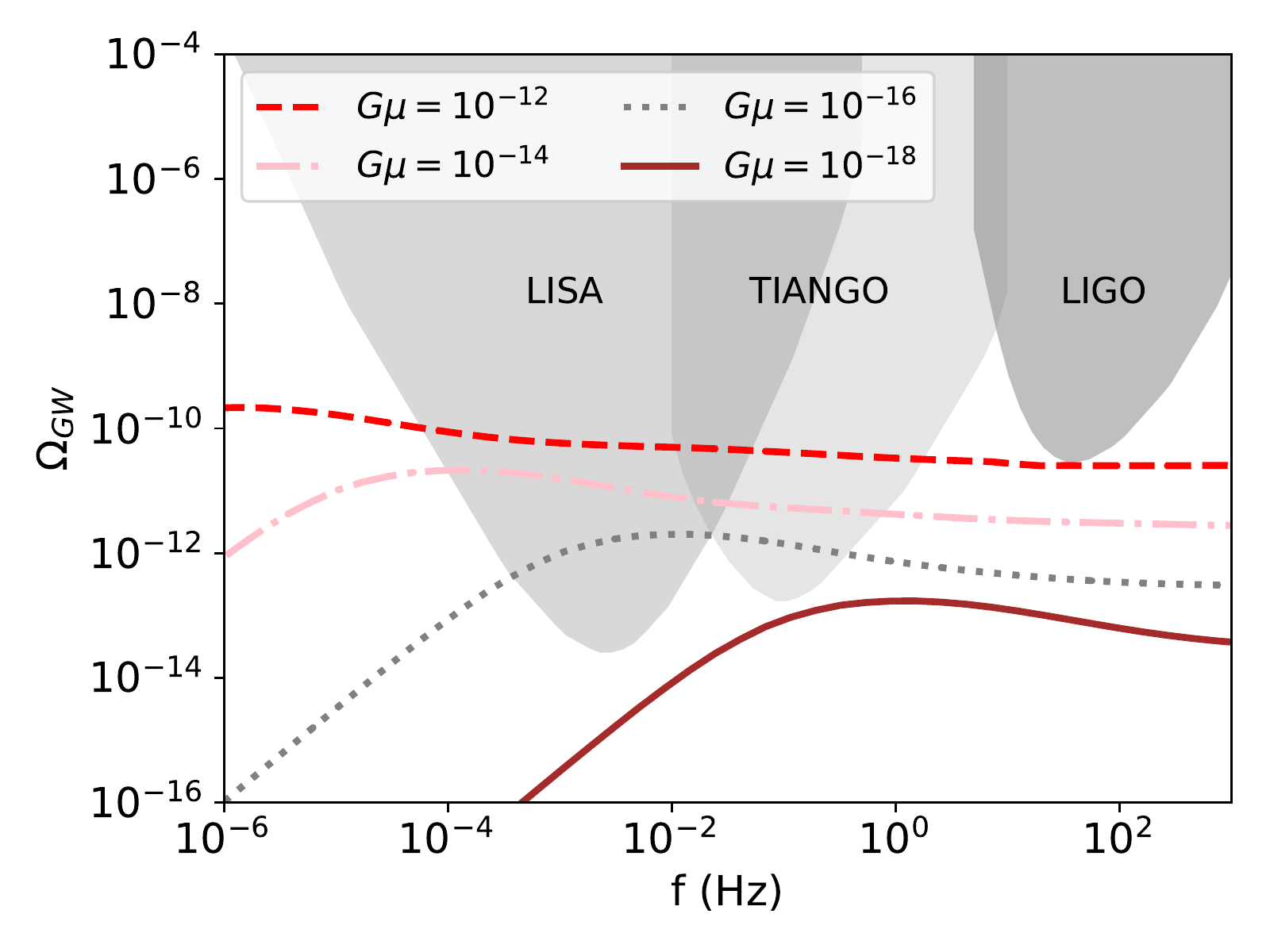}
\caption{Stochastic gravitational wave background signals from cosmic strings. Shown is the expected signal for a variety of cosmic string tensions less than the current upper bound from pulsar timing. Grey shaded regions show experimental power law sensitivity curves with SNR$=1$.}
  \label{fig:strings}
\end{figure}

Figure~\ref{fig:stringsample} shows the results of our forecast for constraining a cosmic string SGWB based on mock data including astrophysical sources only. We compare the likelihood contours with only {\small LISA} and {\small LIGO} to those including TianGo. The midband experiment produces a quantitative improvement in the constraints. With only {\small LIGO} and {\small LISA}, the marginalised $95\%$ upper confidence limit on $G\mu$ was $2.7\times 10^{-17}$, whereas with TianGo it became $9.2 \times 10^{-18}$, an improvement of a factor of $2.9$. We performed chains with the more sensitive {\small B-DECIGO} experiment and found an upper limit of $2.5 \times 10^{-18}$, an improvement of a further factor of $3.7$.

The improvement in the upper limit on $G\mu$ is driven by improved constraints on the SGWB from EMRI and IMRI, which improves following the power law sensitivity of the combined experiments. StMBBH rate constraints do not improve substantially as they are already well constrained by {\small LIGO}. Figure~\ref{fig:strings} explains these results: because the SGWB from cosmic strings is flat between $10^{-3}$ Hz and $1$ Hz, LISA dominates the sensitivity if astrophysical sources are neglected. Improvements in $G\mu$ constraints with TianGo are thus driven primarily by improved component separation.

Note that, since neither IMRIs nor EMRIs emit at {\small LIGO} frequencies, the third generation detectors are unlikely to further improve component separation. However, the raw improvement by a factor of $25-100$ in sensitivity to $\Omega_\mathrm{GW}$ means that the third generation network may be able to directly detect a cosmic string SGWB with $G\mu > 10^{-17}$ \cite{EinsteinTelescope}.

\subsubsection{Discovery Potential}
\begin{figure}
 \includegraphics[width=0.45\textwidth]{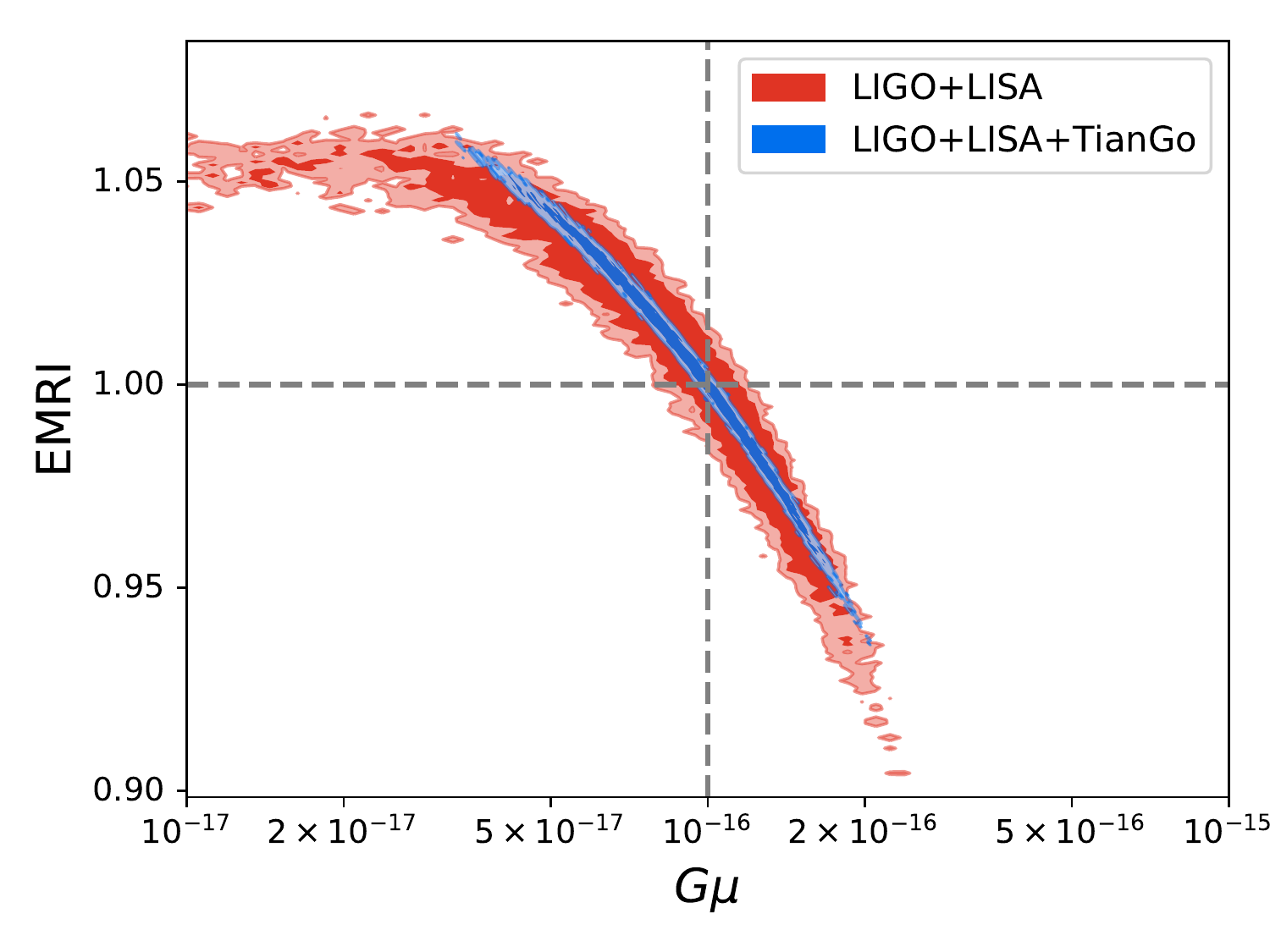}
\caption{Posterior likelihood contour for signal input with astrophysical SGWB sources and a cosmic string model with $G\mu = 10^{-16}$, showing the degeneracy between $G\mu$ and the EMRI merger rate. Red contours include {\small LISA} and {\small LIGO} but no midband, while blue contours also include TianGo. Dashed lines show the true parameters of the mock model. The 2D shaded regions show $1-\sigma$ and $2-\sigma$ marginalised confidence interval contours.}
 \label{fig:stringdiscovery}
\end{figure}

To further assess discovery potential, we ran chains where the simulated data include a cosmic string SGWB with $G\mu = 10^{-16}$, near the edge of the amplitude detectable with {\small LISA}. As expected, without a midband experiment, the string signal was detected at low confidence. Figure~\ref{fig:stringdiscovery} shows our results. A strong curving degeneracy emerged between the amplitude of the EMRI SGWB signal and the cosmic string signal: in the presence of a cosmological signal, {\small LISA} alone was unable to correctly separate astrophysical and cosmological components. The degeneracy ran between $G\mu \sim 0$, and $G\mu = 2 \times 10^{-16}$, while the EMRI merger rate runs between $0.95$ and $1.05$ the fiducial rate. Since we have probably underestimated the uncertainty in the EMRI SGWB by assuming the fiducial model of \cite{Bonetti:2020}, this suggests that {\small LISA} will struggle to perform component separation for these low string tensions. The addition of the extra information from a midband experiment resolved this issue. Cosmic strings were separated from the EMRI SGWB with a $95$\% confidence interval on the tension of $G \mu = 4\times 10^{-17} - 1.7 \times 10^{-16}$ for TianGo. For {\small B-DECIGO} the interval was slightly narrower, $6\times 10^{-17} - 1.65 \times 10^{-16}$.

\subsection{Phase Transitions}
\subsubsection{Constraints}
\begin{figure}
  \includegraphics[width=0.5\textwidth]{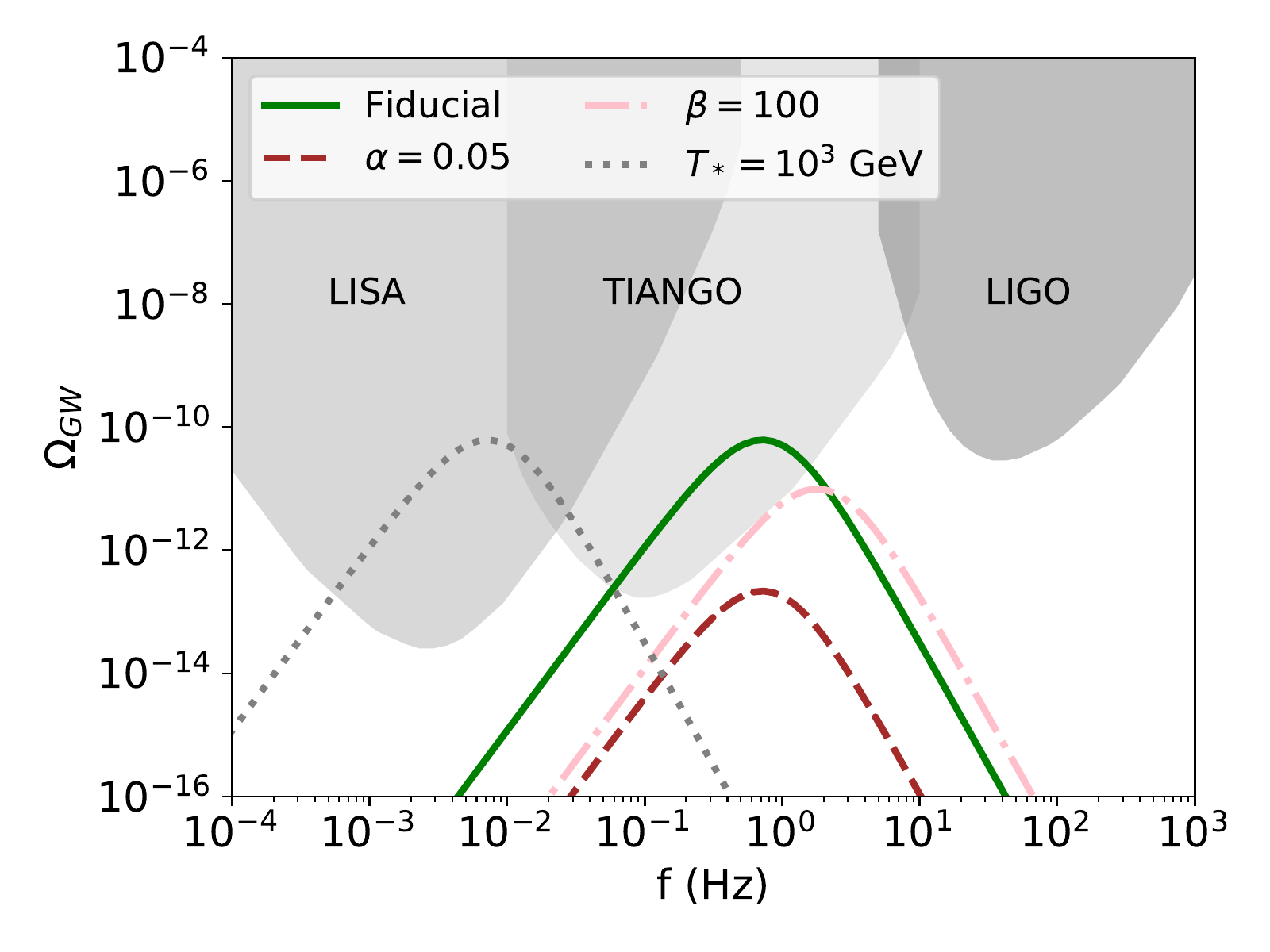}
\caption{Stochastic gravitational wave background signals from phase transitions. The fiducial model (solid, green) has $\beta/H_* = 40$, $\alpha = 0.5$ and $T_* = 10^5$ GeV. The other curves differ from the fiducial model only in the listed parameter. Hence the curve labelled $T_* = 10^3$ GeV has $\beta/H_* = 40$ and $\alpha = 0.5$. Grey shaded regions show experimental power law sensitivity curves with SNR$=1$.}
  \label{fig:phase}
\end{figure}

\begin{figure*}
 \includegraphics[width=0.90\textwidth]{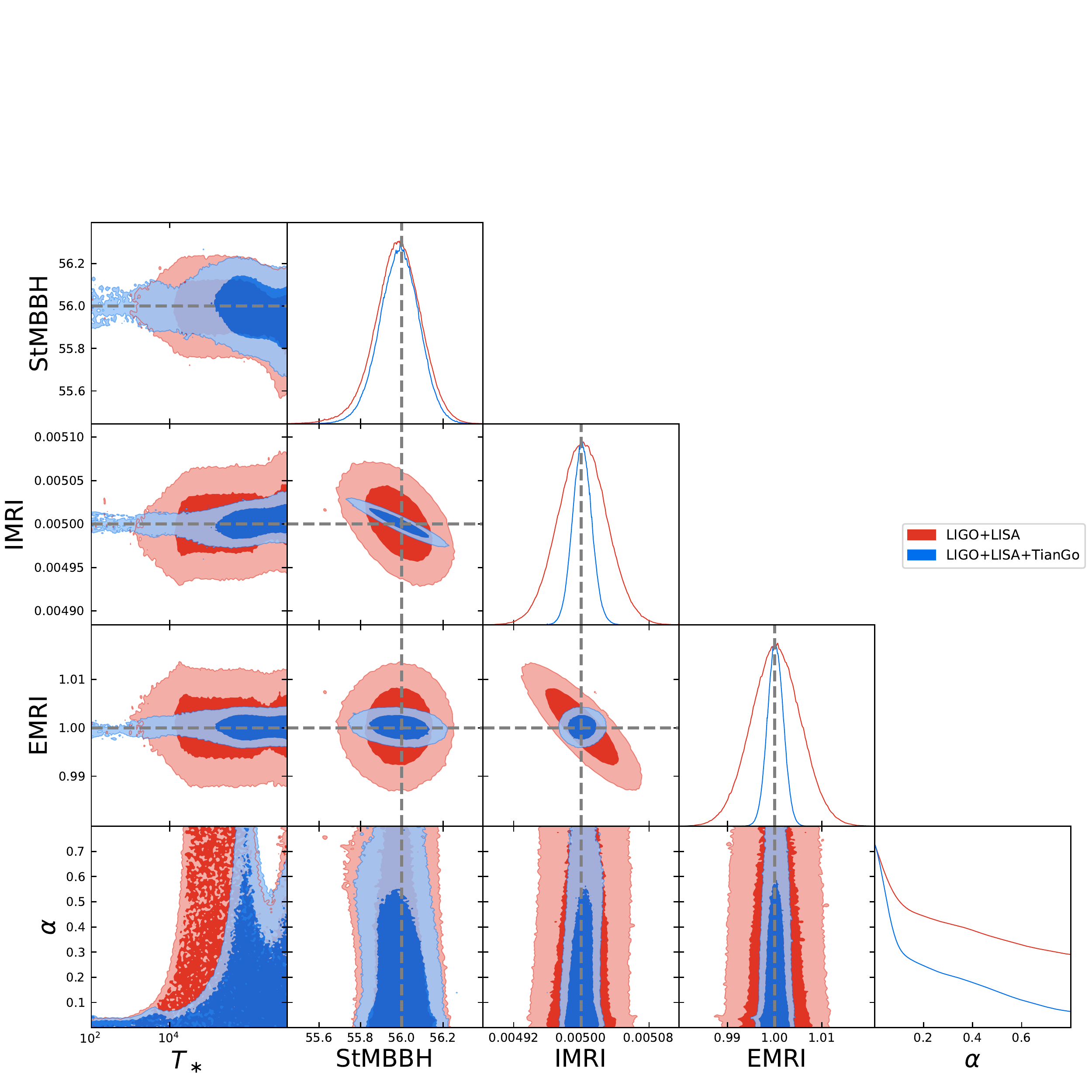}
 \caption{Markov chain samples for the phase transition likelihood function. Red: Including {\small LISA} and LIGO but no midband. Blue: Including {\small LISA}, LIGO and the TianGo midband experiment. IMRI and StMBBH merger rates are shown in units of yr$^{-1}$ Gpc$^{-3}$. The EMRI SGWB amplitude parameter is given as a fraction of the fiducial model. $T_s$ is the phase transition energy in GeV. $\alpha$ is dimensionless. Dashed lines show the true parameters of the mock astrophysical model. The line plots show marginalised one-dimensional likelihoods, while the 2D shaded regions show $1-\sigma$ and $2-\sigma$ marginalised confidence interval contours for each two-parameter combination.}
 \label{fig:phasesample}
\end{figure*}

Figure~\ref{fig:phase} shows the expected SGWB signal from a variety of phase transitions. This SGWB signal is sharply peaked, at a frequency depending on the energy scale and an amplitude directly proportional to the strength of the transition. For our fiducial choice of $\beta/H_* = 40$, transitions peak in the midband region with a temperature (or energy scale) at $T_* \sim 10^4 - 10^6$ GeV. Transitions around the electroweak energy scale at $10^2 - 10^4$ GeV peak in the {\small LISA} band. Finally, strong phase transitions with $T_* = 10^7$ GeV peak in the {\small LIGO} band, although these are only detectable for $\alpha > 0.5$. A future third generation network with a sensitivity improvement of $25-100$ would further close this energy gap and improve constraints on phase transitions in this energy band to $\alpha \lesssim 0.1$.
For completeness, we also show the effect of increasing $\beta/H_*$. This increases the peak frequency by decreasing the effective bubble size $R_*$ as well as decreasing the amplitude of the SGWB.

Figure~\ref{fig:phase} thus suggests that there is a region of parameter space where the midband experiment will sharply constrain the presence of a phase transition, and a region of parameter space where the signal peaks
at lower energies, within the {\small LISA} frequency range. This is confirmed by Figure~\ref{fig:phasesample}, where we shows constraints on the phase transition parameters from our Markov chains, including only astrophysical SGWBs. Again we show {\small LISA} and LIGO only, followed by the results also including TianGo. The midband experiment does not improve constraints for phase transitions with $T_* > 10^7$ GeV, where detectability is dominated by {\small LIGO}. For transitions with $T_* < 10^4$ GeV, {\small LISA} dominates the constraints, and the midband has little effect.

For phase transitions with $T_* = 10^4 - 10^6$ GeV, the midband experiment substantially improves constraints, as these transitions peak in a frequency band where only the midband experiment has sensitivity. The {\small TianGo } experiment leaves a small window around $T_* = 10^6$ GeV where the presence of a phase transition is not well constrained. Our {\small B-DECIGO} chains show that the more sensitive experiment also closes this window.

\subsubsection{Discovery Potential}

\begin{figure}
 \includegraphics[width=0.45\textwidth]{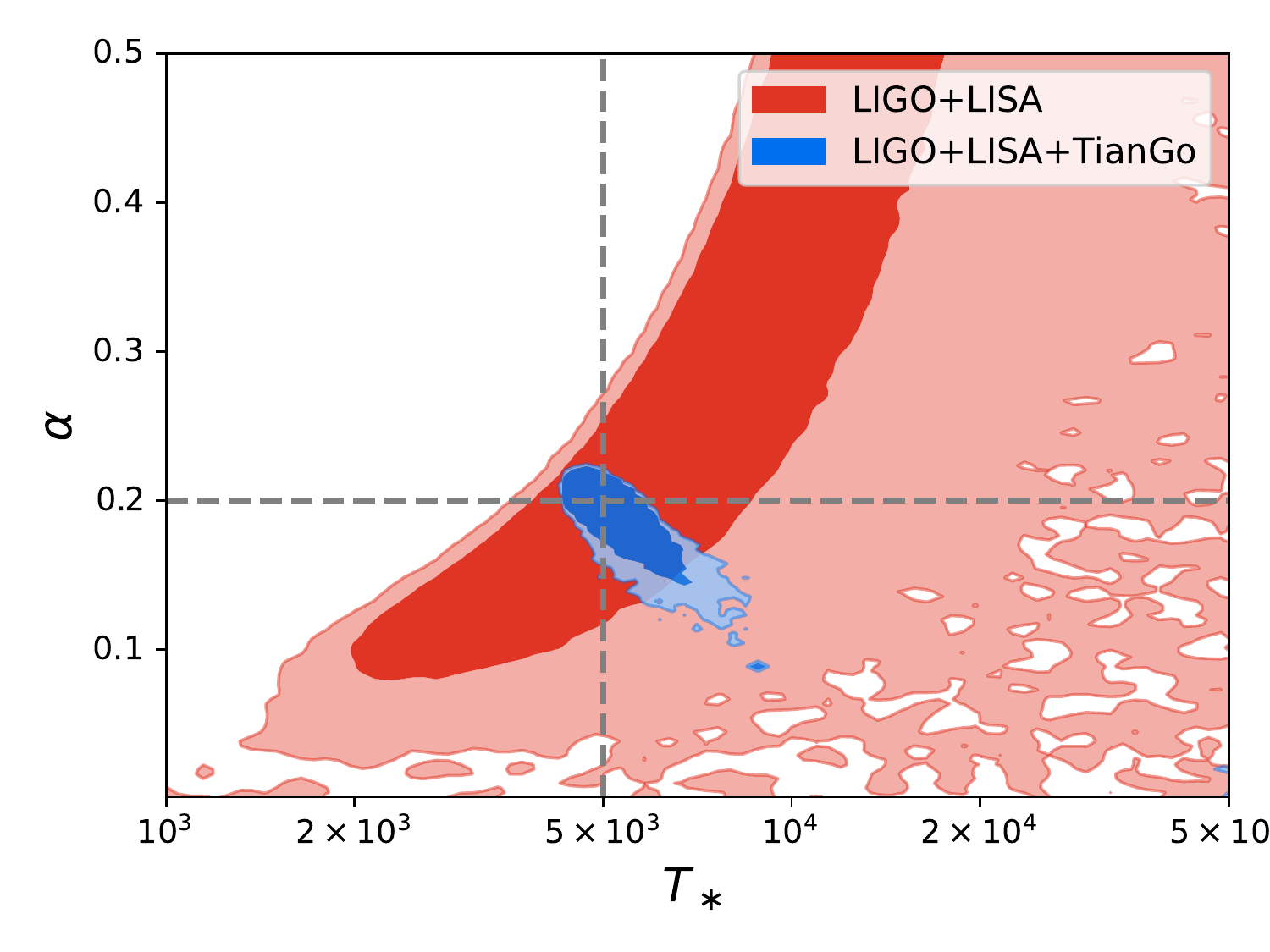}
\caption{Posterior likelihood contour for signal input with astrophysical SGWB sources and a sample phase transition at the electroweak energy scale  . Red contours include {\small LISA} and {\small LIGO} but no midband, while blue contours also include TianGo. Dashed lines show the true parameters of the mock model, $T_* = 5\times 10^3$ GeV and $\alpha = 0.2$. The 2D shaded regions show $1-\sigma$ and $2-\sigma$ marginalised confidence interval contours.}
 \label{fig:phasediscovery}
\end{figure}

To assess discovery potential, we have run chains where the mock signal includes a phase transition with a variety of energies. We set $\alpha = 0.2$. We found that, because there is uncertainty on the parameters of the phase transition, there is an energy region where experiments can detect the transition signal, but not estimate its parameters correctly. For example, a transition with $T_* = 5\times 10^3$ GeV and $\alpha = 0.2$ is within the range detectable by {\small LISA}. However, because {\small LISA} is much less sensitive $\Omega_{\rm GW}$ at higher frequencies, it is not able distinguish a SGWB which peaks within the {\small LISA} band and then diminishes in the midband from one which peaks in the midband. Thus it is difficult for {\small LISA} to estimate the parameters of the phase transition for signals near the edge of its sensitivity as it cannot measure both sides of the peak in the SGWB. Figure~\ref{fig:phasediscovery} shows our results for this parameter choice. With the combination of {\small LISA} and {\small LIGO}\footnote{{\small LIGO} does not probe these scales, but is necessary to constrain the astrophysical signal from StMBBH.}, we can only constrain that $T_* > 10^3$ GeV and $\alpha > 0.1$, with a range of possible signals at higher $T_*$ and $\alpha$ tracing the edge of the {\small LISA} PLS curve. The TianGo midband experiment provides extra frequency coverage and measures $T_*$ between $4.7\times 10^3$ and $10^4$ GeV, with $\alpha = 0.1 - 0.22$. At lower energies in the expected region for an electroweak phase transition, a cosmological signal with $T_* = 10^3$ GeV had parameters which were fairly well localised by {\small LISA} alone, which found $\alpha > 0$ at $> 2-\sigma$.

We further examined the effect of the gap TianGo leaves at $T_*\sim 10^6$ GeV on our constraints. We found that a signal with $T_* = 5\times 10^4$ GeV can be detected with a combination of {\small LISA}, {\small LIGO} and TianGo, producing $95\%$ confidence intervals of $T_* = 4\times 10^4 - 4\times 10^5$ GeV and $\alpha = 0.13 - 0.24$, with slightly smaller parameter ranges for {\small B-DECIGO}. However, a higher energy transition with $T_* = 10^5$ GeV was only reliably separable with {\small B-DECIGO}, as TianGo was unable to localise the transition energy away from the poorly measured $10^6$ GeV region. The more sensitive {\small B-DECIGO} or {\small AEDGE} is thus preferred for the most robust phase transition measurement.

\subsection{Discussion: Uncertainties in the Astrophysical SGWB Models}

Here we assess the likely uncertainty in our conclusions due to our modeling choices for astrophysical SGWB sources. The amplitude of the StMBBH background is currently uncertain by a factor of two, while the EMRI background is uncertain at an order of magnitude level. For the SGWB from IMRI mergers, even the shape is uncertain, although the power law index of the SGWB is likely to be between that of the EMRI and StMBBH backgrounds. Our quantitative forecast limits with B-DECIGO/AEDGE ($G\mu < 2.5 \times 10^{-18}$ and strong constraints on phase transitions in the $T_* = 10^4 - 10^6$ GeV range) thus represent an estimate. Qualitatively, however, the model we have built includes a separate astrophysical SGWB source in each frequency band: {\small LIGO}, {\small LISA} and the midband. As long as the IMRI SGWB is close to a power law with index $2/3$ and the EMRI SGWB close to our assumed shape, our conclusion that a midband experiment improves component separation will be valid.
Over the next decade a great deal of new data will become available. In particular, once {\small LISA} and {\small TianGo} begin taking data they should detect EMRI and IMRI mergers, and thus will better constrain the power law index of the SGWB.

\section{Conclusions}
We have examined the ability of a future midband gravitational wave experiment to improve detection prospects for cosmological SGWB signals, when combined with the existing {\small LISA} and {\small LIGO} detectors. We propose a \textit{combined power law sensitivity (CPLS)} curve as a simple way to quantify the sensitivity to SGWB of detectors covering multiple frequency bands. The CPLS shows that the midband significantly improves sensitivity to $\Omega_{\rm GW}$ in the transitional frequency region between {\small LIGO} and {\small LISA}.

We then conducted a dedicated analysis of the potential of a midband experiment to improve prospects for probing a cosmogenic SGWB signal in the presence of a variety of realistic astrophysical signals from black hole mergers. We consider phase transitions and cosmic string SGWB templates, and either TianGo or {\small B-DECIGO} as prototypical midband experiments. Our results for {\small B-DECIGO} are also valid for {\small AEDGE}, which has a similar sensitivity curve.

We find that combining a midband with existing detectors substantially improves constraints on the cosmic string tension. Upper limits on $G\mu$ strengthen by a factor of $3$ with TianGo and $11$ with {\small B-DECIGO} or {\small AEDGE}. We showed that the addition of an extra frequency channel improves component separation for cosmic string signals. We considered a signal near the lower bound accessible to {\small LISA}, $G\mu = 10^{-16}$, and showed that a midband experiment was necessary for the network to distinguish a cosmic string SGWB from the signal due to extreme mass ratio inspirals.

The phase transition energy scale sets the peak frequency of its SGWB signal. The midband experiment is extremely powerful for understanding phase transitions which peak within its observational frequency band. For our fiducial model choices, it severely constrains the strength of a phase transition in the energy scale $T_* = 10^4 - 10^6$ GeV. With {\small LISA} alone, a phase transition in this energy range is not meaningfully constrained, allowing a phase transition strength $\alpha \sim 1$. TianGo can strongly constrain $T_* = 10^4 - 10^5$ GeV to $\alpha < 0.05$. It does, however, leave an energy gap around $T_* = 10^6$ GeV which requires the more sensitive {\small B-DECIGO} or {\small AEDGE} to fully close. We show that a midband experiment allows improved parameter measurement in the presence of phase transitions at lower energies, by ruling out the possibility that the signal comes from a strong phase transition in the $T_* = 10^4 - 10^6$ GeV range. Note that our analysis fixed some observationally degenerate phase transition parameters.
By varying these parameters, we could choose plausible parameters for electroweak phase transition models for which the midband experiment would be critical for measurement of the GW signal.

For a transition at the upper end of the electroweak energy range with $T_* = 5 \times 10^3$ GeV and $\alpha = 0.2$, {\small LISA} and {\small LIGO} alone show an excess distinguishable from the astrophysical model at about $2-\sigma$. However, with the addition of TianGo to the network, it is possible to measure $\alpha$ and $T_*$ with precision and confidently distinguish them from an astrophysical signal. The midband experiment thus allows the combined detector network to measure the properties of a phase transition, while {\small LISA} alone will only show that it exists. For measuring the properties of a phase transition at a few TeV, {\small B-DECIGO} or {\small AEDGE} provides additional power by completely closing the frequency gap between {\small LISA} and {\small LIGO}. A third generation ground based detector network would further improve constraints at higher energies.

Our approach can be applied to other cosmological SGWB sources and other proposed GW detectors such as MAGIS \cite{Adamson:2018mbw} or BBO \cite{Yagi:2011wg}.
We demonstrated the significant impact of a potential midband GW experiment in boosting detection prospects for a cosmological SGWB. Our modeling code and chains are available at: \url{https://github.com/sbird/grav_midband}. Our results can be further generalized to showcase the advantages for probing new physics obtainable by invigorating a well-coordinated multiple frequency band GW program. This could include not just detectors covering LIGO, LISA and midband frequencies but also other frequency channels. For example, the $\mu$ - nano Hz range is accessible by pulsar timing arrays \cite{Janssen:2014dka, NANOGRAV} and milli - $\mu$ Hz by $\mu$Ares \cite{Sesana:2019vho}.

\section*{Acknowledgments}
\begin{acknowledgments}
SB was supported by NSF grant AST-1817256 and would like to thank his wife, Priya Bird. YC is supported in part by the US Department of Energy under award number DE-SC0008541, and thanks the Kavli Institute for Theoretical Physics (supported by the National Science Foundation under Grant No. NSF PHY-1748958) for support and hospitality while the work was being completed. We thank Mark Hindmarsh, Marek Lewicki and David Weir for helpful discussions and Chia-Feng Chiang for calculating the early time radiation density.
\end{acknowledgments}

\bibliography{gravmidband}

\begin{thebibliography}{134}%
\makeatletter
\providecommand \@ifxundefined [1]{%
 \@ifx{#1\undefined}
}%
\providecommand \@ifnum [1]{%
 \ifnum #1\expandafter \@firstoftwo
 \else \expandafter \@secondoftwo
 \fi
}%
\providecommand \@ifx [1]{%
 \ifx #1\expandafter \@firstoftwo
 \else \expandafter \@secondoftwo
 \fi
}%
\providecommand \natexlab [1]{#1}%
\providecommand \enquote  [1]{``#1''}%
\providecommand \bibnamefont  [1]{#1}%
\providecommand \bibfnamefont [1]{#1}%
\providecommand \citenamefont [1]{#1}%
\providecommand \href@noop [0]{\@secondoftwo}%
\providecommand \href [0]{\begingroup \@sanitize@url \@href}%
\providecommand \@href[1]{\@@startlink{#1}\@@href}%
\providecommand \@@href[1]{\endgroup#1\@@endlink}%
\providecommand \@sanitize@url [0]{\catcode `\\12\catcode `\$12\catcode
  `\&12\catcode `\#12\catcode `\^12\catcode `\_12\catcode `\%12\relax}%
\providecommand \@@startlink[1]{}%
\providecommand \@@endlink[0]{}%
\providecommand \url  [0]{\begingroup\@sanitize@url \@url }%
\providecommand \@url [1]{\endgroup\@href {#1}{\urlprefix }}%
\providecommand \urlprefix  [0]{URL }%
\providecommand \Eprint [0]{\href }%
\providecommand \doibase [0]{https://doi.org/}%
\providecommand \selectlanguage [0]{\@gobble}%
\providecommand \bibinfo  [0]{\@secondoftwo}%
\providecommand \bibfield  [0]{\@secondoftwo}%
\providecommand \translation [1]{[#1]}%
\providecommand \BibitemOpen [0]{}%
\providecommand \bibitemStop [0]{}%
\providecommand \bibitemNoStop [0]{.\EOS\space}%
\providecommand \EOS [0]{\spacefactor3000\relax}%
\providecommand \BibitemShut  [1]{\csname bibitem#1\endcsname}%
\let\auto@bib@innerbib\@empty
%</preamble>
\bibitem [{\citenamefont {{Abbott}}\ \emph {et~al.}(2016)\citenamefont
  {{Abbott}}, \citenamefont {{Abbott}}, \citenamefont {{Abbott}}, \citenamefont
  {{Abernathy}}, \citenamefont {{LIGO Scientific Collaboration}},\ and\
  \citenamefont {{Virgo Collaboration}}}]{Abbott:2016}%
  \BibitemOpen
  \bibfield  {author} {\bibinfo {author} {\bibfnamefont {B.~P.}\ \bibnamefont
  {{Abbott}}}, \bibinfo {author} {\bibfnamefont {R.}~\bibnamefont {{Abbott}}},
  \bibinfo {author} {\bibfnamefont {T.~D.}\ \bibnamefont {{Abbott}}}, \bibinfo
  {author} {\bibfnamefont {M.~R.}\ \bibnamefont {{Abernathy}}}, \bibinfo
  {author} {\bibnamefont {{LIGO Scientific Collaboration}}},\ and\ \bibinfo
  {author} {\bibnamefont {{Virgo Collaboration}}},\ }\bibfield  {title}
  {\bibinfo {title} {{Observation of Gravitational Waves from a Binary Black
  Hole Merger}},\ }\href {https://doi.org/10.1103/PhysRevLett.116.061102}
  {\bibfield  {journal} {\bibinfo  {journal} {\prl}\ }\textbf {\bibinfo
  {volume} {116}},\ \bibinfo {eid} {061102} (\bibinfo {year} {2016})},\ \Eprint
  {https://arxiv.org/abs/1602.03837} {arXiv:1602.03837 [gr-qc]} \BibitemShut
  {NoStop}%
\bibitem [{\citenamefont {{The LISA Collaboration}}()}]{LISASRD}%
  \BibitemOpen
  \bibfield  {author} {\bibinfo {author} {\bibnamefont {{The LISA
  Collaboration}}},\ }\href@noop {} {\bibinfo {title} {{LISA Science
  Requirements Document}}},\ \bibinfo {howpublished}
  {\url{https://www.elisascience.org/files/publications/LISA_L3_20170120.pdf}}\BibitemShut
  {NoStop}%
\bibitem [{\citenamefont {Abbott}\ \emph {et~al.}(2009)\citenamefont {Abbott}
  \emph {et~al.}}]{Abbott:2009ws}%
  \BibitemOpen
  \bibfield  {author} {\bibinfo {author} {\bibfnamefont {B.~P.}\ \bibnamefont
  {Abbott}} \emph {et~al.} (\bibinfo {collaboration} {VIRGO, LIGO
  Scientific}),\ }\bibfield  {title} {\bibinfo {title} {{An Upper Limit on the
  Stochastic Gravitational-Wave Background of Cosmological Origin}},\ }\href
  {https://doi.org/10.1038/nature08278} {\bibfield  {journal} {\bibinfo
  {journal} {Nature}\ }\textbf {\bibinfo {volume} {460}},\ \bibinfo {pages}
  {990} (\bibinfo {year} {2009})},\ \Eprint {https://arxiv.org/abs/0910.5772}
  {arXiv:0910.5772 [astro-ph.CO]} \BibitemShut {NoStop}%
%%CITATION = ARXIV:0910.5772;%%
\bibitem [{\citenamefont {{Kawamura}}\ \emph {et~al.}(2011)\citenamefont
  {{Kawamura}} \emph {et~al.}}]{DECIGO2011}%
  \BibitemOpen
  \bibfield  {author} {\bibinfo {author} {\bibfnamefont {S.}~\bibnamefont
  {{Kawamura}}} \emph {et~al.},\ }\bibfield  {title} {\bibinfo {title} {{The
  Japanese space gravitational wave antenna: DECIGO}},\ }\href
  {https://doi.org/10.1088/0264-9381/28/9/094011} {\bibfield  {journal}
  {\bibinfo  {journal} {Classical and Quantum Gravity}\ }\textbf {\bibinfo
  {volume} {28}},\ \bibinfo {eid} {094011} (\bibinfo {year}
  {2011})}\BibitemShut {NoStop}%
\bibitem [{\citenamefont {{Luo}}\ \emph {et~al.}(2016)\citenamefont {{Luo}}
  \emph {et~al.}}]{TianQin}%
  \BibitemOpen
  \bibfield  {author} {\bibinfo {author} {\bibfnamefont {J.}~\bibnamefont
  {{Luo}}} \emph {et~al.},\ }\bibfield  {title} {\bibinfo {title} {{TianQin: a
  space-borne gravitational wave detector}},\ }\href
  {https://doi.org/10.1088/0264-9381/33/3/035010} {\bibfield  {journal}
  {\bibinfo  {journal} {Classical and Quantum Gravity}\ }\textbf {\bibinfo
  {volume} {33}},\ \bibinfo {eid} {035010} (\bibinfo {year} {2016})},\ \Eprint
  {https://arxiv.org/abs/1512.02076} {arXiv:1512.02076 [astro-ph.IM]}
  \BibitemShut {NoStop}%
\bibitem [{\citenamefont {{Kuns}}\ \emph {et~al.}(2020)\citenamefont {{Kuns}},
  \citenamefont {{Yu}}, \citenamefont {{Chen}},\ and\ \citenamefont
  {{Adhikari}}}]{TianGo}%
  \BibitemOpen
  \bibfield  {author} {\bibinfo {author} {\bibfnamefont {K.~A.}\ \bibnamefont
  {{Kuns}}}, \bibinfo {author} {\bibfnamefont {H.}~\bibnamefont {{Yu}}},
  \bibinfo {author} {\bibfnamefont {Y.}~\bibnamefont {{Chen}}},\ and\ \bibinfo
  {author} {\bibfnamefont {R.~X.}\ \bibnamefont {{Adhikari}}},\ }\bibfield
  {title} {\bibinfo {title} {{Astrophysics and cosmology with a decihertz
  gravitational-wave detector: TianGO}},\ }\href
  {https://doi.org/10.1103/PhysRevD.102.043001} {\bibfield  {journal} {\bibinfo
   {journal} {\prd}\ }\textbf {\bibinfo {volume} {102}},\ \bibinfo {eid}
  {043001} (\bibinfo {year} {2020})},\ \Eprint
  {https://arxiv.org/abs/1908.06004} {arXiv:1908.06004 [gr-qc]} \BibitemShut
  {NoStop}%
\bibitem [{\citenamefont {El-Neaj}\ \emph {et~al.}(2020)\citenamefont {El-Neaj}
  \emph {et~al.}}]{AEDGE}%
  \BibitemOpen
  \bibfield  {author} {\bibinfo {author} {\bibfnamefont {Y.~A.}\ \bibnamefont
  {El-Neaj}} \emph {et~al.} (\bibinfo {collaboration} {AEDGE}),\ }\bibfield
  {title} {\bibinfo {title} {{AEDGE: Atomic Experiment for Dark Matter and
  Gravity Exploration in Space}},\ }\href
  {https://doi.org/10.1140/epjqt/s40507-020-0080-0} {\bibfield  {journal}
  {\bibinfo  {journal} {EPJ Quant. Technol.}\ }\textbf {\bibinfo {volume}
  {7}},\ \bibinfo {pages} {6} (\bibinfo {year} {2020})},\ \Eprint
  {https://arxiv.org/abs/1908.00802} {arXiv:1908.00802 [gr-qc]} \BibitemShut
  {NoStop}%
\bibitem [{\citenamefont {Graham}\ \emph {et~al.}(2017)\citenamefont {Graham},
  \citenamefont {Hogan}, \citenamefont {Kasevich}, \citenamefont {Rajendran},\
  and\ \citenamefont {Romani}}]{Graham:2017pmn}%
  \BibitemOpen
  \bibfield  {author} {\bibinfo {author} {\bibfnamefont {P.~W.}\ \bibnamefont
  {Graham}}, \bibinfo {author} {\bibfnamefont {J.~M.}\ \bibnamefont {Hogan}},
  \bibinfo {author} {\bibfnamefont {M.~A.}\ \bibnamefont {Kasevich}}, \bibinfo
  {author} {\bibfnamefont {S.}~\bibnamefont {Rajendran}},\ and\ \bibinfo
  {author} {\bibfnamefont {R.~W.}\ \bibnamefont {Romani}} (\bibinfo
  {collaboration} {MAGIS}),\ }\bibfield  {title} {\bibinfo {title} {{Mid-band
  gravitational wave detection with precision atomic sensors}},\ }\href@noop {}
  {\  (\bibinfo {year} {2017})},\ \Eprint {https://arxiv.org/abs/1711.02225}
  {arXiv:1711.02225 [astro-ph.IM]} \BibitemShut {NoStop}%
%%CITATION = ARXIV:1711.02225;%%
\bibitem [{\citenamefont {Graham}\ \emph {et~al.}(2016)\citenamefont {Graham},
  \citenamefont {Hogan}, \citenamefont {Kasevich},\ and\ \citenamefont
  {Rajendran}}]{Graham:2016plp}%
  \BibitemOpen
  \bibfield  {author} {\bibinfo {author} {\bibfnamefont {P.~W.}\ \bibnamefont
  {Graham}}, \bibinfo {author} {\bibfnamefont {J.~M.}\ \bibnamefont {Hogan}},
  \bibinfo {author} {\bibfnamefont {M.~A.}\ \bibnamefont {Kasevich}},\ and\
  \bibinfo {author} {\bibfnamefont {S.}~\bibnamefont {Rajendran}},\ }\bibfield
  {title} {\bibinfo {title} {{Resonant mode for gravitational wave detectors
  based on atom interferometry}},\ }\href
  {https://doi.org/10.1103/PhysRevD.94.104022} {\bibfield  {journal} {\bibinfo
  {journal} {Phys. Rev.}\ }\textbf {\bibinfo {volume} {D94}},\ \bibinfo {pages}
  {104022} (\bibinfo {year} {2016})},\ \Eprint
  {https://arxiv.org/abs/1606.01860} {arXiv:1606.01860 [physics.atom-ph]}
  \BibitemShut {NoStop}%
%%CITATION = ARXIV:1606.01860;%%
\bibitem [{\citenamefont {{Caprini}}\ and\ \citenamefont
  {{Figueroa}}(2018)}]{Caprini:2018}%
  \BibitemOpen
  \bibfield  {author} {\bibinfo {author} {\bibfnamefont {C.}~\bibnamefont
  {{Caprini}}}\ and\ \bibinfo {author} {\bibfnamefont {D.~G.}\ \bibnamefont
  {{Figueroa}}},\ }\bibfield  {title} {\bibinfo {title} {{Cosmological
  backgrounds of gravitational waves}},\ }\href
  {https://doi.org/10.1088/1361-6382/aac608} {\bibfield  {journal} {\bibinfo
  {journal} {Classical and Quantum Gravity}\ }\textbf {\bibinfo {volume}
  {35}},\ \bibinfo {eid} {163001} (\bibinfo {year} {2018})},\ \Eprint
  {https://arxiv.org/abs/1801.04268} {arXiv:1801.04268 [astro-ph.CO]}
  \BibitemShut {NoStop}%
\bibitem [{\citenamefont {Vilenkin}(1981)}]{Vilenkin:1981bx}%
  \BibitemOpen
  \bibfield  {author} {\bibinfo {author} {\bibfnamefont {A.}~\bibnamefont
  {Vilenkin}},\ }\bibfield  {title} {\bibinfo {title} {{Gravitational radiation
  from cosmic strings}},\ }\href {https://doi.org/10.1016/0370-2693(81)91144-8}
  {\bibfield  {journal} {\bibinfo  {journal} {Phys. Lett.}\ }\textbf {\bibinfo
  {volume} {107B}},\ \bibinfo {pages} {47} (\bibinfo {year}
  {1981})}\BibitemShut {NoStop}%
%%CITATION = PHLTA,107B,47;%%
\bibitem [{\citenamefont {Turok}(1984)}]{Turok:1984cn}%
  \BibitemOpen
  \bibfield  {author} {\bibinfo {author} {\bibfnamefont {N.}~\bibnamefont
  {Turok}},\ }\bibfield  {title} {\bibinfo {title} {{Grand Unified Strings and
  Galaxy Formation}},\ }\href {https://doi.org/10.1016/0550-3213(84)90407-3}
  {\bibfield  {journal} {\bibinfo  {journal} {Nucl. Phys.}\ }\textbf {\bibinfo
  {volume} {B242}},\ \bibinfo {pages} {520} (\bibinfo {year}
  {1984})}\BibitemShut {NoStop}%
%%CITATION = NUPHA,B242,520;%%
\bibitem [{\citenamefont {Vachaspati}\ and\ \citenamefont
  {Vilenkin}(1985)}]{Vachaspati:1984gt}%
  \BibitemOpen
  \bibfield  {author} {\bibinfo {author} {\bibfnamefont {T.}~\bibnamefont
  {Vachaspati}}\ and\ \bibinfo {author} {\bibfnamefont {A.}~\bibnamefont
  {Vilenkin}},\ }\bibfield  {title} {\bibinfo {title} {{Gravitational Radiation
  from Cosmic Strings}},\ }\href {https://doi.org/10.1103/PhysRevD.31.3052}
  {\bibfield  {journal} {\bibinfo  {journal} {Phys. Rev.}\ }\textbf {\bibinfo
  {volume} {D31}},\ \bibinfo {pages} {3052} (\bibinfo {year}
  {1985})}\BibitemShut {NoStop}%
%%CITATION = PHRVA,D31,3052;%%
\bibitem [{\citenamefont {Burden}(1985)}]{Burden:1985md}%
  \BibitemOpen
  \bibfield  {author} {\bibinfo {author} {\bibfnamefont {C.~J.}\ \bibnamefont
  {Burden}},\ }\bibfield  {title} {\bibinfo {title} {{Gravitational Radiation
  From a Particular Class of Cosmic Strings}},\ }\href
  {https://doi.org/10.1016/0370-2693(85)90326-0} {\bibfield  {journal}
  {\bibinfo  {journal} {Phys. Lett.}\ }\textbf {\bibinfo {volume} {164B}},\
  \bibinfo {pages} {277} (\bibinfo {year} {1985})}\BibitemShut {NoStop}%
%%CITATION = PHLTA,164B,277;%%
\bibitem [{\citenamefont {Olum}\ and\ \citenamefont
  {Blanco-Pillado}(2000)}]{Olum:1999sg}%
  \BibitemOpen
  \bibfield  {author} {\bibinfo {author} {\bibfnamefont {K.~D.}\ \bibnamefont
  {Olum}}\ and\ \bibinfo {author} {\bibfnamefont {J.~J.}\ \bibnamefont
  {Blanco-Pillado}},\ }\bibfield  {title} {\bibinfo {title} {{Radiation from
  cosmic string standing waves}},\ }\href
  {https://doi.org/10.1103/PhysRevLett.84.4288} {\bibfield  {journal} {\bibinfo
   {journal} {Phys. Rev. Lett.}\ }\textbf {\bibinfo {volume} {84}},\ \bibinfo
  {pages} {4288} (\bibinfo {year} {2000})},\ \Eprint
  {https://arxiv.org/abs/astro-ph/9910354} {arXiv:astro-ph/9910354 [astro-ph]}
  \BibitemShut {NoStop}%
%%CITATION = ASTRO-PH/9910354;%%
\bibitem [{\citenamefont {Moore}\ \emph {et~al.}(2001)\citenamefont {Moore},
  \citenamefont {Shellard},\ and\ \citenamefont {Martins}}]{Moore:2001px}%
  \BibitemOpen
  \bibfield  {author} {\bibinfo {author} {\bibfnamefont {J.~N.}\ \bibnamefont
  {Moore}}, \bibinfo {author} {\bibfnamefont {E.~P.~S.}\ \bibnamefont
  {Shellard}},\ and\ \bibinfo {author} {\bibfnamefont {C.~J. A.~P.}\
  \bibnamefont {Martins}},\ }\bibfield  {title} {\bibinfo {title} {{On the
  evolution of Abelian-Higgs string networks}},\ }\href
  {https://doi.org/10.1103/PhysRevD.65.023503} {\bibfield  {journal} {\bibinfo
  {journal} {Phys. Rev.}\ }\textbf {\bibinfo {volume} {D65}},\ \bibinfo {pages}
  {023503} (\bibinfo {year} {2001})},\ \Eprint
  {https://arxiv.org/abs/hep-ph/0107171} {arXiv:hep-ph/0107171 [hep-ph]}
  \BibitemShut {NoStop}%
%%CITATION = HEP-PH/0107171;%%
\bibitem [{\citenamefont {Nielsen}\ and\ \citenamefont
  {Olesen}(1973)}]{Nielsen:1973cs}%
  \BibitemOpen
  \bibfield  {author} {\bibinfo {author} {\bibfnamefont {H.~B.}\ \bibnamefont
  {Nielsen}}\ and\ \bibinfo {author} {\bibfnamefont {P.}~\bibnamefont
  {Olesen}},\ }\bibfield  {title} {\bibinfo {title} {{Vortex Line Models for
  Dual Strings}},\ }\href {https://doi.org/10.1016/0550-3213(73)90350-7}
  {\bibfield  {journal} {\bibinfo  {journal} {Nucl. Phys.}\ }\textbf {\bibinfo
  {volume} {B61}},\ \bibinfo {pages} {45} (\bibinfo {year} {1973})},\ \bibinfo
  {note} {[,302(1973)]}\BibitemShut {NoStop}%
%%CITATION = NUPHA,B61,45;%%
\bibitem [{\citenamefont {Kibble}(1976)}]{Kibble:1976sj}%
  \BibitemOpen
  \bibfield  {author} {\bibinfo {author} {\bibfnamefont {T.~W.~B.}\
  \bibnamefont {Kibble}},\ }\bibfield  {title} {\bibinfo {title} {{Topology of
  Cosmic Domains and Strings}},\ }\href
  {https://doi.org/10.1088/0305-4470/9/8/029} {\bibfield  {journal} {\bibinfo
  {journal} {J. Phys.}\ }\textbf {\bibinfo {volume} {A9}},\ \bibinfo {pages}
  {1387} (\bibinfo {year} {1976})}\BibitemShut {NoStop}%
%%CITATION = JPAGA,A9,1387;%%
\bibitem [{\citenamefont {Jackson}\ \emph {et~al.}(2005)\citenamefont
  {Jackson}, \citenamefont {Jones},\ and\ \citenamefont
  {Polchinski}}]{Jackson:2004zg}%
  \BibitemOpen
  \bibfield  {author} {\bibinfo {author} {\bibfnamefont {M.~G.}\ \bibnamefont
  {Jackson}}, \bibinfo {author} {\bibfnamefont {N.~T.}\ \bibnamefont {Jones}},\
  and\ \bibinfo {author} {\bibfnamefont {J.}~\bibnamefont {Polchinski}},\
  }\bibfield  {title} {\bibinfo {title} {{Collisions of cosmic F and
  D-strings}},\ }\href {https://doi.org/10.1088/1126-6708/2005/10/013}
  {\bibfield  {journal} {\bibinfo  {journal} {JHEP}\ }\textbf {\bibinfo
  {volume} {10}},\ \bibinfo {pages} {013}},\ \Eprint
  {https://arxiv.org/abs/hep-th/0405229} {arXiv:hep-th/0405229 [hep-th]}
  \BibitemShut {NoStop}%
%%CITATION = HEP-TH/0405229;%%
\bibitem [{\citenamefont {Tye}\ \emph {et~al.}(2005)\citenamefont {Tye},
  \citenamefont {Wasserman},\ and\ \citenamefont {Wyman}}]{Tye:2005fn}%
  \BibitemOpen
  \bibfield  {author} {\bibinfo {author} {\bibfnamefont {S.~H.~H.}\
  \bibnamefont {Tye}}, \bibinfo {author} {\bibfnamefont {I.}~\bibnamefont
  {Wasserman}},\ and\ \bibinfo {author} {\bibfnamefont {M.}~\bibnamefont
  {Wyman}},\ }\bibfield  {title} {\bibinfo {title} {{Scaling of multi-tension
  cosmic superstring networks}},\ }\href
  {https://doi.org/10.1103/PhysRevD.71.103508, 10.1103/PhysRevD.71.129906}
  {\bibfield  {journal} {\bibinfo  {journal} {Phys. Rev.}\ }\textbf {\bibinfo
  {volume} {D71}},\ \bibinfo {pages} {103508} (\bibinfo {year} {2005})},\
  \bibinfo {note} {[Erratum: Phys. Rev.D71,129906(2005)]},\ \Eprint
  {https://arxiv.org/abs/astro-ph/0503506} {arXiv:astro-ph/0503506 [astro-ph]}
  \BibitemShut {NoStop}%
%%CITATION = ASTRO-PH/0503506;%%
\bibitem [{\citenamefont {Dubath}\ \emph {et~al.}(2008)\citenamefont {Dubath},
  \citenamefont {Polchinski},\ and\ \citenamefont {Rocha}}]{Dubath:2007mf}%
  \BibitemOpen
  \bibfield  {author} {\bibinfo {author} {\bibfnamefont {F.}~\bibnamefont
  {Dubath}}, \bibinfo {author} {\bibfnamefont {J.}~\bibnamefont {Polchinski}},\
  and\ \bibinfo {author} {\bibfnamefont {J.~V.}\ \bibnamefont {Rocha}},\
  }\bibfield  {title} {\bibinfo {title} {{Cosmic String Loops, Large and
  Small}},\ }\href {https://doi.org/10.1103/PhysRevD.77.123528} {\bibfield
  {journal} {\bibinfo  {journal} {Phys. Rev.}\ }\textbf {\bibinfo {volume}
  {D77}},\ \bibinfo {pages} {123528} (\bibinfo {year} {2008})},\ \Eprint
  {https://arxiv.org/abs/0711.0994} {arXiv:0711.0994 [astro-ph]} \BibitemShut
  {NoStop}%
%%CITATION = ARXIV:0711.0994;%%
\bibitem [{\citenamefont {Figueroa}\ \emph {et~al.}(2020)\citenamefont
  {Figueroa}, \citenamefont {Hindmarsh}, \citenamefont {Lizarraga},\ and\
  \citenamefont {Urrestilla}}]{Figueroa:2020}%
  \BibitemOpen
  \bibfield  {author} {\bibinfo {author} {\bibfnamefont {D.~G.}\ \bibnamefont
  {Figueroa}}, \bibinfo {author} {\bibfnamefont {M.}~\bibnamefont {Hindmarsh}},
  \bibinfo {author} {\bibfnamefont {J.}~\bibnamefont {Lizarraga}},\ and\
  \bibinfo {author} {\bibfnamefont {J.}~\bibnamefont {Urrestilla}},\ }\bibfield
   {title} {\bibinfo {title} {{Irreducible background of gravitational waves
  from a cosmic defect network: update and comparison of numerical
  techniques}},\ }\href {https://doi.org/10.1103/PhysRevD.102.103516}
  {\bibfield  {journal} {\bibinfo  {journal} {Phys. Rev. D}\ }\textbf {\bibinfo
  {volume} {102}},\ \bibinfo {pages} {103516} (\bibinfo {year} {2020})},\
  \Eprint {https://arxiv.org/abs/2007.03337} {arXiv:2007.03337 [astro-ph.CO]}
  \BibitemShut {NoStop}%
\bibitem [{\citenamefont {Caprini}\ \emph {et~al.}(2009)\citenamefont
  {Caprini}, \citenamefont {Durrer}, \citenamefont {Konstandin},\ and\
  \citenamefont {Servant}}]{Caprini:2009fx}%
  \BibitemOpen
  \bibfield  {author} {\bibinfo {author} {\bibfnamefont {C.}~\bibnamefont
  {Caprini}}, \bibinfo {author} {\bibfnamefont {R.}~\bibnamefont {Durrer}},
  \bibinfo {author} {\bibfnamefont {T.}~\bibnamefont {Konstandin}},\ and\
  \bibinfo {author} {\bibfnamefont {G.}~\bibnamefont {Servant}},\ }\bibfield
  {title} {\bibinfo {title} {{General Properties of the Gravitational Wave
  Spectrum from Phase Transitions}},\ }\href
  {https://doi.org/10.1103/PhysRevD.79.083519} {\bibfield  {journal} {\bibinfo
  {journal} {Phys. Rev.}\ }\textbf {\bibinfo {volume} {D79}},\ \bibinfo {pages}
  {083519} (\bibinfo {year} {2009})},\ \Eprint
  {https://arxiv.org/abs/0901.1661} {arXiv:0901.1661 [astro-ph.CO]}
  \BibitemShut {NoStop}%
%%CITATION = ARXIV:0901.1661;%%
\bibitem [{\citenamefont {Schwaller}(2015)}]{Schwaller:2015tja}%
  \BibitemOpen
  \bibfield  {author} {\bibinfo {author} {\bibfnamefont {P.}~\bibnamefont
  {Schwaller}},\ }\bibfield  {title} {\bibinfo {title} {{Gravitational Waves
  from a Dark Phase Transition}},\ }\href
  {https://doi.org/10.1103/PhysRevLett.115.181101} {\bibfield  {journal}
  {\bibinfo  {journal} {Phys. Rev. Lett.}\ }\textbf {\bibinfo {volume} {115}},\
  \bibinfo {pages} {181101} (\bibinfo {year} {2015})},\ \Eprint
  {https://arxiv.org/abs/1504.07263} {arXiv:1504.07263 [hep-ph]} \BibitemShut
  {NoStop}%
%%CITATION = ARXIV:1504.07263;%%
\bibitem [{\citenamefont {Helmboldt}\ \emph {et~al.}(2019)\citenamefont
  {Helmboldt}, \citenamefont {Kubo},\ and\ \citenamefont {van~der
  Woude}}]{Helmboldt:2019pan}%
  \BibitemOpen
  \bibfield  {author} {\bibinfo {author} {\bibfnamefont {A.~J.}\ \bibnamefont
  {Helmboldt}}, \bibinfo {author} {\bibfnamefont {J.}~\bibnamefont {Kubo}},\
  and\ \bibinfo {author} {\bibfnamefont {S.}~\bibnamefont {van~der Woude}},\
  }\bibfield  {title} {\bibinfo {title} {{Observational prospects for
  gravitational waves from hidden or dark chiral phase transitions}},\ }\href
  {https://doi.org/10.1103/PhysRevD.100.055025} {\bibfield  {journal} {\bibinfo
   {journal} {Phys. Rev. D}\ }\textbf {\bibinfo {volume} {100}},\ \bibinfo
  {pages} {055025} (\bibinfo {year} {2019})},\ \Eprint
  {https://arxiv.org/abs/1904.07891} {arXiv:1904.07891 [hep-ph]} \BibitemShut
  {NoStop}%
\bibitem [{\citenamefont {Hall}\ \emph {et~al.}(2020)\citenamefont {Hall},
  \citenamefont {Konstandin}, \citenamefont {McGehee}, \citenamefont
  {Murayama},\ and\ \citenamefont {Servant}}]{Hall:2019ank}%
  \BibitemOpen
  \bibfield  {author} {\bibinfo {author} {\bibfnamefont {E.}~\bibnamefont
  {Hall}}, \bibinfo {author} {\bibfnamefont {T.}~\bibnamefont {Konstandin}},
  \bibinfo {author} {\bibfnamefont {R.}~\bibnamefont {McGehee}}, \bibinfo
  {author} {\bibfnamefont {H.}~\bibnamefont {Murayama}},\ and\ \bibinfo
  {author} {\bibfnamefont {G.}~\bibnamefont {Servant}},\ }\bibfield  {title}
  {\bibinfo {title} {{Baryogenesis From a Dark First-Order Phase Transition}},\
  }\href {https://doi.org/10.1007/JHEP04(2020)042} {\bibfield  {journal}
  {\bibinfo  {journal} {JHEP}\ }\textbf {\bibinfo {volume} {04}},\ \bibinfo
  {pages} {042}},\ \Eprint {https://arxiv.org/abs/1910.08068} {arXiv:1910.08068
  [hep-ph]} \BibitemShut {NoStop}%
%%CITATION = ARXIV:1910.08068;%%
\bibitem [{\citenamefont {Cohen}\ \emph {et~al.}(1991)\citenamefont {Cohen},
  \citenamefont {Kaplan},\ and\ \citenamefont {Nelson}}]{Cohen:1990it}%
  \BibitemOpen
  \bibfield  {author} {\bibinfo {author} {\bibfnamefont {A.~G.}\ \bibnamefont
  {Cohen}}, \bibinfo {author} {\bibfnamefont {D.~B.}\ \bibnamefont {Kaplan}},\
  and\ \bibinfo {author} {\bibfnamefont {A.~E.}\ \bibnamefont {Nelson}},\
  }\bibfield  {title} {\bibinfo {title} {{Baryogenesis at the weak phase
  transition}},\ }\href {https://doi.org/10.1016/0550-3213(91)90395-E}
  {\bibfield  {journal} {\bibinfo  {journal} {Nucl. Phys.}\ }\textbf {\bibinfo
  {volume} {B349}},\ \bibinfo {pages} {727} (\bibinfo {year}
  {1991})}\BibitemShut {NoStop}%
%%CITATION = NUPHA,B349,727;%%
\bibitem [{\citenamefont {Anderson}\ and\ \citenamefont
  {Hall}(1992)}]{Anderson:1991zb}%
  \BibitemOpen
  \bibfield  {author} {\bibinfo {author} {\bibfnamefont {G.~W.}\ \bibnamefont
  {Anderson}}\ and\ \bibinfo {author} {\bibfnamefont {L.~J.}\ \bibnamefont
  {Hall}},\ }\bibfield  {title} {\bibinfo {title} {{The Electroweak phase
  transition and baryogenesis}},\ }\href
  {https://doi.org/10.1103/PhysRevD.45.2685} {\bibfield  {journal} {\bibinfo
  {journal} {Phys. Rev.}\ }\textbf {\bibinfo {volume} {D45}},\ \bibinfo {pages}
  {2685} (\bibinfo {year} {1992})}\BibitemShut {NoStop}%
%%CITATION = PHRVA,D45,2685;%%
\bibitem [{\citenamefont {Cui}\ \emph {et~al.}(2018)\citenamefont {Cui},
  \citenamefont {Lewicki}, \citenamefont {Morrissey},\ and\ \citenamefont
  {Wells}}]{Cui:2017ufi}%
  \BibitemOpen
  \bibfield  {author} {\bibinfo {author} {\bibfnamefont {Y.}~\bibnamefont
  {Cui}}, \bibinfo {author} {\bibfnamefont {M.}~\bibnamefont {Lewicki}},
  \bibinfo {author} {\bibfnamefont {D.~E.}\ \bibnamefont {Morrissey}},\ and\
  \bibinfo {author} {\bibfnamefont {J.~D.}\ \bibnamefont {Wells}},\ }\bibfield
  {title} {\bibinfo {title} {{Cosmic Archaeology with Gravitational Waves from
  Cosmic Strings}},\ }\href {https://doi.org/10.1103/PhysRevD.97.123505}
  {\bibfield  {journal} {\bibinfo  {journal} {Phys. Rev.}\ }\textbf {\bibinfo
  {volume} {D97}},\ \bibinfo {pages} {123505} (\bibinfo {year} {2018})},\
  \Eprint {https://arxiv.org/abs/1711.03104} {arXiv:1711.03104 [hep-ph]}
  \BibitemShut {NoStop}%
%%CITATION = ARXIV:1711.03104;%%
\bibitem [{\citenamefont {Caldwell}\ \emph {et~al.}(2019)\citenamefont
  {Caldwell}, \citenamefont {Smith},\ and\ \citenamefont
  {Walker}}]{Caldwell:2018giq}%
  \BibitemOpen
  \bibfield  {author} {\bibinfo {author} {\bibfnamefont {R.~R.}\ \bibnamefont
  {Caldwell}}, \bibinfo {author} {\bibfnamefont {T.~L.}\ \bibnamefont
  {Smith}},\ and\ \bibinfo {author} {\bibfnamefont {D.~G.}\ \bibnamefont
  {Walker}},\ }\bibfield  {title} {\bibinfo {title} {{Using a Primordial
  Gravitational Wave Background to Illuminate New Physics}},\ }\href
  {https://doi.org/10.1103/PhysRevD.100.043513} {\bibfield  {journal} {\bibinfo
   {journal} {Phys. Rev. D}\ }\textbf {\bibinfo {volume} {100}},\ \bibinfo
  {pages} {043513} (\bibinfo {year} {2019})},\ \Eprint
  {https://arxiv.org/abs/1812.07577} {arXiv:1812.07577 [astro-ph.CO]}
  \BibitemShut {NoStop}%
\bibitem [{\citenamefont {{Cui}}\ \emph {et~al.}(2019)\citenamefont {{Cui}},
  \citenamefont {{Lewicki}}, \citenamefont {{Morrissey}},\ and\ \citenamefont
  {{Wells}}}]{Cui:2018}%
  \BibitemOpen
  \bibfield  {author} {\bibinfo {author} {\bibfnamefont {Y.}~\bibnamefont
  {{Cui}}}, \bibinfo {author} {\bibfnamefont {M.}~\bibnamefont {{Lewicki}}},
  \bibinfo {author} {\bibfnamefont {D.~E.}\ \bibnamefont {{Morrissey}}},\ and\
  \bibinfo {author} {\bibfnamefont {J.~D.}\ \bibnamefont {{Wells}}},\
  }\bibfield  {title} {\bibinfo {title} {{Probing the pre-BBN universe with
  gravitational waves from cosmic strings}},\ }\href
  {https://doi.org/10.1007/JHEP01(2019)081} {\bibfield  {journal} {\bibinfo
  {journal} {Journal of High Energy Physics}\ }\textbf {\bibinfo {volume}
  {2019}},\ \bibinfo {eid} {81} (\bibinfo {year} {2019})},\ \Eprint
  {https://arxiv.org/abs/1808.08968} {arXiv:1808.08968 [hep-ph]} \BibitemShut
  {NoStop}%
\bibitem [{\citenamefont {Cui}\ \emph {et~al.}(2020)\citenamefont {Cui},
  \citenamefont {Lewicki},\ and\ \citenamefont {Morrissey}}]{Cui:2019kkd}%
  \BibitemOpen
  \bibfield  {author} {\bibinfo {author} {\bibfnamefont {Y.}~\bibnamefont
  {Cui}}, \bibinfo {author} {\bibfnamefont {M.}~\bibnamefont {Lewicki}},\ and\
  \bibinfo {author} {\bibfnamefont {D.~E.}\ \bibnamefont {Morrissey}},\
  }\bibfield  {title} {\bibinfo {title} {{Gravitational Wave Bursts as
  Harbingers of Cosmic Strings Diluted by Inflation}},\ }\href
  {https://doi.org/10.1103/PhysRevLett.125.211302} {\bibfield  {journal}
  {\bibinfo  {journal} {Phys. Rev. Lett.}\ }\textbf {\bibinfo {volume} {125}},\
  \bibinfo {pages} {211302} (\bibinfo {year} {2020})},\ \Eprint
  {https://arxiv.org/abs/1912.08832} {arXiv:1912.08832 [hep-ph]} \BibitemShut
  {NoStop}%
\bibitem [{\citenamefont {Chang}\ and\ \citenamefont
  {Cui}(2020)}]{Chang:2019mza}%
  \BibitemOpen
  \bibfield  {author} {\bibinfo {author} {\bibfnamefont {C.-F.}\ \bibnamefont
  {Chang}}\ and\ \bibinfo {author} {\bibfnamefont {Y.}~\bibnamefont {Cui}},\
  }\bibfield  {title} {\bibinfo {title} {{Stochastic Gravitational Wave
  Background from Global Cosmic Strings}},\ }\href
  {https://doi.org/10.1016/j.dark.2020.100604} {\bibfield  {journal} {\bibinfo
  {journal} {Phys. Dark Univ.}\ }\textbf {\bibinfo {volume} {29}},\ \bibinfo
  {pages} {100604} (\bibinfo {year} {2020})},\ \Eprint
  {https://arxiv.org/abs/1910.04781} {arXiv:1910.04781 [hep-ph]} \BibitemShut
  {NoStop}%
%%CITATION = ARXIV:1910.04781;%%
\bibitem [{\citenamefont {Dror}\ \emph {et~al.}(2020)\citenamefont {Dror},
  \citenamefont {Hiramatsu}, \citenamefont {Kohri}, \citenamefont {Murayama},\
  and\ \citenamefont {White}}]{Dror:2019syi}%
  \BibitemOpen
  \bibfield  {author} {\bibinfo {author} {\bibfnamefont {J.~A.}\ \bibnamefont
  {Dror}}, \bibinfo {author} {\bibfnamefont {T.}~\bibnamefont {Hiramatsu}},
  \bibinfo {author} {\bibfnamefont {K.}~\bibnamefont {Kohri}}, \bibinfo
  {author} {\bibfnamefont {H.}~\bibnamefont {Murayama}},\ and\ \bibinfo
  {author} {\bibfnamefont {G.}~\bibnamefont {White}},\ }\bibfield  {title}
  {\bibinfo {title} {{Testing the Seesaw Mechanism and Leptogenesis with
  Gravitational Waves}},\ }\href
  {https://doi.org/10.1103/PhysRevLett.124.041804} {\bibfield  {journal}
  {\bibinfo  {journal} {Phys. Rev. Lett.}\ }\textbf {\bibinfo {volume} {124}},\
  \bibinfo {pages} {041804} (\bibinfo {year} {2020})},\ \Eprint
  {https://arxiv.org/abs/1908.03227} {arXiv:1908.03227 [hep-ph]} \BibitemShut
  {NoStop}%
%%CITATION = ARXIV:1908.03227;%%
\bibitem [{\citenamefont {Buchmuller}\ \emph
  {et~al.}(2020{\natexlab{a}})\citenamefont {Buchmuller}, \citenamefont
  {Domcke}, \citenamefont {Murayama},\ and\ \citenamefont
  {Schmitz}}]{Buchmuller:2019gfy}%
  \BibitemOpen
  \bibfield  {author} {\bibinfo {author} {\bibfnamefont {W.}~\bibnamefont
  {Buchmuller}}, \bibinfo {author} {\bibfnamefont {V.}~\bibnamefont {Domcke}},
  \bibinfo {author} {\bibfnamefont {H.}~\bibnamefont {Murayama}},\ and\
  \bibinfo {author} {\bibfnamefont {K.}~\bibnamefont {Schmitz}},\ }\bibfield
  {title} {\bibinfo {title} {{Probing the scale of grand unification with
  gravitational waves}},\ }\href
  {https://doi.org/10.1016/j.physletb.2020.135764} {\bibfield  {journal}
  {\bibinfo  {journal} {Phys. Lett. B}\ }\textbf {\bibinfo {volume} {809}},\
  \bibinfo {pages} {135764} (\bibinfo {year} {2020}{\natexlab{a}})},\ \Eprint
  {https://arxiv.org/abs/1912.03695} {arXiv:1912.03695 [hep-ph]} \BibitemShut
  {NoStop}%
\bibitem [{\citenamefont {Gouttenoire}\ \emph
  {et~al.}(2020{\natexlab{a}})\citenamefont {Gouttenoire}, \citenamefont
  {Servant},\ and\ \citenamefont {Simakachorn}}]{Gouttenoire:2019kij}%
  \BibitemOpen
  \bibfield  {author} {\bibinfo {author} {\bibfnamefont {Y.}~\bibnamefont
  {Gouttenoire}}, \bibinfo {author} {\bibfnamefont {G.}~\bibnamefont
  {Servant}},\ and\ \bibinfo {author} {\bibfnamefont {P.}~\bibnamefont
  {Simakachorn}},\ }\bibfield  {title} {\bibinfo {title} {{Beyond the Standard
  Models with Cosmic Strings}},\ }\href
  {https://doi.org/10.1088/1475-7516/2020/07/032} {\bibfield  {journal}
  {\bibinfo  {journal} {JCAP}\ }\textbf {\bibinfo {volume} {07}},\ \bibinfo
  {pages} {032}},\ \Eprint {https://arxiv.org/abs/1912.02569} {arXiv:1912.02569
  [hep-ph]} \BibitemShut {NoStop}%
\bibitem [{\citenamefont {Gouttenoire}\ \emph
  {et~al.}(2020{\natexlab{b}})\citenamefont {Gouttenoire}, \citenamefont
  {Servant},\ and\ \citenamefont {Simakachorn}}]{Gouttenoire:2019rtn}%
  \BibitemOpen
  \bibfield  {author} {\bibinfo {author} {\bibfnamefont {Y.}~\bibnamefont
  {Gouttenoire}}, \bibinfo {author} {\bibfnamefont {G.}~\bibnamefont
  {Servant}},\ and\ \bibinfo {author} {\bibfnamefont {P.}~\bibnamefont
  {Simakachorn}},\ }\bibfield  {title} {\bibinfo {title} {{BSM with Cosmic
  Strings: Heavy, up to EeV mass, Unstable Particles}},\ }\href
  {https://doi.org/10.1088/1475-7516/2020/07/016} {\bibfield  {journal}
  {\bibinfo  {journal} {JCAP}\ }\textbf {\bibinfo {volume} {07}},\ \bibinfo
  {pages} {016}},\ \Eprint {https://arxiv.org/abs/1912.03245} {arXiv:1912.03245
  [hep-ph]} \BibitemShut {NoStop}%
\bibitem [{\citenamefont {Dev}\ \emph {et~al.}(2019)\citenamefont {Dev},
  \citenamefont {Ferrer}, \citenamefont {Zhang},\ and\ \citenamefont
  {Zhang}}]{Dev:2019njv}%
  \BibitemOpen
  \bibfield  {author} {\bibinfo {author} {\bibfnamefont {P.~B.}\ \bibnamefont
  {Dev}}, \bibinfo {author} {\bibfnamefont {F.}~\bibnamefont {Ferrer}},
  \bibinfo {author} {\bibfnamefont {Y.}~\bibnamefont {Zhang}},\ and\ \bibinfo
  {author} {\bibfnamefont {Y.}~\bibnamefont {Zhang}},\ }\bibfield  {title}
  {\bibinfo {title} {{Gravitational Waves from First-Order Phase Transition in
  a Simple Axion-Like Particle Model}},\ }\href
  {https://doi.org/10.1088/1475-7516/2019/11/006} {\bibfield  {journal}
  {\bibinfo  {journal} {JCAP}\ }\textbf {\bibinfo {volume} {11}},\ \bibinfo
  {pages} {006}},\ \Eprint {https://arxiv.org/abs/1905.00891} {arXiv:1905.00891
  [hep-ph]} \BibitemShut {NoStop}%
\bibitem [{\citenamefont {Abbott}\ \emph {et~al.}(2018)\citenamefont {Abbott}
  \emph {et~al.}}]{Abbott:2017mem}%
  \BibitemOpen
  \bibfield  {author} {\bibinfo {author} {\bibfnamefont {B.~P.}\ \bibnamefont
  {Abbott}} \emph {et~al.} (\bibinfo {collaboration} {LIGO Scientific,
  Virgo}),\ }\bibfield  {title} {\bibinfo {title} {{Constraints on cosmic
  strings using data from the first Advanced LIGO observing run}},\ }\href
  {https://doi.org/10.1103/PhysRevD.97.102002} {\bibfield  {journal} {\bibinfo
  {journal} {Phys. Rev.}\ }\textbf {\bibinfo {volume} {D97}},\ \bibinfo {pages}
  {102002} (\bibinfo {year} {2018})},\ \Eprint
  {https://arxiv.org/abs/1712.01168} {arXiv:1712.01168 [gr-qc]} \BibitemShut
  {NoStop}%
%%CITATION = ARXIV:1712.01168;%%
\bibitem [{\citenamefont {Caprini}\ \emph {et~al.}(2020)\citenamefont {Caprini}
  \emph {et~al.}}]{Caprini:2019egz}%
  \BibitemOpen
  \bibfield  {author} {\bibinfo {author} {\bibfnamefont {C.}~\bibnamefont
  {Caprini}} \emph {et~al.},\ }\bibfield  {title} {\bibinfo {title} {{Detecting
  gravitational waves from cosmological phase transitions with LISA: an
  update}},\ }\href {https://doi.org/10.1088/1475-7516/2020/03/024} {\bibfield
  {journal} {\bibinfo  {journal} {JCAP}\ }\textbf {\bibinfo {volume} {2003}},\
  \bibinfo {pages} {024}},\ \Eprint {https://arxiv.org/abs/1910.13125}
  {arXiv:1910.13125 [astro-ph.CO]} \BibitemShut {NoStop}%
%%CITATION = ARXIV:1910.13125;%%
\bibitem [{\citenamefont {Auclair}\ \emph {et~al.}(2020)\citenamefont {Auclair}
  \emph {et~al.}}]{Auclair:2019wcv}%
  \BibitemOpen
  \bibfield  {author} {\bibinfo {author} {\bibfnamefont {P.}~\bibnamefont
  {Auclair}} \emph {et~al.},\ }\bibfield  {title} {\bibinfo {title} {{Probing
  the gravitational wave background from cosmic strings with LISA}},\ }\href
  {https://doi.org/10.1088/1475-7516/2020/04/034} {\bibfield  {journal}
  {\bibinfo  {journal} {JCAP}\ }\textbf {\bibinfo {volume} {2004}},\ \bibinfo
  {pages} {034}},\ \Eprint {https://arxiv.org/abs/1909.00819} {arXiv:1909.00819
  [astro-ph.CO]} \BibitemShut {NoStop}%
%%CITATION = ARXIV:1909.00819;%%
\bibitem [{\citenamefont {Arzoumanian}\ \emph {et~al.}(2020)\citenamefont
  {Arzoumanian} \emph {et~al.}}]{NANOGRAV}%
  \BibitemOpen
  \bibfield  {author} {\bibinfo {author} {\bibfnamefont {Z.}~\bibnamefont
  {Arzoumanian}} \emph {et~al.} (\bibinfo {collaboration} {NANOGrav}),\
  }\bibfield  {title} {\bibinfo {title} {{The NANOGrav 12.5 yr Data Set: Search
  for an Isotropic Stochastic Gravitational-wave Background}},\ }\href
  {https://doi.org/10.3847/2041-8213/abd401} {\bibfield  {journal} {\bibinfo
  {journal} {Astrophys. J. Lett.}\ }\textbf {\bibinfo {volume} {905}},\
  \bibinfo {pages} {L34} (\bibinfo {year} {2020})},\ \Eprint
  {https://arxiv.org/abs/2009.04496} {arXiv:2009.04496 [astro-ph.HE]}
  \BibitemShut {NoStop}%
\bibitem [{\citenamefont {Ellis}\ and\ \citenamefont
  {Lewicki}(2021)}]{Ellis:2020ena}%
  \BibitemOpen
  \bibfield  {author} {\bibinfo {author} {\bibfnamefont {J.}~\bibnamefont
  {Ellis}}\ and\ \bibinfo {author} {\bibfnamefont {M.}~\bibnamefont
  {Lewicki}},\ }\bibfield  {title} {\bibinfo {title} {{Cosmic String
  Interpretation of NANOGrav Pulsar Timing Data}},\ }\href
  {https://doi.org/10.1103/PhysRevLett.126.041304} {\bibfield  {journal}
  {\bibinfo  {journal} {Phys. Rev. Lett.}\ }\textbf {\bibinfo {volume} {126}},\
  \bibinfo {pages} {041304} (\bibinfo {year} {2021})},\ \Eprint
  {https://arxiv.org/abs/2009.06555} {arXiv:2009.06555 [astro-ph.CO]}
  \BibitemShut {NoStop}%
\bibitem [{\citenamefont {Addazi}\ \emph {et~al.}(2020)\citenamefont {Addazi},
  \citenamefont {Cai}, \citenamefont {Gan}, \citenamefont {Marciano},\ and\
  \citenamefont {Zeng}}]{Addazi:2020zcj}%
  \BibitemOpen
  \bibfield  {author} {\bibinfo {author} {\bibfnamefont {A.}~\bibnamefont
  {Addazi}}, \bibinfo {author} {\bibfnamefont {Y.-F.}\ \bibnamefont {Cai}},
  \bibinfo {author} {\bibfnamefont {Q.}~\bibnamefont {Gan}}, \bibinfo {author}
  {\bibfnamefont {A.}~\bibnamefont {Marciano}},\ and\ \bibinfo {author}
  {\bibfnamefont {K.}~\bibnamefont {Zeng}},\ }\bibfield  {title} {\bibinfo
  {title} {{NANOGrav results and Dark First Order Phase Transitions}},\
  }\href@noop {} {\  (\bibinfo {year} {2020})},\ \Eprint
  {https://arxiv.org/abs/2009.10327} {arXiv:2009.10327 [hep-ph]} \BibitemShut
  {NoStop}%
\bibitem [{\citenamefont {Ratzinger}\ and\ \citenamefont
  {Schwaller}(2021)}]{Ratzinger:2020koh}%
  \BibitemOpen
  \bibfield  {author} {\bibinfo {author} {\bibfnamefont {W.}~\bibnamefont
  {Ratzinger}}\ and\ \bibinfo {author} {\bibfnamefont {P.}~\bibnamefont
  {Schwaller}},\ }\bibfield  {title} {\bibinfo {title} {{Whispers from the dark
  side: Confronting light new physics with NANOGrav data}},\ }\href
  {https://doi.org/10.21468/SciPostPhys.10.2.047} {\bibfield  {journal}
  {\bibinfo  {journal} {SciPost Phys.}\ }\textbf {\bibinfo {volume} {10}},\
  \bibinfo {pages} {047} (\bibinfo {year} {2021})},\ \Eprint
  {https://arxiv.org/abs/2009.11875} {arXiv:2009.11875 [astro-ph.CO]}
  \BibitemShut {NoStop}%
\bibitem [{\citenamefont {Blasi}\ \emph {et~al.}(2021)\citenamefont {Blasi},
  \citenamefont {Brdar},\ and\ \citenamefont {Schmitz}}]{Blasi:2020mfx}%
  \BibitemOpen
  \bibfield  {author} {\bibinfo {author} {\bibfnamefont {S.}~\bibnamefont
  {Blasi}}, \bibinfo {author} {\bibfnamefont {V.}~\bibnamefont {Brdar}},\ and\
  \bibinfo {author} {\bibfnamefont {K.}~\bibnamefont {Schmitz}},\ }\bibfield
  {title} {\bibinfo {title} {{Has NANOGrav found first evidence for cosmic
  strings?}},\ }\href {https://doi.org/10.1103/PhysRevLett.126.041305}
  {\bibfield  {journal} {\bibinfo  {journal} {Phys. Rev. Lett.}\ }\textbf
  {\bibinfo {volume} {126}},\ \bibinfo {pages} {041305} (\bibinfo {year}
  {2021})},\ \Eprint {https://arxiv.org/abs/2009.06607} {arXiv:2009.06607
  [astro-ph.CO]} \BibitemShut {NoStop}%
\bibitem [{\citenamefont {Buchmuller}\ \emph
  {et~al.}(2020{\natexlab{b}})\citenamefont {Buchmuller}, \citenamefont
  {Domcke},\ and\ \citenamefont {Schmitz}}]{Buchmuller:2020lbh}%
  \BibitemOpen
  \bibfield  {author} {\bibinfo {author} {\bibfnamefont {W.}~\bibnamefont
  {Buchmuller}}, \bibinfo {author} {\bibfnamefont {V.}~\bibnamefont {Domcke}},\
  and\ \bibinfo {author} {\bibfnamefont {K.}~\bibnamefont {Schmitz}},\
  }\bibfield  {title} {\bibinfo {title} {{From NANOGrav to LIGO with metastable
  cosmic strings}},\ }\href {https://doi.org/10.1016/j.physletb.2020.135914}
  {\bibfield  {journal} {\bibinfo  {journal} {Phys. Lett. B}\ }\textbf
  {\bibinfo {volume} {811}},\ \bibinfo {pages} {135914} (\bibinfo {year}
  {2020}{\natexlab{b}})},\ \Eprint {https://arxiv.org/abs/2009.10649}
  {arXiv:2009.10649 [astro-ph.CO]} \BibitemShut {NoStop}%
\bibitem [{\citenamefont {Samanta}\ and\ \citenamefont
  {Datta}(2020)}]{Samanta:2020cdk}%
  \BibitemOpen
  \bibfield  {author} {\bibinfo {author} {\bibfnamefont {R.}~\bibnamefont
  {Samanta}}\ and\ \bibinfo {author} {\bibfnamefont {S.}~\bibnamefont
  {Datta}},\ }\bibfield  {title} {\bibinfo {title} {{Gravitational wave
  complementarity and impact of NANOGrav data on gravitational leptogenesis:
  cosmic strings}},\ }\href@noop {} {\  (\bibinfo {year} {2020})},\ \Eprint
  {https://arxiv.org/abs/2009.13452} {arXiv:2009.13452 [hep-ph]} \BibitemShut
  {NoStop}%
\bibitem [{\citenamefont {Nakai}\ \emph {et~al.}(2021)\citenamefont {Nakai},
  \citenamefont {Suzuki}, \citenamefont {Takahashi},\ and\ \citenamefont
  {Yamada}}]{Nakai:2020oit}%
  \BibitemOpen
  \bibfield  {author} {\bibinfo {author} {\bibfnamefont {Y.}~\bibnamefont
  {Nakai}}, \bibinfo {author} {\bibfnamefont {M.}~\bibnamefont {Suzuki}},
  \bibinfo {author} {\bibfnamefont {F.}~\bibnamefont {Takahashi}},\ and\
  \bibinfo {author} {\bibfnamefont {M.}~\bibnamefont {Yamada}},\ }\bibfield
  {title} {\bibinfo {title} {{Gravitational Waves and Dark Radiation from Dark
  Phase Transition: Connecting NANOGrav Pulsar Timing Data and Hubble
  Tension}},\ }\href {https://doi.org/10.1016/j.physletb.2021.136238}
  {\bibfield  {journal} {\bibinfo  {journal} {Phys. Lett. B}\ }\textbf
  {\bibinfo {volume} {816}},\ \bibinfo {pages} {136238} (\bibinfo {year}
  {2021})},\ \Eprint {https://arxiv.org/abs/2009.09754} {arXiv:2009.09754
  [astro-ph.CO]} \BibitemShut {NoStop}%
\bibitem [{\citenamefont {Neronov}\ \emph {et~al.}(2021)\citenamefont
  {Neronov}, \citenamefont {Roper~Pol}, \citenamefont {Caprini},\ and\
  \citenamefont {Semikoz}}]{Neronov:2020qrl}%
  \BibitemOpen
  \bibfield  {author} {\bibinfo {author} {\bibfnamefont {A.}~\bibnamefont
  {Neronov}}, \bibinfo {author} {\bibfnamefont {A.}~\bibnamefont {Roper~Pol}},
  \bibinfo {author} {\bibfnamefont {C.}~\bibnamefont {Caprini}},\ and\ \bibinfo
  {author} {\bibfnamefont {D.}~\bibnamefont {Semikoz}},\ }\bibfield  {title}
  {\bibinfo {title} {{NANOGrav signal from magnetohydrodynamic turbulence at
  the QCD phase transition in the early Universe}},\ }\href
  {https://doi.org/10.1103/PhysRevD.103.L041302} {\bibfield  {journal}
  {\bibinfo  {journal} {Phys. Rev. D}\ }\textbf {\bibinfo {volume} {103}},\
  \bibinfo {pages} {L041302} (\bibinfo {year} {2021})},\ \Eprint
  {https://arxiv.org/abs/2009.14174} {arXiv:2009.14174 [astro-ph.CO]}
  \BibitemShut {NoStop}%
\bibitem [{\citenamefont {Smith}\ and\ \citenamefont
  {Thrane}(2018)}]{Smith:2017vfk}%
  \BibitemOpen
  \bibfield  {author} {\bibinfo {author} {\bibfnamefont {R.}~\bibnamefont
  {Smith}}\ and\ \bibinfo {author} {\bibfnamefont {E.}~\bibnamefont {Thrane}},\
  }\bibfield  {title} {\bibinfo {title} {{Optimal Search for an Astrophysical
  Gravitational-Wave Background}},\ }\href
  {https://doi.org/10.1103/PhysRevX.8.021019} {\bibfield  {journal} {\bibinfo
  {journal} {Phys. Rev. X}\ }\textbf {\bibinfo {volume} {8}},\ \bibinfo {pages}
  {021019} (\bibinfo {year} {2018})},\ \Eprint
  {https://arxiv.org/abs/1712.00688} {arXiv:1712.00688 [gr-qc]} \BibitemShut
  {NoStop}%
\bibitem [{\citenamefont {Bartolo}\ \emph {et~al.}(2018)\citenamefont
  {Bartolo}, \citenamefont {Domcke}, \citenamefont {Figueroa}, \citenamefont
  {Garc\'\i{}a-Bellido}, \citenamefont {Peloso}, \citenamefont {Pieroni},
  \citenamefont {Ricciardone}, \citenamefont {Sakellariadou}, \citenamefont
  {Sorbo},\ and\ \citenamefont {Tasinato}}]{Bartolo:2018qqn}%
  \BibitemOpen
  \bibfield  {author} {\bibinfo {author} {\bibfnamefont {N.}~\bibnamefont
  {Bartolo}}, \bibinfo {author} {\bibfnamefont {V.}~\bibnamefont {Domcke}},
  \bibinfo {author} {\bibfnamefont {D.~G.}\ \bibnamefont {Figueroa}}, \bibinfo
  {author} {\bibfnamefont {J.}~\bibnamefont {Garc\'\i{}a-Bellido}}, \bibinfo
  {author} {\bibfnamefont {M.}~\bibnamefont {Peloso}}, \bibinfo {author}
  {\bibfnamefont {M.}~\bibnamefont {Pieroni}}, \bibinfo {author} {\bibfnamefont
  {A.}~\bibnamefont {Ricciardone}}, \bibinfo {author} {\bibfnamefont
  {M.}~\bibnamefont {Sakellariadou}}, \bibinfo {author} {\bibfnamefont
  {L.}~\bibnamefont {Sorbo}},\ and\ \bibinfo {author} {\bibfnamefont
  {G.}~\bibnamefont {Tasinato}},\ }\bibfield  {title} {\bibinfo {title}
  {{Probing non-Gaussian Stochastic Gravitational Wave Backgrounds with
  LISA}},\ }\href {https://doi.org/10.1088/1475-7516/2018/11/034} {\bibfield
  {journal} {\bibinfo  {journal} {JCAP}\ }\textbf {\bibinfo {volume} {11}},\
  \bibinfo {pages} {034}},\ \Eprint {https://arxiv.org/abs/1806.02819}
  {arXiv:1806.02819 [astro-ph.CO]} \BibitemShut {NoStop}%
\bibitem [{\citenamefont {Ginat}\ \emph {et~al.}(2020)\citenamefont {Ginat},
  \citenamefont {Desjacques}, \citenamefont {Reischke},\ and\ \citenamefont
  {Perets}}]{Ginat:2019aed}%
  \BibitemOpen
  \bibfield  {author} {\bibinfo {author} {\bibfnamefont {Y.~B.}\ \bibnamefont
  {Ginat}}, \bibinfo {author} {\bibfnamefont {V.}~\bibnamefont {Desjacques}},
  \bibinfo {author} {\bibfnamefont {R.}~\bibnamefont {Reischke}},\ and\
  \bibinfo {author} {\bibfnamefont {H.~B.}\ \bibnamefont {Perets}},\ }\bibfield
   {title} {\bibinfo {title} {{Probability distribution of astrophysical
  gravitational-wave background fluctuations}},\ }\href
  {https://doi.org/10.1103/PhysRevD.102.083501} {\bibfield  {journal} {\bibinfo
   {journal} {Phys. Rev. D}\ }\textbf {\bibinfo {volume} {102}},\ \bibinfo
  {pages} {083501} (\bibinfo {year} {2020})},\ \Eprint
  {https://arxiv.org/abs/1910.04587} {arXiv:1910.04587 [astro-ph.CO]}
  \BibitemShut {NoStop}%
\bibitem [{\citenamefont {Cui}\ \emph {et~al.}(2019)\citenamefont {Cui},
  \citenamefont {Lewicki}, \citenamefont {Morrissey},\ and\ \citenamefont
  {Wells}}]{Cui:2018rwi}%
  \BibitemOpen
  \bibfield  {author} {\bibinfo {author} {\bibfnamefont {Y.}~\bibnamefont
  {Cui}}, \bibinfo {author} {\bibfnamefont {M.}~\bibnamefont {Lewicki}},
  \bibinfo {author} {\bibfnamefont {D.~E.}\ \bibnamefont {Morrissey}},\ and\
  \bibinfo {author} {\bibfnamefont {J.~D.}\ \bibnamefont {Wells}},\ }\bibfield
  {title} {\bibinfo {title} {{Probing the pre-BBN universe with gravitational
  waves from cosmic strings}},\ }\href
  {https://doi.org/10.1007/JHEP01(2019)081} {\bibfield  {journal} {\bibinfo
  {journal} {JHEP}\ }\textbf {\bibinfo {volume} {01}},\ \bibinfo {pages}
  {081}},\ \Eprint {https://arxiv.org/abs/1808.08968} {arXiv:1808.08968
  [hep-ph]} \BibitemShut {NoStop}%
%%CITATION = ARXIV:1808.08968;%%
\bibitem [{\citenamefont {{Romano}}\ and\ \citenamefont
  {{Cornish}}(2017)}]{Romano:2017}%
  \BibitemOpen
  \bibfield  {author} {\bibinfo {author} {\bibfnamefont {J.~D.}\ \bibnamefont
  {{Romano}}}\ and\ \bibinfo {author} {\bibfnamefont {N.~J.}\ \bibnamefont
  {{Cornish}}},\ }\bibfield  {title} {\bibinfo {title} {{Detection methods for
  stochastic gravitational-wave backgrounds: a unified treatment}},\ }\href
  {https://doi.org/10.1007/s41114-017-0004-1} {\bibfield  {journal} {\bibinfo
  {journal} {Living Reviews in Relativity}\ }\textbf {\bibinfo {volume} {20}},\
  \bibinfo {eid} {2} (\bibinfo {year} {2017})},\ \Eprint
  {https://arxiv.org/abs/1608.06889} {arXiv:1608.06889 [gr-qc]} \BibitemShut
  {NoStop}%
\bibitem [{\citenamefont {{Romano}}(2019)}]{Romano:2019}%
  \BibitemOpen
  \bibfield  {author} {\bibinfo {author} {\bibfnamefont {J.~D.}\ \bibnamefont
  {{Romano}}},\ }\bibfield  {title} {\bibinfo {title} {{Searches for stochastic
  gravitational-wave backgrounds}},\ }\href@noop {} {\bibfield  {journal}
  {\bibinfo  {journal} {arXiv e-prints}\ } (\bibinfo {year} {2019})},\ \Eprint
  {https://arxiv.org/abs/1909.00269} {arXiv:1909.00269 [gr-qc]} \BibitemShut
  {NoStop}%
\bibitem [{\citenamefont {{Cutler}}\ and\ \citenamefont
  {{Harms}}(2006)}]{Cutler:2006}%
  \BibitemOpen
  \bibfield  {author} {\bibinfo {author} {\bibfnamefont {C.}~\bibnamefont
  {{Cutler}}}\ and\ \bibinfo {author} {\bibfnamefont {J.}~\bibnamefont
  {{Harms}}},\ }\bibfield  {title} {\bibinfo {title} {{Big Bang Observer and
  the neutron-star-binary subtraction problem}},\ }\href
  {https://doi.org/10.1103/PhysRevD.73.042001} {\bibfield  {journal} {\bibinfo
  {journal} {\prd}\ }\textbf {\bibinfo {volume} {73}},\ \bibinfo {eid} {042001}
  (\bibinfo {year} {2006})},\ \Eprint {https://arxiv.org/abs/gr-qc/0511092}
  {arXiv:gr-qc/0511092 [gr-qc]} \BibitemShut {NoStop}%
\bibitem [{\citenamefont {Pan}\ and\ \citenamefont {Yang}(2020)}]{Pan:2019}%
  \BibitemOpen
  \bibfield  {author} {\bibinfo {author} {\bibfnamefont {Z.}~\bibnamefont
  {Pan}}\ and\ \bibinfo {author} {\bibfnamefont {H.}~\bibnamefont {Yang}},\
  }\bibfield  {title} {\bibinfo {title} {{Probing Primordial Stochastic
  Gravitational Wave Background with Multi-band Astrophysical Foreground
  Cleaning}},\ }\href {https://doi.org/10.1088/1361-6382/abb074} {\bibfield
  {journal} {\bibinfo  {journal} {Class. Quant. Grav.}\ }\textbf {\bibinfo
  {volume} {37}},\ \bibinfo {pages} {195020} (\bibinfo {year} {2020})},\
  \Eprint {https://arxiv.org/abs/1910.09637} {arXiv:1910.09637 [astro-ph.CO]}
  \BibitemShut {NoStop}%
\bibitem [{\citenamefont {{Pieroni}}\ and\ \citenamefont
  {{Barausse}}(2020)}]{Pieroni:2020}%
  \BibitemOpen
  \bibfield  {author} {\bibinfo {author} {\bibfnamefont {M.}~\bibnamefont
  {{Pieroni}}}\ and\ \bibinfo {author} {\bibfnamefont {E.}~\bibnamefont
  {{Barausse}}},\ }\bibfield  {title} {\bibinfo {title} {{Foreground cleaning
  and template-free stochastic background extraction for LISA}},\ }\href
  {https://doi.org/10.1088/1475-7516/2020/07/021} {\bibfield  {journal}
  {\bibinfo  {journal} {\jcap}\ }\textbf {\bibinfo {volume} {2020}},\ \bibinfo
  {eid} {021} (\bibinfo {year} {2020})},\ \Eprint
  {https://arxiv.org/abs/2004.01135} {arXiv:2004.01135 [astro-ph.CO]}
  \BibitemShut {NoStop}%
\bibitem [{\citenamefont {Boileau}\ \emph {et~al.}(2021)\citenamefont
  {Boileau}, \citenamefont {Christensen}, \citenamefont {Meyer},\ and\
  \citenamefont {Cornish}}]{Boileau:2020rpg}%
  \BibitemOpen
  \bibfield  {author} {\bibinfo {author} {\bibfnamefont {G.}~\bibnamefont
  {Boileau}}, \bibinfo {author} {\bibfnamefont {N.}~\bibnamefont
  {Christensen}}, \bibinfo {author} {\bibfnamefont {R.}~\bibnamefont {Meyer}},\
  and\ \bibinfo {author} {\bibfnamefont {N.~J.}\ \bibnamefont {Cornish}},\
  }\bibfield  {title} {\bibinfo {title} {{Spectral separation of the stochastic
  gravitational-wave background for LISA: Observing both cosmological and
  astrophysical backgrounds}},\ }\href
  {https://doi.org/10.1103/PhysRevD.103.103529} {\bibfield  {journal} {\bibinfo
   {journal} {Phys. Rev. D}\ }\textbf {\bibinfo {volume} {103}},\ \bibinfo
  {pages} {103529} (\bibinfo {year} {2021})},\ \Eprint
  {https://arxiv.org/abs/2011.05055} {arXiv:2011.05055 [gr-qc]} \BibitemShut
  {NoStop}%
\bibitem [{\citenamefont {Sedda}\ \emph {et~al.}(2020)\citenamefont {Sedda}
  \emph {et~al.}}]{Sedda:2019}%
  \BibitemOpen
  \bibfield  {author} {\bibinfo {author} {\bibfnamefont {M.~A.}\ \bibnamefont
  {Sedda}} \emph {et~al.},\ }\bibfield  {title} {\bibinfo {title} {{The missing
  link in gravitational-wave astronomy: discoveries waiting in the decihertz
  range}},\ }\href {https://doi.org/10.1088/1361-6382/abb5c1} {\bibfield
  {journal} {\bibinfo  {journal} {Class. Quant. Grav.}\ }\textbf {\bibinfo
  {volume} {37}},\ \bibinfo {pages} {215011} (\bibinfo {year} {2020})},\
  \Eprint {https://arxiv.org/abs/1908.11375} {arXiv:1908.11375 [gr-qc]}
  \BibitemShut {NoStop}%
\bibitem [{\citenamefont {Thrane}\ and\ \citenamefont
  {Romano}(2013)}]{Thrane:2013oya}%
  \BibitemOpen
  \bibfield  {author} {\bibinfo {author} {\bibfnamefont {E.}~\bibnamefont
  {Thrane}}\ and\ \bibinfo {author} {\bibfnamefont {J.~D.}\ \bibnamefont
  {Romano}},\ }\bibfield  {title} {\bibinfo {title} {{Sensitivity curves for
  searches for gravitational-wave backgrounds}},\ }\href
  {https://doi.org/10.1103/PhysRevD.88.124032} {\bibfield  {journal} {\bibinfo
  {journal} {Phys. Rev.}\ }\textbf {\bibinfo {volume} {D88}},\ \bibinfo {pages}
  {124032} (\bibinfo {year} {2013})},\ \Eprint
  {https://arxiv.org/abs/1310.5300} {arXiv:1310.5300 [astro-ph.IM]}
  \BibitemShut {NoStop}%
%%CITATION = ARXIV:1310.5300;%%
\bibitem [{\citenamefont {{Moore}}\ \emph {et~al.}(2015)\citenamefont
  {{Moore}}, \citenamefont {{Cole}},\ and\ \citenamefont
  {{Berry}}}]{Moore:2015}%
  \BibitemOpen
  \bibfield  {author} {\bibinfo {author} {\bibfnamefont {C.~J.}\ \bibnamefont
  {{Moore}}}, \bibinfo {author} {\bibfnamefont {R.~H.}\ \bibnamefont
  {{Cole}}},\ and\ \bibinfo {author} {\bibfnamefont {C.~P.~L.}\ \bibnamefont
  {{Berry}}},\ }\bibfield  {title} {\bibinfo {title} {{Gravitational-wave
  sensitivity curves}},\ }\href {https://doi.org/10.1088/0264-9381/32/1/015014}
  {\bibfield  {journal} {\bibinfo  {journal} {Classical and Quantum Gravity}\
  }\textbf {\bibinfo {volume} {32}},\ \bibinfo {eid} {015014} (\bibinfo {year}
  {2015})},\ \Eprint {https://arxiv.org/abs/1408.0740} {arXiv:1408.0740
  [gr-qc]} \BibitemShut {NoStop}%
\bibitem [{\citenamefont {Caprini}\ \emph {et~al.}(2019)\citenamefont
  {Caprini}, \citenamefont {Figueroa}, \citenamefont {Flauger}, \citenamefont
  {Nardini}, \citenamefont {Peloso}, \citenamefont {Pieroni}, \citenamefont
  {Ricciardone},\ and\ \citenamefont {Tasinato}}]{Caprini:2019}%
  \BibitemOpen
  \bibfield  {author} {\bibinfo {author} {\bibfnamefont {C.}~\bibnamefont
  {Caprini}}, \bibinfo {author} {\bibfnamefont {D.~G.}\ \bibnamefont
  {Figueroa}}, \bibinfo {author} {\bibfnamefont {R.}~\bibnamefont {Flauger}},
  \bibinfo {author} {\bibfnamefont {G.}~\bibnamefont {Nardini}}, \bibinfo
  {author} {\bibfnamefont {M.}~\bibnamefont {Peloso}}, \bibinfo {author}
  {\bibfnamefont {M.}~\bibnamefont {Pieroni}}, \bibinfo {author} {\bibfnamefont
  {A.}~\bibnamefont {Ricciardone}},\ and\ \bibinfo {author} {\bibfnamefont
  {G.}~\bibnamefont {Tasinato}},\ }\bibfield  {title} {\bibinfo {title}
  {{Reconstructing the spectral shape of a stochastic gravitational wave
  background with LISA}},\ }\href
  {https://doi.org/10.1088/1475-7516/2019/11/017} {\bibfield  {journal}
  {\bibinfo  {journal} {JCAP}\ }\textbf {\bibinfo {volume} {11}},\ \bibinfo
  {pages} {017}},\ \Eprint {https://arxiv.org/abs/1906.09244} {arXiv:1906.09244
  [astro-ph.CO]} \BibitemShut {NoStop}%
\bibitem [{\citenamefont {Reitze}\ \emph {et~al.}(2019)\citenamefont {Reitze}
  \emph {et~al.}}]{CosmicExplorer}%
  \BibitemOpen
  \bibfield  {author} {\bibinfo {author} {\bibfnamefont {D.}~\bibnamefont
  {Reitze}} \emph {et~al.},\ }\bibfield  {title} {\bibinfo {title} {{Cosmic
  Explorer: The U.S. Contribution to Gravitational-Wave Astronomy beyond
  LIGO}},\ }\href@noop {} {\bibfield  {journal} {\bibinfo  {journal} {Bull. Am.
  Astron. Soc.}\ }\textbf {\bibinfo {volume} {51}},\ \bibinfo {pages} {035}
  (\bibinfo {year} {2019})},\ \Eprint {https://arxiv.org/abs/1907.04833}
  {arXiv:1907.04833 [astro-ph.IM]} \BibitemShut {NoStop}%
\bibitem [{\citenamefont {Maggiore}\ \emph {et~al.}(2020)\citenamefont
  {Maggiore} \emph {et~al.}}]{EinsteinTelescope}%
  \BibitemOpen
  \bibfield  {author} {\bibinfo {author} {\bibfnamefont {M.}~\bibnamefont
  {Maggiore}} \emph {et~al.},\ }\bibfield  {title} {\bibinfo {title} {{Science
  Case for the Einstein Telescope}},\ }\href
  {https://doi.org/10.1088/1475-7516/2020/03/050} {\bibfield  {journal}
  {\bibinfo  {journal} {JCAP}\ }\textbf {\bibinfo {volume} {03}},\ \bibinfo
  {pages} {050}},\ \Eprint {https://arxiv.org/abs/1912.02622} {arXiv:1912.02622
  [astro-ph.CO]} \BibitemShut {NoStop}%
\bibitem [{\citenamefont {Ruan}\ \emph {et~al.}(2020)\citenamefont {Ruan},
  \citenamefont {Guo}, \citenamefont {Cai},\ and\ \citenamefont
  {Zhang}}]{Taiji}%
  \BibitemOpen
  \bibfield  {author} {\bibinfo {author} {\bibfnamefont {W.-H.}\ \bibnamefont
  {Ruan}}, \bibinfo {author} {\bibfnamefont {Z.-K.}\ \bibnamefont {Guo}},
  \bibinfo {author} {\bibfnamefont {R.-G.}\ \bibnamefont {Cai}},\ and\ \bibinfo
  {author} {\bibfnamefont {Y.-Z.}\ \bibnamefont {Zhang}},\ }\bibfield  {title}
  {\bibinfo {title} {{Taiji program: Gravitational-wave sources}},\ }\href
  {https://doi.org/10.1142/S0217751X2050075X} {\bibfield  {journal} {\bibinfo
  {journal} {Int. J. Mod. Phys. A}\ }\textbf {\bibinfo {volume} {35}},\
  \bibinfo {pages} {2050075} (\bibinfo {year} {2020})},\ \Eprint
  {https://arxiv.org/abs/1807.09495} {arXiv:1807.09495 [gr-qc]} \BibitemShut
  {NoStop}%
\bibitem [{\citenamefont {Flauger}\ \emph {et~al.}(2021)\citenamefont
  {Flauger}, \citenamefont {Karnesis}, \citenamefont {Nardini}, \citenamefont
  {Pieroni}, \citenamefont {Ricciardone},\ and\ \citenamefont
  {Torrado}}]{Flauger:2020qyi}%
  \BibitemOpen
  \bibfield  {author} {\bibinfo {author} {\bibfnamefont {R.}~\bibnamefont
  {Flauger}}, \bibinfo {author} {\bibfnamefont {N.}~\bibnamefont {Karnesis}},
  \bibinfo {author} {\bibfnamefont {G.}~\bibnamefont {Nardini}}, \bibinfo
  {author} {\bibfnamefont {M.}~\bibnamefont {Pieroni}}, \bibinfo {author}
  {\bibfnamefont {A.}~\bibnamefont {Ricciardone}},\ and\ \bibinfo {author}
  {\bibfnamefont {J.}~\bibnamefont {Torrado}},\ }\bibfield  {title} {\bibinfo
  {title} {{Improved reconstruction of a stochastic gravitational wave
  background with LISA}},\ }\href
  {https://doi.org/10.1088/1475-7516/2021/01/059} {\bibfield  {journal}
  {\bibinfo  {journal} {JCAP}\ }\textbf {\bibinfo {volume} {01}},\ \bibinfo
  {pages} {059}},\ \Eprint {https://arxiv.org/abs/2009.11845} {arXiv:2009.11845
  [astro-ph.CO]} \BibitemShut {NoStop}%
\bibitem [{\citenamefont {{Larson}}\ \emph {et~al.}(2000)\citenamefont
  {{Larson}}, \citenamefont {{Hiscock}},\ and\ \citenamefont
  {{Hellings}}}]{Larson:2000}%
  \BibitemOpen
  \bibfield  {author} {\bibinfo {author} {\bibfnamefont {S.~L.}\ \bibnamefont
  {{Larson}}}, \bibinfo {author} {\bibfnamefont {W.~A.}\ \bibnamefont
  {{Hiscock}}},\ and\ \bibinfo {author} {\bibfnamefont {R.~W.}\ \bibnamefont
  {{Hellings}}},\ }\bibfield  {title} {\bibinfo {title} {{Sensitivity curves
  for spaceborne gravitational wave interferometers}},\ }\href
  {https://doi.org/10.1103/PhysRevD.62.062001} {\bibfield  {journal} {\bibinfo
  {journal} {\prd}\ }\textbf {\bibinfo {volume} {62}},\ \bibinfo {eid} {062001}
  (\bibinfo {year} {2000})},\ \Eprint {https://arxiv.org/abs/gr-qc/9909080}
  {arXiv:gr-qc/9909080 [gr-qc]} \BibitemShut {NoStop}%
\bibitem [{\citenamefont {{Cornish}}\ and\ \citenamefont
  {{Rubbo}}(2003)}]{Cornish:2003}%
  \BibitemOpen
  \bibfield  {author} {\bibinfo {author} {\bibfnamefont {N.~J.}\ \bibnamefont
  {{Cornish}}}\ and\ \bibinfo {author} {\bibfnamefont {L.~J.}\ \bibnamefont
  {{Rubbo}}},\ }\bibfield  {title} {\bibinfo {title} {{Publisher's Note: LISA
  response function [Phys. Rev. D 67, 022001 (2003)]}},\ }\href
  {https://doi.org/10.1103/PhysRevD.67.029905} {\bibfield  {journal} {\bibinfo
  {journal} {\prd}\ }\textbf {\bibinfo {volume} {67}},\ \bibinfo {eid}
  {029905(E)} (\bibinfo {year} {2003})},\ \Eprint
  {https://arxiv.org/abs/gr-qc/0209011} {arXiv:gr-qc/0209011 [gr-qc]}
  \BibitemShut {NoStop}%
\bibitem [{\citenamefont {Sato}\ \emph {et~al.}(2017)\citenamefont {Sato} \emph
  {et~al.}}]{Sato_2017}%
  \BibitemOpen
  \bibfield  {author} {\bibinfo {author} {\bibfnamefont {S.}~\bibnamefont
  {Sato}} \emph {et~al.},\ }\bibfield  {title} {\bibinfo {title} {The status of
  {DECIGO}},\ }\href {https://doi.org/10.1088/1742-6596/840/1/012010}
  {\bibfield  {journal} {\bibinfo  {journal} {Journal of Physics: Conference
  Series}\ }\textbf {\bibinfo {volume} {840}},\ \bibinfo {pages} {012010}
  (\bibinfo {year} {2017})}\BibitemShut {NoStop}%
\bibitem [{\citenamefont {{Nakamura}}\ \emph {et~al.}(2016)\citenamefont
  {{Nakamura}} \emph {et~al.}}]{PreDECIGO}%
  \BibitemOpen
  \bibfield  {author} {\bibinfo {author} {\bibfnamefont {T.}~\bibnamefont
  {{Nakamura}}} \emph {et~al.},\ }\bibfield  {title} {\bibinfo {title}
  {{Pre-DECIGO can get the smoking gun to decide the astrophysical or
  cosmological origin of GW150914-like binary black holes}},\ }\href
  {https://doi.org/10.1093/ptep/ptw127} {\bibfield  {journal} {\bibinfo
  {journal} {Progress of Theoretical and Experimental Physics}\ }\textbf
  {\bibinfo {volume} {2016}},\ \bibinfo {eid} {093E01} (\bibinfo {year}
  {2016})},\ \Eprint {https://arxiv.org/abs/1607.00897} {arXiv:1607.00897
  [astro-ph.HE]} \BibitemShut {NoStop}%
\bibitem [{\citenamefont {{National Science Foundation}}(2019)}]{Adetector}%
  \BibitemOpen
  \bibfield  {author} {\bibinfo {author} {\bibnamefont {{National Science
  Foundation}}},\ }\href@noop {} {\bibinfo {title} {Press statement: Upgraded
  {LIGO} to search for universe's most extreme events.}},\ \bibinfo
  {howpublished} {\url{https://www.nsf.gov/news/news_summ.jsp?cntn_id=297414}}
  (\bibinfo {year} {2019})\BibitemShut {NoStop}%
\bibitem [{\citenamefont {Huang}\ \emph {et~al.}(2016)\citenamefont {Huang},
  \citenamefont {Long},\ and\ \citenamefont {Wang}}]{Huang:2016cjm}%
  \BibitemOpen
  \bibfield  {author} {\bibinfo {author} {\bibfnamefont {P.}~\bibnamefont
  {Huang}}, \bibinfo {author} {\bibfnamefont {A.~J.}\ \bibnamefont {Long}},\
  and\ \bibinfo {author} {\bibfnamefont {L.-T.}\ \bibnamefont {Wang}},\
  }\bibfield  {title} {\bibinfo {title} {{Probing the Electroweak Phase
  Transition with Higgs Factories and Gravitational Waves}},\ }\href
  {https://doi.org/10.1103/PhysRevD.94.075008} {\bibfield  {journal} {\bibinfo
  {journal} {Phys. Rev. D}\ }\textbf {\bibinfo {volume} {94}},\ \bibinfo
  {pages} {075008} (\bibinfo {year} {2016})},\ \Eprint
  {https://arxiv.org/abs/1608.06619} {arXiv:1608.06619 [hep-ph]} \BibitemShut
  {NoStop}%
\bibitem [{\citenamefont {Gould}\ \emph {et~al.}(2019)\citenamefont {Gould},
  \citenamefont {Kozaczuk}, \citenamefont {Niemi}, \citenamefont
  {Ramsey-Musolf}, \citenamefont {Tenkanen},\ and\ \citenamefont
  {Weir}}]{Gould:2019qek}%
  \BibitemOpen
  \bibfield  {author} {\bibinfo {author} {\bibfnamefont {O.}~\bibnamefont
  {Gould}}, \bibinfo {author} {\bibfnamefont {J.}~\bibnamefont {Kozaczuk}},
  \bibinfo {author} {\bibfnamefont {L.}~\bibnamefont {Niemi}}, \bibinfo
  {author} {\bibfnamefont {M.~J.}\ \bibnamefont {Ramsey-Musolf}}, \bibinfo
  {author} {\bibfnamefont {T.~V.}\ \bibnamefont {Tenkanen}},\ and\ \bibinfo
  {author} {\bibfnamefont {D.~J.}\ \bibnamefont {Weir}},\ }\bibfield  {title}
  {\bibinfo {title} {{Nonperturbative analysis of the gravitational waves from
  a first-order electroweak phase transition}},\ }\href
  {https://doi.org/10.1103/PhysRevD.100.115024} {\bibfield  {journal} {\bibinfo
   {journal} {Phys. Rev. D}\ }\textbf {\bibinfo {volume} {100}},\ \bibinfo
  {pages} {115024} (\bibinfo {year} {2019})},\ \Eprint
  {https://arxiv.org/abs/1903.11604} {arXiv:1903.11604 [hep-ph]} \BibitemShut
  {NoStop}%
\bibitem [{\citenamefont {Vincent}\ \emph {et~al.}(1998)\citenamefont
  {Vincent}, \citenamefont {Antunes},\ and\ \citenamefont
  {Hindmarsh}}]{Vincent:1997cx}%
  \BibitemOpen
  \bibfield  {author} {\bibinfo {author} {\bibfnamefont {G.}~\bibnamefont
  {Vincent}}, \bibinfo {author} {\bibfnamefont {N.~D.}\ \bibnamefont
  {Antunes}},\ and\ \bibinfo {author} {\bibfnamefont {M.}~\bibnamefont
  {Hindmarsh}},\ }\bibfield  {title} {\bibinfo {title} {{Numerical simulations
  of string networks in the Abelian Higgs model}},\ }\href
  {https://doi.org/10.1103/PhysRevLett.80.2277} {\bibfield  {journal} {\bibinfo
   {journal} {Phys. Rev. Lett.}\ }\textbf {\bibinfo {volume} {80}},\ \bibinfo
  {pages} {2277} (\bibinfo {year} {1998})},\ \Eprint
  {https://arxiv.org/abs/hep-ph/9708427} {arXiv:hep-ph/9708427 [hep-ph]}
  \BibitemShut {NoStop}%
%%CITATION = HEP-PH/9708427;%%
\bibitem [{\citenamefont {Bevis}\ \emph {et~al.}(2007)\citenamefont {Bevis},
  \citenamefont {Hindmarsh}, \citenamefont {Kunz},\ and\ \citenamefont
  {Urrestilla}}]{Bevis:2006mj}%
  \BibitemOpen
  \bibfield  {author} {\bibinfo {author} {\bibfnamefont {N.}~\bibnamefont
  {Bevis}}, \bibinfo {author} {\bibfnamefont {M.}~\bibnamefont {Hindmarsh}},
  \bibinfo {author} {\bibfnamefont {M.}~\bibnamefont {Kunz}},\ and\ \bibinfo
  {author} {\bibfnamefont {J.}~\bibnamefont {Urrestilla}},\ }\bibfield  {title}
  {\bibinfo {title} {{CMB power spectrum contribution from cosmic strings using
  field-evolution simulations of the Abelian Higgs model}},\ }\href
  {https://doi.org/10.1103/PhysRevD.75.065015} {\bibfield  {journal} {\bibinfo
  {journal} {Phys. Rev.}\ }\textbf {\bibinfo {volume} {D75}},\ \bibinfo {pages}
  {065015} (\bibinfo {year} {2007})},\ \Eprint
  {https://arxiv.org/abs/astro-ph/0605018} {arXiv:astro-ph/0605018 [astro-ph]}
  \BibitemShut {NoStop}%
%%CITATION = ASTRO-PH/0605018;%%
\bibitem [{\citenamefont {Blanco-Pillado}\ \emph {et~al.}(2014)\citenamefont
  {Blanco-Pillado}, \citenamefont {Olum},\ and\ \citenamefont
  {Shlaer}}]{Blanco-Pillado:2013qja}%
  \BibitemOpen
  \bibfield  {author} {\bibinfo {author} {\bibfnamefont {J.~J.}\ \bibnamefont
  {Blanco-Pillado}}, \bibinfo {author} {\bibfnamefont {K.~D.}\ \bibnamefont
  {Olum}},\ and\ \bibinfo {author} {\bibfnamefont {B.}~\bibnamefont {Shlaer}},\
  }\bibfield  {title} {\bibinfo {title} {{The number of cosmic string loops}},\
  }\href {https://doi.org/10.1103/PhysRevD.89.023512} {\bibfield  {journal}
  {\bibinfo  {journal} {Phys. Rev.}\ }\textbf {\bibinfo {volume} {D89}},\
  \bibinfo {pages} {023512} (\bibinfo {year} {2014})},\ \Eprint
  {https://arxiv.org/abs/1309.6637} {arXiv:1309.6637 [astro-ph.CO]}
  \BibitemShut {NoStop}%
%%CITATION = ARXIV:1309.6637;%%
\bibitem [{\citenamefont {Ringeval}\ \emph {et~al.}(2007)\citenamefont
  {Ringeval}, \citenamefont {Sakellariadou},\ and\ \citenamefont
  {Bouchet}}]{Ringeval:2005kr}%
  \BibitemOpen
  \bibfield  {author} {\bibinfo {author} {\bibfnamefont {C.}~\bibnamefont
  {Ringeval}}, \bibinfo {author} {\bibfnamefont {M.}~\bibnamefont
  {Sakellariadou}},\ and\ \bibinfo {author} {\bibfnamefont {F.}~\bibnamefont
  {Bouchet}},\ }\bibfield  {title} {\bibinfo {title} {{Cosmological evolution
  of cosmic string loops}},\ }\href
  {https://doi.org/10.1088/1475-7516/2007/02/023} {\bibfield  {journal}
  {\bibinfo  {journal} {JCAP}\ }\textbf {\bibinfo {volume} {0702}},\ \bibinfo
  {pages} {023}},\ \Eprint {https://arxiv.org/abs/astro-ph/0511646}
  {arXiv:astro-ph/0511646 [astro-ph]} \BibitemShut {NoStop}%
%%CITATION = ASTRO-PH/0511646;%%
\bibitem [{\citenamefont {D'Eramo}\ and\ \citenamefont
  {Schmitz}(2019)}]{DEramo:2019tit}%
  \BibitemOpen
  \bibfield  {author} {\bibinfo {author} {\bibfnamefont {F.}~\bibnamefont
  {D'Eramo}}\ and\ \bibinfo {author} {\bibfnamefont {K.}~\bibnamefont
  {Schmitz}},\ }\bibfield  {title} {\bibinfo {title} {{Imprint of a scalar era
  on the primordial spectrum of gravitational waves}},\ }\href
  {https://doi.org/10.1103/PhysRevResearch.1.013010} {\bibfield  {journal}
  {\bibinfo  {journal} {Phys. Rev. Research.}\ }\textbf {\bibinfo {volume}
  {1}},\ \bibinfo {pages} {013010} (\bibinfo {year} {2019})},\ \Eprint
  {https://arxiv.org/abs/1904.07870} {arXiv:1904.07870 [hep-ph]} \BibitemShut
  {NoStop}%
\bibitem [{\citenamefont {Blasi}\ \emph {et~al.}(2020)\citenamefont {Blasi},
  \citenamefont {Brdar},\ and\ \citenamefont {Schmitz}}]{Blasi:2020wpy}%
  \BibitemOpen
  \bibfield  {author} {\bibinfo {author} {\bibfnamefont {S.}~\bibnamefont
  {Blasi}}, \bibinfo {author} {\bibfnamefont {V.}~\bibnamefont {Brdar}},\ and\
  \bibinfo {author} {\bibfnamefont {K.}~\bibnamefont {Schmitz}},\ }\bibfield
  {title} {\bibinfo {title} {{Fingerprint of low-scale leptogenesis in the
  primordial gravitational-wave spectrum}},\ }\href
  {https://doi.org/10.1103/PhysRevResearch.2.043321} {\bibfield  {journal}
  {\bibinfo  {journal} {Phys. Rev. Res.}\ }\textbf {\bibinfo {volume} {2}},\
  \bibinfo {pages} {043321} (\bibinfo {year} {2020})},\ \Eprint
  {https://arxiv.org/abs/2004.02889} {arXiv:2004.02889 [hep-ph]} \BibitemShut
  {NoStop}%
\bibitem [{\citenamefont {{van Haasteren}}\ \emph {et~al.}(2011)\citenamefont
  {{van Haasteren}} \emph {et~al.}}]{EPTA}%
  \BibitemOpen
  \bibfield  {author} {\bibinfo {author} {\bibfnamefont {R.}~\bibnamefont {{van
  Haasteren}}} \emph {et~al.},\ }\bibfield  {title} {\bibinfo {title} {{Placing
  limits on the stochastic gravitational-wave background using European Pulsar
  Timing Array data}},\ }\href
  {https://doi.org/10.1111/j.1365-2966.2011.18613.x} {\bibfield  {journal}
  {\bibinfo  {journal} {\mnras}\ }\textbf {\bibinfo {volume} {414}},\ \bibinfo
  {pages} {3117} (\bibinfo {year} {2011})},\ \Eprint
  {https://arxiv.org/abs/1103.0576} {arXiv:1103.0576 [astro-ph.CO]}
  \BibitemShut {NoStop}%
\bibitem [{\citenamefont {{Caprini}}\ \emph {et~al.}(2016)\citenamefont
  {{Caprini}} \emph {et~al.}}]{Caprini:2016}%
  \BibitemOpen
  \bibfield  {author} {\bibinfo {author} {\bibfnamefont {C.}~\bibnamefont
  {{Caprini}}} \emph {et~al.},\ }\bibfield  {title} {\bibinfo {title} {{Science
  with the space-based interferometer eLISA. II: gravitational waves from
  cosmological phase transitions}},\ }\href
  {https://doi.org/10.1088/1475-7516/2016/04/001} {\bibfield  {journal}
  {\bibinfo  {journal} {\jcap}\ }\textbf {\bibinfo {volume} {2016}},\ \bibinfo
  {eid} {001} (\bibinfo {year} {2016})},\ \Eprint
  {https://arxiv.org/abs/1512.06239} {arXiv:1512.06239 [astro-ph.CO]}
  \BibitemShut {NoStop}%
\bibitem [{\citenamefont {{Alanne}}\ \emph {et~al.}(2020)\citenamefont
  {{Alanne}}, \citenamefont {{Hugle}}, \citenamefont {{Platscher}},\ and\
  \citenamefont {{Schmitz}}}]{Alanne:2020}%
  \BibitemOpen
  \bibfield  {author} {\bibinfo {author} {\bibfnamefont {T.}~\bibnamefont
  {{Alanne}}}, \bibinfo {author} {\bibfnamefont {T.}~\bibnamefont {{Hugle}}},
  \bibinfo {author} {\bibfnamefont {M.}~\bibnamefont {{Platscher}}},\ and\
  \bibinfo {author} {\bibfnamefont {K.}~\bibnamefont {{Schmitz}}},\ }\bibfield
  {title} {\bibinfo {title} {{A fresh look at the gravitational-wave signal
  from cosmological phase transitions}},\ }\href
  {https://doi.org/10.1007/JHEP03(2020)004} {\bibfield  {journal} {\bibinfo
  {journal} {Journal of High Energy Physics}\ }\textbf {\bibinfo {volume}
  {2020}},\ \bibinfo {eid} {4} (\bibinfo {year} {2020})},\ \Eprint
  {https://arxiv.org/abs/1909.11356} {arXiv:1909.11356 [hep-ph]} \BibitemShut
  {NoStop}%
\bibitem [{\citenamefont {Schmitz}(2021)}]{Schmitz:2020}%
  \BibitemOpen
  \bibfield  {author} {\bibinfo {author} {\bibfnamefont {K.}~\bibnamefont
  {Schmitz}},\ }\bibfield  {title} {\bibinfo {title} {{New Sensitivity Curves
  for Gravitational-Wave Signals from Cosmological Phase Transitions}},\ }\href
  {https://doi.org/10.1007/JHEP01(2021)097} {\bibfield  {journal} {\bibinfo
  {journal} {JHEP}\ }\textbf {\bibinfo {volume} {01}},\ \bibinfo {pages}
  {097}},\ \Eprint {https://arxiv.org/abs/2002.04615} {arXiv:2002.04615
  [hep-ph]} \BibitemShut {NoStop}%
\bibitem [{\citenamefont {{Caprini}}\ \emph {et~al.}(2020)\citenamefont
  {{Caprini}} \emph {et~al.}}]{Caprini:2020}%
  \BibitemOpen
  \bibfield  {author} {\bibinfo {author} {\bibfnamefont {C.}~\bibnamefont
  {{Caprini}}} \emph {et~al.},\ }\bibfield  {title} {\bibinfo {title}
  {{Detecting gravitational waves from cosmological phase transitions with
  LISA: an update}},\ }\href {https://doi.org/10.1088/1475-7516/2020/03/024}
  {\bibfield  {journal} {\bibinfo  {journal} {\jcap}\ }\textbf {\bibinfo
  {volume} {2020}},\ \bibinfo {eid} {024} (\bibinfo {year} {2020})},\ \Eprint
  {https://arxiv.org/abs/1910.13125} {arXiv:1910.13125 [astro-ph.CO]}
  \BibitemShut {NoStop}%
\bibitem [{\citenamefont {Ellis}\ \emph {et~al.}(2019)\citenamefont {Ellis},
  \citenamefont {Lewicki}, \citenamefont {No},\ and\ \citenamefont
  {Vaskonen}}]{Ellis:2019oqb}%
  \BibitemOpen
  \bibfield  {author} {\bibinfo {author} {\bibfnamefont {J.}~\bibnamefont
  {Ellis}}, \bibinfo {author} {\bibfnamefont {M.}~\bibnamefont {Lewicki}},
  \bibinfo {author} {\bibfnamefont {J.~M.}\ \bibnamefont {No}},\ and\ \bibinfo
  {author} {\bibfnamefont {V.}~\bibnamefont {Vaskonen}},\ }\bibfield  {title}
  {\bibinfo {title} {{Gravitational wave energy budget in strongly supercooled
  phase transitions}},\ }\href {https://doi.org/10.1088/1475-7516/2019/06/024}
  {\bibfield  {journal} {\bibinfo  {journal} {JCAP}\ }\textbf {\bibinfo
  {volume} {06}},\ \bibinfo {pages} {024}},\ \Eprint
  {https://arxiv.org/abs/1903.09642} {arXiv:1903.09642 [hep-ph]} \BibitemShut
  {NoStop}%
\bibitem [{\citenamefont {Hindmarsh}\ \emph {et~al.}(2021)\citenamefont
  {Hindmarsh}, \citenamefont {L\"uben}, \citenamefont {Lumma},\ and\
  \citenamefont {Pauly}}]{Hindmarsh:2020}%
  \BibitemOpen
  \bibfield  {author} {\bibinfo {author} {\bibfnamefont {M.~B.}\ \bibnamefont
  {Hindmarsh}}, \bibinfo {author} {\bibfnamefont {M.}~\bibnamefont {L\"uben}},
  \bibinfo {author} {\bibfnamefont {J.}~\bibnamefont {Lumma}},\ and\ \bibinfo
  {author} {\bibfnamefont {M.}~\bibnamefont {Pauly}},\ }\bibfield  {title}
  {\bibinfo {title} {{Phase transitions in the early universe}},\ }\href
  {https://doi.org/10.21468/SciPostPhysLectNotes.24} {\bibfield  {journal}
  {\bibinfo  {journal} {SciPost Phys. Lect. Notes}\ }\textbf {\bibinfo {volume}
  {24}},\ \bibinfo {pages} {1} (\bibinfo {year} {2021})},\ \Eprint
  {https://arxiv.org/abs/2008.09136} {arXiv:2008.09136 [astro-ph.CO]}
  \BibitemShut {NoStop}%
\bibitem [{\citenamefont {{Cutting}}\ \emph {et~al.}(2020)\citenamefont
  {{Cutting}}, \citenamefont {{Hindmarsh}},\ and\ \citenamefont
  {{Weir}}}]{Cutting:2020}%
  \BibitemOpen
  \bibfield  {author} {\bibinfo {author} {\bibfnamefont {D.}~\bibnamefont
  {{Cutting}}}, \bibinfo {author} {\bibfnamefont {M.}~\bibnamefont
  {{Hindmarsh}}},\ and\ \bibinfo {author} {\bibfnamefont {D.~J.}\ \bibnamefont
  {{Weir}}},\ }\bibfield  {title} {\bibinfo {title} {{Vorticity, Kinetic
  Energy, and Suppressed Gravitational-Wave Production in Strong First-Order
  Phase Transitions}},\ }\href {https://doi.org/10.1103/PhysRevLett.125.021302}
  {\bibfield  {journal} {\bibinfo  {journal} {\prl}\ }\textbf {\bibinfo
  {volume} {125}},\ \bibinfo {eid} {021302} (\bibinfo {year} {2020})},\ \Eprint
  {https://arxiv.org/abs/1906.00480} {arXiv:1906.00480 [hep-ph]} \BibitemShut
  {NoStop}%
\bibitem [{\citenamefont {{Hindmarsh}}\ \emph {et~al.}(2017)\citenamefont
  {{Hindmarsh}}, \citenamefont {{Huber}}, \citenamefont {{Rummukainen}},\ and\
  \citenamefont {{Weir}}}]{Hindmarsh:2017}%
  \BibitemOpen
  \bibfield  {author} {\bibinfo {author} {\bibfnamefont {M.}~\bibnamefont
  {{Hindmarsh}}}, \bibinfo {author} {\bibfnamefont {S.~J.}\ \bibnamefont
  {{Huber}}}, \bibinfo {author} {\bibfnamefont {K.}~\bibnamefont
  {{Rummukainen}}},\ and\ \bibinfo {author} {\bibfnamefont {D.~J.}\
  \bibnamefont {{Weir}}},\ }\bibfield  {title} {\bibinfo {title} {{Shape of the
  acoustic gravitational wave power spectrum from a first order phase
  transition}},\ }\href {https://doi.org/10.1103/PhysRevD.96.103520} {\bibfield
   {journal} {\bibinfo  {journal} {\prd}\ }\textbf {\bibinfo {volume} {96}},\
  \bibinfo {eid} {103520} (\bibinfo {year} {2017})},\ \Eprint
  {https://arxiv.org/abs/1704.05871} {arXiv:1704.05871 [astro-ph.CO]}
  \BibitemShut {NoStop}%
\bibitem [{\citenamefont {Guo}\ \emph {et~al.}(2021)\citenamefont {Guo},
  \citenamefont {Sinha}, \citenamefont {Vagie},\ and\ \citenamefont
  {White}}]{Guo:2020}%
  \BibitemOpen
  \bibfield  {author} {\bibinfo {author} {\bibfnamefont {H.-K.}\ \bibnamefont
  {Guo}}, \bibinfo {author} {\bibfnamefont {K.}~\bibnamefont {Sinha}}, \bibinfo
  {author} {\bibfnamefont {D.}~\bibnamefont {Vagie}},\ and\ \bibinfo {author}
  {\bibfnamefont {G.}~\bibnamefont {White}},\ }\bibfield  {title} {\bibinfo
  {title} {{Phase Transitions in an Expanding Universe: Stochastic
  Gravitational Waves in Standard and Non-Standard Histories}},\ }\href
  {https://doi.org/10.1088/1475-7516/2021/01/001} {\bibfield  {journal}
  {\bibinfo  {journal} {JCAP}\ }\textbf {\bibinfo {volume} {01}},\ \bibinfo
  {pages} {001}},\ \Eprint {https://arxiv.org/abs/2007.08537} {arXiv:2007.08537
  [hep-ph]} \BibitemShut {NoStop}%
\bibitem [{\citenamefont {{Weir}}(2020)}]{PTPlot}%
  \BibitemOpen
  \bibfield  {author} {\bibinfo {author} {\bibfnamefont {D.~J.}\ \bibnamefont
  {{Weir}}},\ }\href@noop {} {\bibinfo {title} {Ptplot: a tool for exploring
  the gravitational wave power spectrum from first-order phase transitions}}
  (\bibinfo {year} {2020})\BibitemShut {NoStop}%
\bibitem [{\citenamefont {Aghanim}\ \emph {et~al.}(2020)\citenamefont {Aghanim}
  \emph {et~al.}}]{Planck2018}%
  \BibitemOpen
  \bibfield  {author} {\bibinfo {author} {\bibfnamefont {N.}~\bibnamefont
  {Aghanim}} \emph {et~al.} (\bibinfo {collaboration} {Planck}),\ }\bibfield
  {title} {\bibinfo {title} {{Planck 2018 results. VI. Cosmological
  parameters}},\ }\href {https://doi.org/10.1051/0004-6361/201833910}
  {\bibfield  {journal} {\bibinfo  {journal} {Astron. Astrophys.}\ }\textbf
  {\bibinfo {volume} {641}},\ \bibinfo {pages} {A6} (\bibinfo {year} {2020})},\
  \Eprint {https://arxiv.org/abs/1807.06209} {arXiv:1807.06209 [astro-ph.CO]}
  \BibitemShut {NoStop}%
\bibitem [{\citenamefont {Ellis}\ \emph {et~al.}(2020)\citenamefont {Ellis},
  \citenamefont {Lewicki},\ and\ \citenamefont {Vaskonen}}]{Ellis:2020}%
  \BibitemOpen
  \bibfield  {author} {\bibinfo {author} {\bibfnamefont {J.}~\bibnamefont
  {Ellis}}, \bibinfo {author} {\bibfnamefont {M.}~\bibnamefont {Lewicki}},\
  and\ \bibinfo {author} {\bibfnamefont {V.}~\bibnamefont {Vaskonen}},\
  }\bibfield  {title} {\bibinfo {title} {{Updated predictions for gravitational
  waves produced in a strongly supercooled phase transition}},\ }\href
  {https://doi.org/10.1088/1475-7516/2020/11/020} {\bibfield  {journal}
  {\bibinfo  {journal} {JCAP}\ }\textbf {\bibinfo {volume} {11}},\ \bibinfo
  {pages} {020}},\ \Eprint {https://arxiv.org/abs/2007.15586} {arXiv:2007.15586
  [astro-ph.CO]} \BibitemShut {NoStop}%
\bibitem [{\citenamefont {{Hindmarsh}}\ \emph {et~al.}(2015)\citenamefont
  {{Hindmarsh}}, \citenamefont {{Huber}}, \citenamefont {{Rummukainen}},\ and\
  \citenamefont {{Weir}}}]{Hindmarsh:2015}%
  \BibitemOpen
  \bibfield  {author} {\bibinfo {author} {\bibfnamefont {M.}~\bibnamefont
  {{Hindmarsh}}}, \bibinfo {author} {\bibfnamefont {S.~J.}\ \bibnamefont
  {{Huber}}}, \bibinfo {author} {\bibfnamefont {K.}~\bibnamefont
  {{Rummukainen}}},\ and\ \bibinfo {author} {\bibfnamefont {D.~J.}\
  \bibnamefont {{Weir}}},\ }\bibfield  {title} {\bibinfo {title} {{Numerical
  simulations of acoustically generated gravitational waves at a first order
  phase transition}},\ }\href {https://doi.org/10.1103/PhysRevD.92.123009}
  {\bibfield  {journal} {\bibinfo  {journal} {\prd}\ }\textbf {\bibinfo
  {volume} {92}},\ \bibinfo {eid} {123009} (\bibinfo {year} {2015})},\ \Eprint
  {https://arxiv.org/abs/1504.03291} {arXiv:1504.03291 [astro-ph.CO]}
  \BibitemShut {NoStop}%
\bibitem [{\citenamefont {{Ellis}}\ \emph {et~al.}(2019)\citenamefont
  {{Ellis}}, \citenamefont {{Lewicki}},\ and\ \citenamefont
  {{No}}}]{Ellis:2019}%
  \BibitemOpen
  \bibfield  {author} {\bibinfo {author} {\bibfnamefont {J.}~\bibnamefont
  {{Ellis}}}, \bibinfo {author} {\bibfnamefont {M.}~\bibnamefont {{Lewicki}}},\
  and\ \bibinfo {author} {\bibfnamefont {J.~M.}\ \bibnamefont {{No}}},\
  }\bibfield  {title} {\bibinfo {title} {{On the maximal strength of a
  first-order electroweak phase transition and its gravitational wave
  signal}},\ }\href {https://doi.org/10.1088/1475-7516/2019/04/003} {\bibfield
  {journal} {\bibinfo  {journal} {\jcap}\ }\textbf {\bibinfo {volume} {2019}},\
  \bibinfo {eid} {003} (\bibinfo {year} {2019})},\ \Eprint
  {https://arxiv.org/abs/1809.08242} {arXiv:1809.08242 [hep-ph]} \BibitemShut
  {NoStop}%
\bibitem [{\citenamefont {{Ellis}}\ \emph {et~al.}(2020)\citenamefont
  {{Ellis}}, \citenamefont {{Lewicki}},\ and\ \citenamefont
  {{No}}}]{Ellis:2020a}%
  \BibitemOpen
  \bibfield  {author} {\bibinfo {author} {\bibfnamefont {J.}~\bibnamefont
  {{Ellis}}}, \bibinfo {author} {\bibfnamefont {M.}~\bibnamefont {{Lewicki}}},\
  and\ \bibinfo {author} {\bibfnamefont {J.~M.}\ \bibnamefont {{No}}},\
  }\bibfield  {title} {\bibinfo {title} {{Gravitational waves from first-order
  cosmological phase transitions: lifetime of the sound wave source}},\ }\href
  {https://doi.org/10.1088/1475-7516/2020/07/050} {\bibfield  {journal}
  {\bibinfo  {journal} {\jcap}\ }\textbf {\bibinfo {volume} {2020}},\ \bibinfo
  {eid} {050} (\bibinfo {year} {2020})},\ \Eprint
  {https://arxiv.org/abs/2003.07360} {arXiv:2003.07360 [hep-ph]} \BibitemShut
  {NoStop}%
\bibitem [{\citenamefont {Agashe}\ \emph {et~al.}(2021)\citenamefont {Agashe},
  \citenamefont {Du}, \citenamefont {Ekhterachian}, \citenamefont {Kumar},\
  and\ \citenamefont {Sundrum}}]{Agashe:2020lfz}%
  \BibitemOpen
  \bibfield  {author} {\bibinfo {author} {\bibfnamefont {K.}~\bibnamefont
  {Agashe}}, \bibinfo {author} {\bibfnamefont {P.}~\bibnamefont {Du}}, \bibinfo
  {author} {\bibfnamefont {M.}~\bibnamefont {Ekhterachian}}, \bibinfo {author}
  {\bibfnamefont {S.}~\bibnamefont {Kumar}},\ and\ \bibinfo {author}
  {\bibfnamefont {R.}~\bibnamefont {Sundrum}},\ }\bibfield  {title} {\bibinfo
  {title} {{Phase Transitions from the Fifth Dimension}},\ }\href
  {https://doi.org/10.1007/JHEP02(2021)051} {\bibfield  {journal} {\bibinfo
  {journal} {JHEP}\ }\textbf {\bibinfo {volume} {02}},\ \bibinfo {pages}
  {051}},\ \Eprint {https://arxiv.org/abs/2010.04083} {arXiv:2010.04083
  [hep-th]} \BibitemShut {NoStop}%
\bibitem [{\citenamefont {Craig}(2009)}]{Craig:2009zx}%
  \BibitemOpen
  \bibfield  {author} {\bibinfo {author} {\bibfnamefont {N.~J.}\ \bibnamefont
  {Craig}},\ }\bibfield  {title} {\bibinfo {title} {{Gravitational Waves from
  Supersymmetry Breaking}},\ }\href@noop {} {\  (\bibinfo {year} {2009})},\
  \Eprint {https://arxiv.org/abs/0902.1990} {arXiv:0902.1990 [hep-ph]}
  \BibitemShut {NoStop}%
\bibitem [{\citenamefont {{Finn}}\ and\ \citenamefont
  {{Thorne}}(2000)}]{Finn:2000}%
  \BibitemOpen
  \bibfield  {author} {\bibinfo {author} {\bibfnamefont {L.~S.}\ \bibnamefont
  {{Finn}}}\ and\ \bibinfo {author} {\bibfnamefont {K.~S.}\ \bibnamefont
  {{Thorne}}},\ }\bibfield  {title} {\bibinfo {title} {{Gravitational waves
  from a compact star in a circular, inspiral orbit, in the equatorial plane of
  a massive, spinning black hole, as observed by LISA}},\ }\href
  {https://doi.org/10.1103/PhysRevD.62.124021} {\bibfield  {journal} {\bibinfo
  {journal} {\prd}\ }\textbf {\bibinfo {volume} {62}},\ \bibinfo {eid} {124021}
  (\bibinfo {year} {2000})},\ \Eprint {https://arxiv.org/abs/gr-qc/0007074}
  {arXiv:gr-qc/0007074 [gr-qc]} \BibitemShut {NoStop}%
\bibitem [{\citenamefont {{Cholis}}(2017)}]{Cholis:2016}%
  \BibitemOpen
  \bibfield  {author} {\bibinfo {author} {\bibfnamefont {I.}~\bibnamefont
  {{Cholis}}},\ }\bibfield  {title} {\bibinfo {title} {{On the gravitational
  wave background from black hole binaries after the first LIGO detections}},\
  }\href {https://doi.org/10.1088/1475-7516/2017/06/037} {\bibfield  {journal}
  {\bibinfo  {journal} {\jcap}\ }\textbf {\bibinfo {volume} {2017}},\ \bibinfo
  {eid} {037} (\bibinfo {year} {2017})},\ \Eprint
  {https://arxiv.org/abs/1609.03565} {arXiv:1609.03565 [astro-ph.HE]}
  \BibitemShut {NoStop}%
\bibitem [{\citenamefont {Abbott}\ \emph {et~al.}(2019)\citenamefont {Abbott}
  \emph {et~al.}}]{LIGOSGWB}%
  \BibitemOpen
  \bibfield  {author} {\bibinfo {author} {\bibfnamefont {B.~P.}\ \bibnamefont
  {Abbott}} \emph {et~al.} (\bibinfo {collaboration} {LIGO Scientific,
  Virgo}),\ }\bibfield  {title} {\bibinfo {title} {{Search for the isotropic
  stochastic background using data from Advanced LIGO\textquoteright{}s second
  observing run}},\ }\href {https://doi.org/10.1103/PhysRevD.100.061101}
  {\bibfield  {journal} {\bibinfo  {journal} {Phys. Rev. D}\ }\textbf {\bibinfo
  {volume} {100}},\ \bibinfo {pages} {061101} (\bibinfo {year} {2019})},\
  \Eprint {https://arxiv.org/abs/1903.02886} {arXiv:1903.02886 [gr-qc]}
  \BibitemShut {NoStop}%
\bibitem [{\citenamefont {Mandic}\ \emph {et~al.}(2016)\citenamefont {Mandic},
  \citenamefont {Bird},\ and\ \citenamefont {Cholis}}]{Mandic:2016lcn}%
  \BibitemOpen
  \bibfield  {author} {\bibinfo {author} {\bibfnamefont {V.}~\bibnamefont
  {Mandic}}, \bibinfo {author} {\bibfnamefont {S.}~\bibnamefont {Bird}},\ and\
  \bibinfo {author} {\bibfnamefont {I.}~\bibnamefont {Cholis}},\ }\bibfield
  {title} {\bibinfo {title} {{Stochastic Gravitational-Wave Background due to
  Primordial Binary Black Hole Mergers}},\ }\href
  {https://doi.org/10.1103/PhysRevLett.117.201102} {\bibfield  {journal}
  {\bibinfo  {journal} {Phys. Rev. Lett.}\ }\textbf {\bibinfo {volume} {117}},\
  \bibinfo {pages} {201102} (\bibinfo {year} {2016})},\ \Eprint
  {https://arxiv.org/abs/1608.06699} {arXiv:1608.06699 [astro-ph.CO]}
  \BibitemShut {NoStop}%
\bibitem [{\citenamefont {Abbott}\ \emph
  {et~al.}(2020{\natexlab{a}})\citenamefont {Abbott} \emph
  {et~al.}}]{LIGOO2pop}%
  \BibitemOpen
  \bibfield  {author} {\bibinfo {author} {\bibfnamefont {R.}~\bibnamefont
  {Abbott}} \emph {et~al.} (\bibinfo {collaboration} {LIGO Scientific,
  Virgo}),\ }\bibfield  {title} {\bibinfo {title} {{Population Properties of
  Compact Objects from the Second LIGO-Virgo Gravitational-Wave Transient
  Catalog}},\ }\href@noop {} {\  (\bibinfo {year} {2020}{\natexlab{a}})},\
  \Eprint {https://arxiv.org/abs/2010.14533} {arXiv:2010.14533 [astro-ph.HE]}
  \BibitemShut {NoStop}%
\bibitem [{\citenamefont {Ajith}\ \emph {et~al.}(2008)\citenamefont {Ajith},
  \citenamefont {Babak}, \citenamefont {Chen}, \citenamefont {Hewitson},
  \citenamefont {Krishnan}, \citenamefont {Sintes} \emph
  {et~al.}}]{Ajith:2008}%
  \BibitemOpen
  \bibfield  {author} {\bibinfo {author} {\bibfnamefont {P.}~\bibnamefont
  {Ajith}}, \bibinfo {author} {\bibfnamefont {S.}~\bibnamefont {Babak}},
  \bibinfo {author} {\bibfnamefont {Y.}~\bibnamefont {Chen}}, \bibinfo {author}
  {\bibfnamefont {M.}~\bibnamefont {Hewitson}}, \bibinfo {author}
  {\bibfnamefont {B.}~\bibnamefont {Krishnan}}, \bibinfo {author}
  {\bibfnamefont {A.}~\bibnamefont {Sintes}}, \emph {et~al.},\ }\bibfield
  {title} {\bibinfo {title} {{A Template bank for gravitational waveforms from
  coalescing binary black holes. I. Non-spinning binaries}},\ }\href
  {https://doi.org/10.1103/PhysRevD.77.104017} {\bibfield  {journal} {\bibinfo
  {journal} {Phys. Rev. D}\ }\textbf {\bibinfo {volume} {77}},\ \bibinfo
  {pages} {104017} (\bibinfo {year} {2008})},\ \bibinfo {note} {[Erratum:
  Phys.Rev.D 79, 129901 (2009)]},\ \Eprint {https://arxiv.org/abs/0710.2335}
  {arXiv:0710.2335 [gr-qc]} \BibitemShut {NoStop}%
\bibitem [{\citenamefont {{Amaro-Seoane}}\ \emph {et~al.}(2007)\citenamefont
  {{Amaro-Seoane}} \emph {et~al.}}]{Amaro:2007}%
  \BibitemOpen
  \bibfield  {author} {\bibinfo {author} {\bibfnamefont {P.}~\bibnamefont
  {{Amaro-Seoane}}} \emph {et~al.},\ }\bibfield  {title} {\bibinfo {title}
  {{TOPICAL REVIEW: Intermediate and extreme mass-ratio
  inspirals{\textemdash}astrophysics, science applications and detection using
  LISA}},\ }\href {https://doi.org/10.1088/0264-9381/24/17/R01} {\bibfield
  {journal} {\bibinfo  {journal} {Classical and Quantum Gravity}\ }\textbf
  {\bibinfo {volume} {24}},\ \bibinfo {pages} {R113} (\bibinfo {year}
  {2007})},\ \Eprint {https://arxiv.org/abs/astro-ph/0703495}
  {arXiv:astro-ph/0703495 [astro-ph]} \BibitemShut {NoStop}%
\bibitem [{\citenamefont {{Babak}}\ \emph {et~al.}(2017)\citenamefont {{Babak}}
  \emph {et~al.}}]{Babak:2017}%
  \BibitemOpen
  \bibfield  {author} {\bibinfo {author} {\bibfnamefont {S.}~\bibnamefont
  {{Babak}}} \emph {et~al.},\ }\bibfield  {title} {\bibinfo {title} {{Science
  with the space-based interferometer LISA. V. Extreme mass-ratio inspirals}},\
  }\href {https://doi.org/10.1103/PhysRevD.95.103012} {\bibfield  {journal}
  {\bibinfo  {journal} {\prd}\ }\textbf {\bibinfo {volume} {95}},\ \bibinfo
  {eid} {103012} (\bibinfo {year} {2017})},\ \Eprint
  {https://arxiv.org/abs/1703.09722} {arXiv:1703.09722 [gr-qc]} \BibitemShut
  {NoStop}%
\bibitem [{\citenamefont {{Amaro-Seoane}}(2018{\natexlab{a}})}]{Amaro:2018}%
  \BibitemOpen
  \bibfield  {author} {\bibinfo {author} {\bibfnamefont {P.}~\bibnamefont
  {{Amaro-Seoane}}},\ }\bibfield  {title} {\bibinfo {title} {{Relativistic
  dynamics and extreme mass ratio inspirals}},\ }\href
  {https://doi.org/10.1007/s41114-018-0013-8} {\bibfield  {journal} {\bibinfo
  {journal} {Living Reviews in Relativity}\ }\textbf {\bibinfo {volume} {21}},\
  \bibinfo {eid} {4} (\bibinfo {year} {2018}{\natexlab{a}})},\ \Eprint
  {https://arxiv.org/abs/1205.5240} {arXiv:1205.5240 [astro-ph.CO]}
  \BibitemShut {NoStop}%
\bibitem [{\citenamefont {{Gebhardt}}\ \emph {et~al.}(2001)\citenamefont
  {{Gebhardt}} \emph {et~al.}}]{Gebhardt:2001}%
  \BibitemOpen
  \bibfield  {author} {\bibinfo {author} {\bibfnamefont {K.}~\bibnamefont
  {{Gebhardt}}} \emph {et~al.},\ }\bibfield  {title} {\bibinfo {title} {{M33: A
  Galaxy with No Supermassive Black Hole}},\ }\href
  {https://doi.org/10.1086/323481} {\bibfield  {journal} {\bibinfo  {journal}
  {\aj}\ }\textbf {\bibinfo {volume} {122}},\ \bibinfo {pages} {2469} (\bibinfo
  {year} {2001})},\ \Eprint {https://arxiv.org/abs/astro-ph/0107135}
  {arXiv:astro-ph/0107135 [astro-ph]} \BibitemShut {NoStop}%
\bibitem [{\citenamefont {{Gair}}\ \emph {et~al.}(2004)\citenamefont {{Gair}}
  \emph {et~al.}}]{Gair:2004}%
  \BibitemOpen
  \bibfield  {author} {\bibinfo {author} {\bibfnamefont {J.~R.}\ \bibnamefont
  {{Gair}}} \emph {et~al.},\ }\bibfield  {title} {\bibinfo {title} {{Event rate
  estimates for LISA extreme mass ratio capture sources}},\ }\href
  {https://doi.org/10.1088/0264-9381/21/20/003} {\bibfield  {journal} {\bibinfo
   {journal} {Classical and Quantum Gravity}\ }\textbf {\bibinfo {volume}
  {21}},\ \bibinfo {pages} {S1595} (\bibinfo {year} {2004})},\ \Eprint
  {https://arxiv.org/abs/gr-qc/0405137} {arXiv:gr-qc/0405137 [gr-qc]}
  \BibitemShut {NoStop}%
\bibitem [{\citenamefont {{Babak}}\ \emph {et~al.}(2010)\citenamefont {{Babak}}
  \emph {et~al.}}]{Babak:2010}%
  \BibitemOpen
  \bibfield  {author} {\bibinfo {author} {\bibfnamefont {S.}~\bibnamefont
  {{Babak}}} \emph {et~al.},\ }\bibfield  {title} {\bibinfo {title} {{The Mock
  LISA Data Challenges: from challenge 3 to challenge 4}},\ }\href
  {https://doi.org/10.1088/0264-9381/27/8/084009} {\bibfield  {journal}
  {\bibinfo  {journal} {Classical and Quantum Gravity}\ }\textbf {\bibinfo
  {volume} {27}},\ \bibinfo {eid} {084009} (\bibinfo {year} {2010})},\ \Eprint
  {https://arxiv.org/abs/0912.0548} {arXiv:0912.0548 [gr-qc]} \BibitemShut
  {NoStop}%
\bibitem [{\citenamefont {Bonetti}\ and\ \citenamefont
  {Sesana}(2020)}]{Bonetti:2020}%
  \BibitemOpen
  \bibfield  {author} {\bibinfo {author} {\bibfnamefont {M.}~\bibnamefont
  {Bonetti}}\ and\ \bibinfo {author} {\bibfnamefont {A.}~\bibnamefont
  {Sesana}},\ }\bibfield  {title} {\bibinfo {title} {{Gravitational wave
  background from extreme mass ratio inspirals}},\ }\href
  {https://doi.org/10.1103/PhysRevD.102.103023} {\bibfield  {journal} {\bibinfo
   {journal} {Phys. Rev. D}\ }\textbf {\bibinfo {volume} {102}},\ \bibinfo
  {pages} {103023} (\bibinfo {year} {2020})},\ \Eprint
  {https://arxiv.org/abs/2007.14403} {arXiv:2007.14403 [astro-ph.GA]}
  \BibitemShut {NoStop}%
\bibitem [{\citenamefont {{Amaro-Seoane}}\ and\ \citenamefont
  {{Preto}}(2011)}]{Amaro:2011}%
  \BibitemOpen
  \bibfield  {author} {\bibinfo {author} {\bibfnamefont {P.}~\bibnamefont
  {{Amaro-Seoane}}}\ and\ \bibinfo {author} {\bibfnamefont {M.}~\bibnamefont
  {{Preto}}},\ }\bibfield  {title} {\bibinfo {title} {{The impact of realistic
  models of mass segregation on the event rate of extreme-mass ratio inspirals
  and cusp re-growth}},\ }\href {https://doi.org/10.1088/0264-9381/28/9/094017}
  {\bibfield  {journal} {\bibinfo  {journal} {Classical and Quantum Gravity}\
  }\textbf {\bibinfo {volume} {28}},\ \bibinfo {eid} {094017} (\bibinfo {year}
  {2011})},\ \Eprint {https://arxiv.org/abs/1010.5781} {arXiv:1010.5781
  [astro-ph.CO]} \BibitemShut {NoStop}%
\bibitem [{\citenamefont {{Salcido}}\ \emph {et~al.}(2016)\citenamefont
  {{Salcido}} \emph {et~al.}}]{Salcido:2016}%
  \BibitemOpen
  \bibfield  {author} {\bibinfo {author} {\bibfnamefont {J.}~\bibnamefont
  {{Salcido}}} \emph {et~al.},\ }\bibfield  {title} {\bibinfo {title} {{Music
  from the heavens - gravitational waves from supermassive black hole mergers
  in the EAGLE simulations}},\ }\href {https://doi.org/10.1093/mnras/stw2048}
  {\bibfield  {journal} {\bibinfo  {journal} {\mnras}\ }\textbf {\bibinfo
  {volume} {463}},\ \bibinfo {pages} {870} (\bibinfo {year} {2016})},\ \Eprint
  {https://arxiv.org/abs/1601.06156} {arXiv:1601.06156 [astro-ph.GA]}
  \BibitemShut {NoStop}%
\bibitem [{\citenamefont {{Ebisuzaki}}\ \emph {et~al.}(2001)\citenamefont
  {{Ebisuzaki}} \emph {et~al.}}]{Ebisuzaki:2001}%
  \BibitemOpen
  \bibfield  {author} {\bibinfo {author} {\bibfnamefont {T.}~\bibnamefont
  {{Ebisuzaki}}} \emph {et~al.},\ }\bibfield  {title} {\bibinfo {title}
  {{Missing Link Found? The ``Runaway'' Path to Supermassive Black Holes}},\
  }\href {https://doi.org/10.1086/338118} {\bibfield  {journal} {\bibinfo
  {journal} {\apjl}\ }\textbf {\bibinfo {volume} {562}},\ \bibinfo {pages}
  {L19} (\bibinfo {year} {2001})},\ \Eprint
  {https://arxiv.org/abs/astro-ph/0106252} {arXiv:astro-ph/0106252 [astro-ph]}
  \BibitemShut {NoStop}%
\bibitem [{\citenamefont {{Miller}}(2005)}]{Miller:2005}%
  \BibitemOpen
  \bibfield  {author} {\bibinfo {author} {\bibfnamefont {M.~C.}\ \bibnamefont
  {{Miller}}},\ }\bibfield  {title} {\bibinfo {title} {{Probing General
  Relativity with Mergers of Supermassive and Intermediate-Mass Black Holes}},\
  }\href {https://doi.org/10.1086/425910} {\bibfield  {journal} {\bibinfo
  {journal} {\apj}\ }\textbf {\bibinfo {volume} {618}},\ \bibinfo {pages} {426}
  (\bibinfo {year} {2005})},\ \Eprint {https://arxiv.org/abs/astro-ph/0409331}
  {arXiv:astro-ph/0409331 [astro-ph]} \BibitemShut {NoStop}%
\bibitem [{\citenamefont {Abbott}\ \emph
  {et~al.}(2020{\natexlab{b}})\citenamefont {Abbott} \emph
  {et~al.}}]{GW190521}%
  \BibitemOpen
  \bibfield  {author} {\bibinfo {author} {\bibfnamefont {R.}~\bibnamefont
  {Abbott}} \emph {et~al.} (\bibinfo {collaboration} {LIGO Scientific,
  Virgo}),\ }\bibfield  {title} {\bibinfo {title} {{GW190521: A Binary Black
  Hole Merger with a Total Mass of 150\,\,M\ensuremath{\odot}}},\ }\href
  {https://doi.org/10.1103/PhysRevLett.125.101102} {\bibfield  {journal}
  {\bibinfo  {journal} {Phys. Rev. Lett.}\ }\textbf {\bibinfo {volume} {125}},\
  \bibinfo {pages} {101102} (\bibinfo {year} {2020}{\natexlab{b}})},\ \Eprint
  {https://arxiv.org/abs/2009.01075} {arXiv:2009.01075 [gr-qc]} \BibitemShut
  {NoStop}%
\bibitem [{\citenamefont {Ezquiaga}\ and\ \citenamefont
  {Holz}(2021)}]{Ezquiaga:2020tns}%
  \BibitemOpen
  \bibfield  {author} {\bibinfo {author} {\bibfnamefont {J.~M.}\ \bibnamefont
  {Ezquiaga}}\ and\ \bibinfo {author} {\bibfnamefont {D.~E.}\ \bibnamefont
  {Holz}},\ }\bibfield  {title} {\bibinfo {title} {{Jumping the Gap: Searching
  for LIGO\textquoteright{}s Biggest Black Holes}},\ }\href
  {https://doi.org/10.3847/2041-8213/abe638} {\bibfield  {journal} {\bibinfo
  {journal} {Astrophys. J. Lett.}\ }\textbf {\bibinfo {volume} {909}},\
  \bibinfo {pages} {L23} (\bibinfo {year} {2021})},\ \Eprint
  {https://arxiv.org/abs/2006.02211} {arXiv:2006.02211 [astro-ph.HE]}
  \BibitemShut {NoStop}%
\bibitem [{\citenamefont
  {{Amaro-Seoane}}(2018{\natexlab{b}})}]{AmaroSeoaneIMRI}%
  \BibitemOpen
  \bibfield  {author} {\bibinfo {author} {\bibfnamefont {P.}~\bibnamefont
  {{Amaro-Seoane}}},\ }\bibfield  {title} {\bibinfo {title} {{Detecting
  intermediate-mass ratio inspirals from the ground and space}},\ }\href
  {https://doi.org/10.1103/PhysRevD.98.063018} {\bibfield  {journal} {\bibinfo
  {journal} {\prd}\ }\textbf {\bibinfo {volume} {98}},\ \bibinfo {eid} {063018}
  (\bibinfo {year} {2018}{\natexlab{b}})},\ \Eprint
  {https://arxiv.org/abs/1807.03824} {arXiv:1807.03824 [astro-ph.HE]}
  \BibitemShut {NoStop}%
\bibitem [{\citenamefont {{Bender}}\ and\ \citenamefont
  {{Hils}}(1997)}]{Bender:1997}%
  \BibitemOpen
  \bibfield  {author} {\bibinfo {author} {\bibfnamefont {P.~L.}\ \bibnamefont
  {{Bender}}}\ and\ \bibinfo {author} {\bibfnamefont {D.}~\bibnamefont
  {{Hils}}},\ }\bibfield  {title} {\bibinfo {title} {{Confusion noise level due
  to galactic and extragalactic binaries}},\ }\href
  {https://doi.org/10.1088/0264-9381/14/6/008} {\bibfield  {journal} {\bibinfo
  {journal} {Classical and Quantum Gravity}\ }\textbf {\bibinfo {volume}
  {14}},\ \bibinfo {pages} {1439} (\bibinfo {year} {1997})}\BibitemShut
  {NoStop}%
\bibitem [{\citenamefont {{Thrane}}\ \emph {et~al.}(2009)\citenamefont
  {{Thrane}}, \citenamefont {{Ballmer}}, \citenamefont {{Romano}},
  \citenamefont {{Mitra}}, \citenamefont {{Talukder}}, \citenamefont {{Bose}},\
  and\ \citenamefont {{Mandic}}}]{Thrane:2009}%
  \BibitemOpen
  \bibfield  {author} {\bibinfo {author} {\bibfnamefont {E.}~\bibnamefont
  {{Thrane}}}, \bibinfo {author} {\bibfnamefont {S.}~\bibnamefont {{Ballmer}}},
  \bibinfo {author} {\bibfnamefont {J.~D.}\ \bibnamefont {{Romano}}}, \bibinfo
  {author} {\bibfnamefont {S.}~\bibnamefont {{Mitra}}}, \bibinfo {author}
  {\bibfnamefont {D.}~\bibnamefont {{Talukder}}}, \bibinfo {author}
  {\bibfnamefont {S.}~\bibnamefont {{Bose}}},\ and\ \bibinfo {author}
  {\bibfnamefont {V.}~\bibnamefont {{Mandic}}},\ }\bibfield  {title} {\bibinfo
  {title} {{Probing the anisotropies of a stochastic gravitational-wave
  background using a network of ground-based laser interferometers}},\ }\href
  {https://doi.org/10.1103/PhysRevD.80.122002} {\bibfield  {journal} {\bibinfo
  {journal} {\prd}\ }\textbf {\bibinfo {volume} {80}},\ \bibinfo {eid} {122002}
  (\bibinfo {year} {2009})},\ \Eprint {https://arxiv.org/abs/0910.0858}
  {arXiv:0910.0858 [astro-ph.IM]} \BibitemShut {NoStop}%
\bibitem [{\citenamefont {{Adams}}\ and\ \citenamefont
  {{Cornish}}(2014)}]{Adams:2014}%
  \BibitemOpen
  \bibfield  {author} {\bibinfo {author} {\bibfnamefont {M.~R.}\ \bibnamefont
  {{Adams}}}\ and\ \bibinfo {author} {\bibfnamefont {N.~J.}\ \bibnamefont
  {{Cornish}}},\ }\bibfield  {title} {\bibinfo {title} {{Detecting a stochastic
  gravitational wave background in the presence of a galactic foreground and
  instrument noise}},\ }\href {https://doi.org/10.1103/PhysRevD.89.022001}
  {\bibfield  {journal} {\bibinfo  {journal} {\prd}\ }\textbf {\bibinfo
  {volume} {89}},\ \bibinfo {eid} {022001} (\bibinfo {year} {2014})},\ \Eprint
  {https://arxiv.org/abs/1307.4116} {arXiv:1307.4116 [gr-qc]} \BibitemShut
  {NoStop}%
\bibitem [{\citenamefont {Press}\ and\ \citenamefont
  {Thorne}(1972)}]{Press:1972}%
  \BibitemOpen
  \bibfield  {author} {\bibinfo {author} {\bibfnamefont {W.~H.}\ \bibnamefont
  {Press}}\ and\ \bibinfo {author} {\bibfnamefont {K.~S.}\ \bibnamefont
  {Thorne}},\ }\bibfield  {title} {\bibinfo {title} {Gravitational-wave
  astronomy},\ }\href {https://doi.org/10.1146/annurev.aa.10.090172.002003}
  {\bibfield  {journal} {\bibinfo  {journal} {Annual Review of Astronomy and
  Astrophysics}\ }\textbf {\bibinfo {volume} {10}},\ \bibinfo {pages} {335}
  (\bibinfo {year} {1972})},\ \Eprint
  {https://arxiv.org/abs/https://doi.org/10.1146/annurev.aa.10.090172.002003}
  {https://doi.org/10.1146/annurev.aa.10.090172.002003} \BibitemShut {NoStop}%
\bibitem [{\citenamefont {Riles}(2013)}]{Riles:2012yw}%
  \BibitemOpen
  \bibfield  {author} {\bibinfo {author} {\bibfnamefont {K.}~\bibnamefont
  {Riles}},\ }\bibfield  {title} {\bibinfo {title} {{Gravitational Waves:
  Sources, Detectors and Searches}},\ }\href
  {https://doi.org/10.1016/j.ppnp.2012.08.001} {\bibfield  {journal} {\bibinfo
  {journal} {Prog. Part. Nucl. Phys.}\ }\textbf {\bibinfo {volume} {68}},\
  \bibinfo {pages} {1} (\bibinfo {year} {2013})},\ \Eprint
  {https://arxiv.org/abs/1209.0667} {arXiv:1209.0667 [hep-ex]} \BibitemShut
  {NoStop}%
\bibitem [{\citenamefont {Marassi}\ \emph {et~al.}(2011)\citenamefont
  {Marassi}, \citenamefont {Ciolfi}, \citenamefont {Schneider}, \citenamefont
  {Stella},\ and\ \citenamefont {Ferrari}}]{Marassi:2010wj}%
  \BibitemOpen
  \bibfield  {author} {\bibinfo {author} {\bibfnamefont {S.}~\bibnamefont
  {Marassi}}, \bibinfo {author} {\bibfnamefont {R.}~\bibnamefont {Ciolfi}},
  \bibinfo {author} {\bibfnamefont {R.}~\bibnamefont {Schneider}}, \bibinfo
  {author} {\bibfnamefont {L.}~\bibnamefont {Stella}},\ and\ \bibinfo {author}
  {\bibfnamefont {V.}~\bibnamefont {Ferrari}},\ }\bibfield  {title} {\bibinfo
  {title} {{Stochastic background of gravitational waves emitted by
  magnetars}},\ }\href {https://doi.org/10.1111/j.1365-2966.2010.17861.x}
  {\bibfield  {journal} {\bibinfo  {journal} {Mon. Not. Roy. Astron. Soc.}\
  }\textbf {\bibinfo {volume} {411}},\ \bibinfo {pages} {2549} (\bibinfo {year}
  {2011})},\ \Eprint {https://arxiv.org/abs/1009.1240} {arXiv:1009.1240
  [astro-ph.CO]} \BibitemShut {NoStop}%
\bibitem [{\citenamefont {Rosado}(2012)}]{Rosado:2012bk}%
  \BibitemOpen
  \bibfield  {author} {\bibinfo {author} {\bibfnamefont {P.~A.}\ \bibnamefont
  {Rosado}},\ }\bibfield  {title} {\bibinfo {title} {{Gravitational wave
  background from rotating neutron stars}},\ }\href
  {https://doi.org/10.1103/PhysRevD.86.104007} {\bibfield  {journal} {\bibinfo
  {journal} {Phys. Rev. D}\ }\textbf {\bibinfo {volume} {86}},\ \bibinfo
  {pages} {104007} (\bibinfo {year} {2012})},\ \Eprint
  {https://arxiv.org/abs/1206.1330} {arXiv:1206.1330 [gr-qc]} \BibitemShut
  {NoStop}%
\bibitem [{\citenamefont {Christensen}(2019)}]{Christensen:2018iqi}%
  \BibitemOpen
  \bibfield  {author} {\bibinfo {author} {\bibfnamefont {N.}~\bibnamefont
  {Christensen}},\ }\bibfield  {title} {\bibinfo {title} {{Stochastic
  Gravitational Wave Backgrounds}},\ }\href
  {https://doi.org/10.1088/1361-6633/aae6b5} {\bibfield  {journal} {\bibinfo
  {journal} {Rept. Prog. Phys.}\ }\textbf {\bibinfo {volume} {82}},\ \bibinfo
  {pages} {016903} (\bibinfo {year} {2019})},\ \Eprint
  {https://arxiv.org/abs/1811.08797} {arXiv:1811.08797 [gr-qc]} \BibitemShut
  {NoStop}%
\bibitem [{\citenamefont {{Seitenzahl}}\ \emph {et~al.}(2015)\citenamefont
  {{Seitenzahl}}, \citenamefont {{Herzog}}, \citenamefont {{Ruiter}},
  \citenamefont {{Marquardt}}, \citenamefont {{Ohlmann}},\ and\ \citenamefont
  {{R{\"o}pke}}}]{Seitenzahl:2015}%
  \BibitemOpen
  \bibfield  {author} {\bibinfo {author} {\bibfnamefont {I.~R.}\ \bibnamefont
  {{Seitenzahl}}}, \bibinfo {author} {\bibfnamefont {M.}~\bibnamefont
  {{Herzog}}}, \bibinfo {author} {\bibfnamefont {A.~J.}\ \bibnamefont
  {{Ruiter}}}, \bibinfo {author} {\bibfnamefont {K.}~\bibnamefont
  {{Marquardt}}}, \bibinfo {author} {\bibfnamefont {S.~T.}\ \bibnamefont
  {{Ohlmann}}},\ and\ \bibinfo {author} {\bibfnamefont {F.~K.}\ \bibnamefont
  {{R{\"o}pke}}},\ }\bibfield  {title} {\bibinfo {title} {{Neutrino and
  gravitational wave signal of a delayed-detonation model of type Ia
  supernovae}},\ }\href {https://doi.org/10.1103/PhysRevD.92.124013} {\bibfield
   {journal} {\bibinfo  {journal} {\prd}\ }\textbf {\bibinfo {volume} {92}},\
  \bibinfo {eid} {124013} (\bibinfo {year} {2015})},\ \Eprint
  {https://arxiv.org/abs/1511.02542} {arXiv:1511.02542 [astro-ph.SR]}
  \BibitemShut {NoStop}%
\bibitem [{\citenamefont {{Bonaparte}}\ \emph {et~al.}(2013)\citenamefont
  {{Bonaparte}}, \citenamefont {{Matteucci}}, \citenamefont {{Recchi}},
  \citenamefont {{Spitoni}}, \citenamefont {{Pipino}},\ and\ \citenamefont
  {{Grieco}}}]{Bonaparte:2013}%
  \BibitemOpen
  \bibfield  {author} {\bibinfo {author} {\bibfnamefont {I.}~\bibnamefont
  {{Bonaparte}}}, \bibinfo {author} {\bibfnamefont {F.}~\bibnamefont
  {{Matteucci}}}, \bibinfo {author} {\bibfnamefont {S.}~\bibnamefont
  {{Recchi}}}, \bibinfo {author} {\bibfnamefont {E.}~\bibnamefont {{Spitoni}}},
  \bibinfo {author} {\bibfnamefont {A.}~\bibnamefont {{Pipino}}},\ and\
  \bibinfo {author} {\bibfnamefont {V.}~\bibnamefont {{Grieco}}},\ }\bibfield
  {title} {\bibinfo {title} {{Galactic and cosmic Type Ia supernova (SNIa)
  rates: is it possible to impose constraints on SNIa progenitors?}},\ }\href
  {https://doi.org/10.1093/mnras/stt1457} {\bibfield  {journal} {\bibinfo
  {journal} {\mnras}\ }\textbf {\bibinfo {volume} {435}},\ \bibinfo {pages}
  {2460} (\bibinfo {year} {2013})},\ \Eprint {https://arxiv.org/abs/1308.0137}
  {arXiv:1308.0137 [astro-ph.CO]} \BibitemShut {NoStop}%
\bibitem [{\citenamefont {{Foreman-Mackey}}\ \emph {et~al.}(2013)\citenamefont
  {{Foreman-Mackey}}, \citenamefont {{Hogg}}, \citenamefont {{Lang}},\ and\
  \citenamefont {{Goodman}}}]{ForemanMackey:2013}%
  \BibitemOpen
  \bibfield  {author} {\bibinfo {author} {\bibfnamefont {D.}~\bibnamefont
  {{Foreman-Mackey}}}, \bibinfo {author} {\bibfnamefont {D.~W.}\ \bibnamefont
  {{Hogg}}}, \bibinfo {author} {\bibfnamefont {D.}~\bibnamefont {{Lang}}},\
  and\ \bibinfo {author} {\bibfnamefont {J.}~\bibnamefont {{Goodman}}},\
  }\bibfield  {title} {\bibinfo {title} {{emcee: The MCMC Hammer}},\ }\href
  {https://doi.org/10.1086/670067} {\bibfield  {journal} {\bibinfo  {journal}
  {\pasp}\ }\textbf {\bibinfo {volume} {125}},\ \bibinfo {pages} {306}
  (\bibinfo {year} {2013})},\ \Eprint {https://arxiv.org/abs/1202.3665}
  {arXiv:1202.3665 [astro-ph.IM]} \BibitemShut {NoStop}%
\bibitem [{\citenamefont {Adamson}\ \emph {et~al.}(2018)\citenamefont {Adamson}
  \emph {et~al.}}]{Adamson:2018mbw}%
  \BibitemOpen
  \bibfield  {author} {\bibinfo {author} {\bibfnamefont {P.}~\bibnamefont
  {Adamson}} \emph {et~al.},\ }\bibfield  {title} {\bibinfo {title} {{PROPOSAL:
  P-1101 Matter-wave Atomic Gradiometer Interferometric Sensor (MAGIS-100)}}\
  }\href {https://doi.org/10.2172/1605586} {10.2172/1605586} (\bibinfo {year}
  {2018})\BibitemShut {NoStop}%
\bibitem [{\citenamefont {Yagi}\ and\ \citenamefont
  {Seto}(2011)}]{Yagi:2011wg}%
  \BibitemOpen
  \bibfield  {author} {\bibinfo {author} {\bibfnamefont {K.}~\bibnamefont
  {Yagi}}\ and\ \bibinfo {author} {\bibfnamefont {N.}~\bibnamefont {Seto}},\
  }\bibfield  {title} {\bibinfo {title} {{Detector configuration of DECIGO/BBO
  and identification of cosmological neutron-star binaries}},\ }\href
  {https://doi.org/10.1103/PhysRevD.83.044011} {\bibfield  {journal} {\bibinfo
  {journal} {Phys. Rev. D}\ }\textbf {\bibinfo {volume} {83}},\ \bibinfo
  {pages} {044011} (\bibinfo {year} {2011})},\ \bibinfo {note} {[Erratum:
  Phys.Rev.D 95, 109901 (2017)]},\ \Eprint {https://arxiv.org/abs/1101.3940}
  {arXiv:1101.3940 [astro-ph.CO]} \BibitemShut {NoStop}%
\bibitem [{\citenamefont {Janssen}\ \emph {et~al.}(2015)\citenamefont {Janssen}
  \emph {et~al.}}]{Janssen:2014dka}%
  \BibitemOpen
  \bibfield  {author} {\bibinfo {author} {\bibfnamefont {G.}~\bibnamefont
  {Janssen}} \emph {et~al.},\ }\bibfield  {title} {\bibinfo {title}
  {{Gravitational wave astronomy with the SKA}},\ }\href
  {https://doi.org/10.22323/1.215.0037} {\bibfield  {journal} {\bibinfo
  {journal} {PoS}\ }\textbf {\bibinfo {volume} {AASKA14}},\ \bibinfo {pages}
  {037} (\bibinfo {year} {2015})},\ \Eprint {https://arxiv.org/abs/1501.00127}
  {arXiv:1501.00127 [astro-ph.IM]} \BibitemShut {NoStop}%
\bibitem [{\citenamefont {Sesana}\ \emph {et~al.}(2019)\citenamefont {Sesana}
  \emph {et~al.}}]{Sesana:2019vho}%
  \BibitemOpen
  \bibfield  {author} {\bibinfo {author} {\bibfnamefont {A.}~\bibnamefont
  {Sesana}} \emph {et~al.},\ }\bibfield  {title} {\bibinfo {title} {{Unveiling
  the Gravitational Universe at $\mu$-Hz Frequencies}},\ }\href@noop {} {\
  (\bibinfo {year} {2019})},\ \Eprint {https://arxiv.org/abs/1908.11391}
  {arXiv:1908.11391 [astro-ph.IM]} \BibitemShut {NoStop}%
\end{thebibliography}%

\end{document}